\documentclass[prd,reprint,a4paper,showpacs,superscriptaddress,11pt,onecolumn,nofootinbib]{revtex4-1}
\usepackage{setspace}
\usepackage[utf8]{inputenc}
\usepackage{bm}
\usepackage{graphics}
\usepackage{amsmath, amssymb, amsfonts}
\usepackage{natbib}
\usepackage{hyperref}
\usepackage{color}
\hypersetup{colorlinks=true}

\usepackage{graphicx}
\usepackage{sidecap}
\usepackage{subfigure}
\usepackage[T1]{fontenc}
\usepackage{stackrel}
\usepackage{float}
\usepackage{dsfont}

\newcommand{\be}{\begin{equation}}
\newcommand{\ee}{\end{equation}}
\newcommand{\bse}{\begin{subequations}}
\newcommand{\ese}{\end{subequations}}
\newcommand{\ba}{\begin{eqnarray}}
\newcommand{\ea}{\end{eqnarray}}

\newcommand{\bea}{\begin{eqnarray}}
\newcommand{\eea}{\end{eqnarray}}

\usepackage{array}
\usepackage{lineno}
\usepackage{hyperref}
\usepackage{cleveref}
\newcolumntype{L}[1]{>{\raggedright\let\newline\\\arraybackslash\hspace{0pt}}m{#1}}
\newcolumntype{C}[1]{>{\centering\let\newline\\\arraybackslash\hspace{0pt}}m{#1}}
\newcolumntype{R}[1]{>{\raggedleft\let\newline\\\arraybackslash\hspace{0pt}}m{#1}}
\usepackage{setspace}
\usepackage[utf8]{inputenc}
\usepackage{bm}
\usepackage{graphics}
\usepackage{amsmath, amssymb, amsfonts}
\usepackage{natbib}
\usepackage{hyperref}
\usepackage{color}
\hypersetup{colorlinks=true}

\usepackage{graphicx}
\usepackage{sidecap}
\usepackage{subfigure}
\usepackage[T1]{fontenc}
\usepackage{stackrel}
\usepackage{float}
\usepackage{dsfont}

\usepackage{array}
\usepackage{lineno}
\usepackage{hyperref}
\usepackage{cleveref}
\newcolumntype{L}[1]{>{\raggedright\let\newline\\\arraybackslash\hspace{0pt}}m{#1}}
\newcolumntype{C}[1]{>{\centering\let\newline\\\arraybackslash\hspace{0pt}}m{#1}}
\newcolumntype{R}[1]{>{\raggedleft\let\newline\\\arraybackslash\hspace{0pt}}m{#1}}
\usepackage{color}

\usepackage[normalem]{ulem}  

\begin{document}
\title{Density response and collective modes of semi-holographic non-Fermi liquids}

\author{Benoit Dou\c{c}ot}
\email{doucot@lpthe.jussieu.fr}
\affiliation{LPTHE, CNRS-Universit\'{e} Pierre et Marie Curie, Sorbonne Universit\'{e}s, 4 Place Jussieu, 75252 Paris Cedex 05, France}
\author{Christian Ecker}
\email{ecker@hep.itp.tuwien.ac.at}
\affiliation{Institut f\"{u}r Theoretische Physik, Technische Universit\"{a}t Wien, Wiedner Hauptstr.~8-10, A-1040 Vienna, Austria}
\author{Ayan Mukhopadhyay}
\email{ayan@hep.itp.tuwien.ac.at}
\affiliation{Institut f\"{u}r Theoretische Physik, Technische Universit\"{a}t Wien, Wiedner Hauptstr.~8-10, A-1040 Vienna, Austria}
\affiliation{CERN, Theoretical Physics Department, 1211 Geneva 23, Switzerland}
\author{Giuseppe Policastro}
\email{policast@lpt.ens.fr}
\affiliation{Laboratoire de physique th\'eorique, D\'epartement de physique de l'ENS, \'Ecole normale sup\'erieure, PSL Research University, Sorbonne Universit\'es, UPMC Univ. Paris 06, CNRS, 75005 Paris, France}

\begin{abstract}
Semi-holographic models of non-Fermi liquids have been shown to have generically stable generalised quasi-particles on the Fermi surface. Although these excitations are broad and exhibit particle-hole asymmetry, they were argued to be stable from interactions at the Fermi surface. In this work, we use this observation to compute the density response and collective behaviour in these systems. 

Compared to the Fermi liquid case, 
we find that the boundaries of the particle-hole continuum are blurred by incoherent contributions. 
However, there is a region inside this continuum, that we call  inner core, within which salient features of the Fermi liquid case are preserved.  
A particularly striking prediction of our work is that these systems support a plasmonic collective excitation which is well-defined at large momenta, 
has an approximately linear dispersion relation and is located in the low-energy tail of the particle-hole continuum.  

Furthermore, the dynamic screening potential shows deep attractive regions as a function of the distance at higher frequencies which might lead to long-lived pair formation depending on the behaviour of the pair susceptibility. 
We also find that Friedel oscillations are present in these systems but are highly suppressed.

\end{abstract}

\maketitle

\begin{spacing}{1.8}
\tableofcontents
\end{spacing}
\section{Introduction}

There is by now a large amount of experimental evidence for the existence of systems that have a metallic phase but manifest an anomalous behavior, in the sense that their properties are not accounted for by the Landau theory of Fermi liquids, and for this reason have been called Non-Fermi liquids \cite{stewart}. The main examples are the 1D Luttinger liquids \cite{haldane}, heavy fermion metallic systems exhibiting quantum criticality \cite{heavyferm}, and the \emph{strange metal} phase of cuprate superconductors \cite{strangemetal}. 

Such systems pose a big challenge for theorists, since it turns out to be difficult to escape the paradigm of Landau. Several models have been proposed (see e.g. \cite{varma} for a review), but they are either not fully consistent or cannot be studied in a controlled approximation, and the problem is far from being settled. 

It is commonly believed that the anomalous behaviour in many such systems is a consequence of the vicinity to a quantum critical point, and could be the result of the interplay between the fermionic degrees of freedom and the modes associated with the critical behavior. It can be argued that the mixing with  the long-range fluctuations turns the ground state into a non-Fermi liquid \cite{NFL}. 
A natural idea is then to couple an ordinary Fermi liquid to a gapless system.  
The subject of fermion systems coupled by long range interactions,
mediated either by scalar potential or transverse gauge fields,
has been intensively revisited in the 1990's, thanks to its possible connection with high temperature 
superconductivity~\cite{Baskaran88,Anderson90}, and also with the anomalous Fermi liquid in 
a strong magnetic field at density corresponding to a half-filled Landau level~\cite{Halperin93}.     
Many theoretical works have, to a large extent, confirmed the validity of the Random Phase Approximation (RPA) for a large class of systems whose
interactions are singular at small momentum transfer (a detailed review of these works is given by ~\cite{Metzner98}). A key ingredient here is the fact that the 
contributions associated to various orderings of density or current insertions attached to any fermionic loops in a Feynman graph tends to cancel, 
when more than two such insertions are present, in the limit when all momentum transfers are small~\cite{Kopietz95,Froehlich95}. 
Note that it had been known for two decades already that this cancellation is exact for the 1D Luttinger model~\cite{Dzyaloshinskii74}.

It remains to insert this RPA dressing of either the long range scalar potential interaction or the transverse gauge field propagator,
into a calculation of the fermion propagator. This rather difficult task has been addressed by various methods, including eikonal approximation~\cite{Stamp94},
Ward identities~\cite{Castellani94}, higher dimensional bosonization~\cite{Kwon95}, renormalization group approaches~\cite{Gan93,Nayak94,Lee17}, 
large and small $N$ expansions~\cite{Altshuler94,Fitzpatrick:2013rfa}.
Although some discrepancies exist at this stage, these works have confirmed the possibility to destroy the conventional Fermi liquid
fixed point in any dimension, for singular enough interactions. 

The holographic correspondence \cite{adscft} has been employed extensively in recent years as a tool to study a variety of strongly-coupled systems. 
A system has a holographic description if it can be equivalently described by a dual theory that lives in a higher dimensional space and usually contains gravity. In some very particular cases of theories with extended supersymmetry the equivalence is well established, though it cannot be rigorously proven. Its usefulness comes from the fact that it is a weak-strong coupling duality, so that perturbative calculations on the dual theory give insight into strong coupling behaviour of the system. The geometric properties of the dual space reflect dynamical properties of the system, for instance the behavior under renormalization group flow is encoded in the scale factor and one can describe systems that are scale-invariant (they correspond to an anti-de Sitter space of constant curvature), as well as systems with Lifshitz scaling, or hyperscaling violation. 
One can also easily introduce a finite temperature and density by looking at charged black hole solutions in the dual geometry. 
 In particular, the application to systems at finite density, and so with  potential interest for condensed matter systems, was pioneered in \cite{Hartnoll:2007ih} (see the reviews \cite{reviews, Hartnoll:2016apf}). One can obtain some non-trivial predictions on  the features of systems at a quantum critical point (e.g. quantum dissipation,  charge fractionalization) that can be matched to those that have been observed in real materials or in field-theoretic models  \cite{sachdev}. 
However the dual (gravitational) description typically hides the information about the microscopic degrees of freedom and gives access only to a few observables such as conserved currents or order parameters. 
Moreover, as a model building tool, holography is somewhat rigid and does not easily allow for tweaks to accommodate some features that one would like to add to a given model (it is of course, at the same time, part of the strength of the setup, in that one is arguably always obtaining consistent results, purely theoretical as they may be). 

As a remedy to these drawbacks, Faulkner and Polchinski \cite{Faulkner:2010tq} (drawing on previous works  \cite{Cubrovic:2009ye,Hartnoll:2009ns, Faulkner:2009wj,Liu:2009dm})  have proposed that for many purposes  a semi-holographic model, 
in which free fermions living on the boundary of the space are coupled with fields living in the bulk, would be sufficient to capture the low-energy physics of the fully holographic constructions, and at the same time allows for some extensions, for instance the interior geometry can be taken to be $AdS_4$ or $AdS_2 \times {\mathbf R^2}$ or a geometry with Lifshitz scaling, corresponding to coupling the fermions to different types of scale invariant field theories. In all these cases the model describe an IR fixed point that is expected to encode the universal properties of a large class of interacting fermion systems (see also \cite{Iqbal:2011ae}). 

In this class of models, the fermions are interacting with a strongly coupled sector that has a large number $N$ of degrees of freedom. This can be exploited to obtain a small expansion parameter. 
In the geometrical picture, the leading term corresponds to considering the effect of the strongly coupled sector on the fermions, but neglecting the backreaction of the fermions on the 
critical modes. 
In the model of free fermion coupled to a massless boson, the large-N expansion is known to break down \cite{Lee:2009epi}. In our case we can argue that the situation is better, so we assume that the expansion is valid, though this has not been completely proven yet.  

We have then a class of non-Fermi liquids depending on some parameters: a real number $\nu$ depending on the conformal dimension of the lowest CFT operator that couples to the fermions, which controls the deviation from FL behaviour; and a complex coupling constant $\zeta$, whose absolute value determine a cutoff scale below which the semi-holographic theory is an effective description, and whose phase determine the breaking of particle-hole symmetry. However the phase can only take values in a bounded range, and we will take it to be close to the upper limiting value as it can be argued that this will be realised in the generic case.

In a previous paper \cite{Mukhopadhyay:2013dqa} two of the present authors (AM and GP) started to explore the phenomenology of  the semi-holographic model. We attempted to generalise the framework of Landau's theory, and showed that, in analogy with the Fermi liquid case, one can describe the properties of the system in terms of the Landau parameters that essentially contain the information about four-fermion scattering at the Fermi surface. 
We also attempted to solve the generalised Landau-Silin equations in order to find collective excitations, employing a particular ansatz for the solution which  however we could not completely justify from first principles. The key point of the work \cite{Mukhopadhyay:2013dqa} is that the semi-holographic non-Fermi liquids preserve the notion of generalised quasi-particle excitations which are broad and which exhibit particle-hole asymmetry but are stable from interactions at the Fermi surface in the low-energy limit. We elaborate on this in Section \ref{FLgen}.

In the present paper we continue this exploration. 
We consider the Lindhard function, which is nothing but  the density response function, and we consider in particular the case of 2D systems (although non-Fermi liquids can exist also in other dimensions, the 2D case is the most interesting phenomenologically). From this function one obtains information about the continuum spectrum of particle-hole excitations (in the Fermi liquid language, though strictly speaking we do not have quasiparticles), and under the assumption that an RPA  resummation is valid, also about the collective modes, the screening of external charges, and the effective interaction potential. 
We compute the function explicitly by numerically performing the corresponding integrals, for several values of the parameters of the model. In Section \ref{Lindhard Function}, we discuss how to impose the energy cutoff (beyond which our semi-holographic effective theory cannot be trusted) in the one-loop integrals in a consistent way.

We summarise here the main results of this paper: 
\begin{enumerate}
\item The imaginary part of the Lindhard function is not supported only in some region of the $(\Omega, q)$ plane as in the FL case. Rather it has a broad distribution. However, the shape 
of $\textrm{Im} \mathcal{L}$ as a function of the momentum has features that resemble those appearing in the Fermi liquid and can be traced back to kinematic/geometric properties and the existence of a Fermi surface, as we explain in detail in Section \ref{GenL}.  In particular, one can show that a part of the continuum of the Fermi liquid \footnote{The continuum is the region in $\Omega-q$ plane where on-shell gapless particle-hole excitations can exist on the Fermi surface.}, which we refer to as the \textit{inner core} is preserved in the semi-holographic non-Fermi liquids. The features corresponding to the boundaries of the continuum are also present in the Lindhard function although they lie outside of the inner core region. However, these features are blurred.

\item The dressed (i.e. RPA-resummed) Lindhard function describes the response of a system of charged electrons, when the effect of the Coulomb interaction is taken into account. In the FL case, one finds a pole corresponding to plasma oscillations, which becomes damped when it enters the region of the particle-hole continuum. In our case, we have a different behaviour: the response is very incoherent for low frequency and momentum, and only after a threshold we see a well-defined reasonably sharp peak developing standing out very clearly on top of the incoherent background with an approximately linear dispersion relation. The presence of a threshold can be related to the fact that the inner core of the continuum has conventional FL features disallowing any well-defined collective excitation to exist. Based on the behaviour of the Lindhard function, we can give a robust explanation for the existence of these plasmonic excitations in the relatively low frequency and high momentum ($> 2k_F$) regime.

\item The effective potential, that is the dressed Coulomb interaction, is modified and is frequency-dependent. As a function of the distance, it has attractive regions and the depth of the well increases with the frequency. This unexpected behaviour raises the possibility of a pairing mechanism that would lead to a new type of instability, but 
pairing particles of different frequencies. Since there is an appreciable spectral weight even quite far from the Fermi energy, it is possible that this mechanism is operative and  leads to  superconductivity. This is reminiscent of plasmonic mechanisms for superconductivity \cite{Sham}. 
We should note that the role of Coulomb interaction has been emphasized by Leggett in his phenomenological scenario for high-temperature superconductivity in cuprates \cite{Leggett}. According to this scenario, the main energy gain in crossing the superconducting transition is due to an improved screening in the superconducting phase.  If the superconducting instability is indeed present in our models, it would be one possible theoretical realization of Leggett's proposal. 
\end{enumerate}
We would like to point out that semi-holography has been recently proposed as a general method for constructing an effective non-perturbative description of some physical systems in a wide range of energy scales \cite{Iancu:2014ava,Mukhopadhyay:2015smb,Banerjee:2017ozx}. In particular, in the case of QCD it has been proposed that the classical gravity theory capturing the strongly coupled degrees of freedom should be constructed by demanding that it should cure the absence of Borel resummability of perturbation theory. A derivation on these lines in the context of large $N$ QCD has been discussed in \cite{Banerjee:2017ozx}. We will discuss how these developments are relevant for understanding the microscopic origin of the semi-holographic models briefly in Section \ref{gen-mod}.

The plan of the paper is as follows. In Section \ref{mod}, we review a class of semi-holographic non-Fermi liquid models and the arguments why they lead to a generalisation of Landau's Fermi liquid theory. In Section \ref{Lindhard Function}, we discuss how we can consistently impose the energy cut-off in the calculation of the Lindhard function in the semi-holographic non-Fermi liquid models. In Section \ref{GenL}, we discuss the Lindhard function in detail and in Section \ref{collective}, we find the RPA-resummed  Lindhard function and discuss the unconventional plasmonic pole, the possible superconducting instability and the presence of Freidel oscillations. In Section \ref{conclude}, we end with an outlook. The appendix \ref{nu1by2} contains some details 
on the special case of $\nu=1/2$, wich is a limiting case for this class of mdoels. 

\section{Semi-holographic models}\label{mod}
\subsection{Generalisation of Landau's Fermi liquid theory}\label{FLgen}

The class of models that we consider are constructed using a fermionic field $\chi$ and an additional sector that, as explained in the Introduction, represents the critical modes, so we take it to be an emergent infrared conformal field theory (IR-CFT). We assume that in the spectrum of the IR-CFT there is a fermionic operator $\psi$ that has the same quantum numbers as $\chi$ and couples to it linearly leading to tree-level mixing. The most generic form of the action is then: 
 
\begin{widetext}
 \begin{eqnarray}\label{model}
&&  S = \int dt \Bigg[ \sum_k \Big(\chi^{\dagger}_{\mathbf{k}}(i\partial_t - \epsilon_{\mathbf{k}}+\mu)
\chi_{\mathbf{k}} + N^2 S_{\textrm{CFT}} \nonumber \\ 
&& + N  \sum_{k} ( g_{\mathbf{k}} \chi^{\dagger}_{\mathbf{k}}
\psi_{\mathbf{k}} + c.c.  \Big)  +\frac{1}{2}   \sum_{k,k_1,q} \chi^\dagger_{\mathbf{k}}
\chi_{\mathbf{k}-\mathbf{q}}V(\mathbf{q})
\chi^\dagger_{\mathbf{k}_1}
\chi_{\mathbf{k}_1-\mathbf{q}} \nonumber\\
&&
+  \sum_{k,k_1,q}\lambda_{\mathbf{k}, \mathbf{k}_1, \mathbf{q}}\chi^\dagger_{\mathbf{k}}
\chi_{\mathbf{k}-\mathbf{q}}
\chi^\dagger_{\mathbf{k}_1}
\chi_{\mathbf{k}_1-\mathbf{q}}  
+N   \sum_{k,k',q}  \eta_{\mathbf{k},\mathbf{k}'}\chi^{\dagger}_{\mathbf{k}}
\chi_{\mathbf{k}'}
\phi_{\mathbf{k}-\mathbf{k}'}\nonumber\\
&&
+ N  \sum_{k,k_1,k_2}
\Big(\tilde{g}_{\mathbf{k}, \mathbf{k}_1, \mathbf{k}_2}\chi^{\dagger}_{\mathbf{k}} 
\chi_{\mathbf{k}_1}
\chi^{\dagger}_{\mathbf{k}_2}
\psi_{\mathbf{k}-\mathbf{k}_1+\mathbf{k}_2}
+c.c. \Big) \Bigg]  \,. \nonumber
\end{eqnarray}
\end{widetext}

The first line  gives the action for the two decoupled theories; the second line contains the quadratic coupling, and a potential interaction term (we will consider a Coulomb interaction).  
The last two lines contain higher order interactions, possibly with other CFT operators denoted by $\phi$; these terms  were important in establishing the generalisation of Landau's theory in \cite{Mukhopadhyay:2013dqa}, but will not play a role in the present paper. 

We have also included a parameter $N$ that allows us to have a parametric control of the diagrammatic expansion. The most important thing to note is that here in the large $N$ limit (i) all terms in the fermionic sector $\chi$ scale as $\mathcal{O}(1)$, (ii) all terms involving interactions of $\chi$ with the IR-CFT operators (including $\Psi$) scale as $\mathcal{O}(N)$, and (iii) $S_{\rm CFT}$ scales as $\mathcal{O}(N^2)$. This large-$N$ scaling, as we will see presently, is crucial to have a modified propagator at $=\mathcal{O}(1)$, as we will see presently. However, it does not suppresses radiative corrections, e.g. to the vertex coupling the current to the electromagnetic field $A_\mu \bar \chi \gamma^\mu \chi$, which enters in the electromagnetic response. In this paper we will ignore such corrections, adopting what is known as the RPA approximation, though it would be important to investigate them as well.

Resumming the quadratic interaction with $\psi$ leads to the following retarded propagator:\footnote{This propagator is obtained by diagonalising the quadratic action and therefore describes the propagation of a superposition of $\chi$ and $\psi$. However, the component of $\psi$ is $\mathcal{O}(1/N)$. The other piece in the diagonalised quadratic action involves a propagator that vanishes on the Fermi surface, and plays no role in the low energy effective theory.}

\begin{equation}\label{GR-fund}
G_R(\omega, \mathbf{k}) = \frac{1}{\zeta \omega^\nu - \epsilon_{\mathbf{k}}}, \quad  \epsilon_{\mathbf{k}} = \frac{\mathbf{k}^2}{2m} - \frac{k_F^2}{2m}, \quad 0<\nu < 1.
\end{equation}
Above, we have ignored the sub-leading $\omega$ term which arises from the from the free fermionic action. The exponent $\nu$ characterises the deviation from FL behavior; apart from it, the model also has some additional parameters: the complex number $\zeta$ and the Fermi momentum $k_F$. 

From the propagator we deduce the spectral function $\rho = - 2 {\rm Im}G_R$ : 
\begin{equation}\label{spectral}
{\rm Im}G_R(\omega, \mathbf{k}) = -\frac{\zeta_I \omega^\nu}{(\zeta_R \omega^\nu -\epsilon_{\mathbf{k}})^2 + \zeta_I^2  \omega^{2\nu}}\theta(\omega) -\frac{\tilde{\zeta}_I  \lvert\omega\rvert^\nu}{(\tilde{\zeta}_R \lvert\omega\rvert^\nu -\epsilon_{\mathbf{k}})^2 + \tilde{\zeta}_I^2 \lvert\omega\rvert^{2\nu}}\theta(-\omega),
\end{equation}
where $\zeta_{R(I)}$ are the real (imaginary) parts of $\zeta$, and likewise\footnote{Note since $G_R$ is analytic in $\omega$ in the upper half complex plane, it follows that $(-\omega)^\nu = e^{i\pi\nu}\lvert\omega\rvert^{\nu}$. This is why $\tilde{\zeta}$ appears in ${\rm Im}G_R$ for negative values of $\omega$.} $\tilde{\zeta}_{R(I)}$ are the real (imaginary) parts of $\tilde{\zeta}$, with
$$
\tilde{\zeta} = \zeta e^{i\pi\nu}.
$$
Notice that the spectral function manifestly has \textit{particle-hole asymmetry}.
In order for  the spectral function to be positive we must require 
$ \zeta_I >0 , \tilde{\zeta}_I >0$, and this implies 
\begin{equation}\label{zetalimits}
0 < \phi < \pi(1-\nu).
\end{equation}
where $\phi :={\rm arg}(\zeta)$.

When the propagator is derived from a holographic model, as in \cite{Liu:2009dm}, the phase of $\zeta$ depends on the parameters of the model and is given by 
\begin{equation}
\textrm{arg}(\zeta) = \textrm{arg}(\Gamma(-\nu) (e^{-i \pi \nu} - e^{-2\pi q}))
\end{equation} 
where $q$ is the fermion charge in appropriate units; the relation (\ref{zetalimits}) is then automatically satisfied, and the upper bound in \eqref{zetalimits} is saturated when $q \to \infty$. In the other limit $q\rightarrow 0$, which is the probe limit where the backreaction of the bulk fermion on the bulk gauge field can be ignored, we obtain $\phi = (\pi/2)(1-\nu)$, i.e. half of the extremal value. In this work, we will consider a generic case where $\phi$ is closer to the extremal value since otherwise $q$ needs to be very small.\footnote{In practice, choosing the extremal value leads to numerical instability and qualitative features of our model does not depend much on the precise value of $\phi$ as long as we avoid the extremal value.}

The spectral function (\ref{spectral}) is not integrable, since it does not decay sufficiently fast at infinity. This feature is necessary for the spectral function to satisfy the sum rule and therefore is an indication that the model as it stands is not complete -- it requires a UV completion. The simplest 
way to deal with this problem is to consider the theory with a UV cutoff. A better way is to re-introduce the $\omega$ term coming from the free propagator which leads to a crossover to a FL behaviour at high energies with a propagator of the form 

\begin{equation}\label{crossover}
G_R(\omega, \mathbf{k}) = \frac{1}{\zeta \omega^\nu + \omega  - \epsilon_{\mathbf{k}}} \,;
\end{equation}
The crossover to FL behaviour happens at the scale $\omega_c = |\zeta|^{1/(1-\nu)}$. 
Since it will be more difficult to compute using this propagator, we will instead use the crossover scale as a cutoff, and argue that this introduces errors that are small at 
low energies. 

The reasoning behind why these semi-holographic models lead to a generalisation of Landau's Fermi liquid theory has been presented in \cite{Mukhopadhyay:2013dqa}. The crux of this argument is that all the interaction terms beyond the quadratic part of the action which lead to the propagator \eqref{GR-fund} are irrelevant for the low energy non-Fermi liquid in the sense that both the real and imaginary parts of self-energy corrections to \eqref{GR-fund} are smaller than $\omega^\nu$ when $\omega$ is small. This can also be argued on the basis of the scaling behavior, extending the argument from the Fermi liquid case presented e.g. in \cite{sachdev}.  The argument given in \cite{Mukhopadhyay:2013dqa} started from the model defined by (\ref{model}), but one can also start directly from the model corresponding to the propagator (\ref{GR-fund}); the quadratic action is invariant under the rescaling $\{\omega, \mathbf{k}_\perp, \mathbf{k}_{\vert \vert}\} \rightarrow \{\lambda^{2 / \nu} \omega, \lambda^2 \mathbf{k}_\perp, \lambda \mathbf{k}_{\vert \vert}\}$\footnote{The large $N$ scaling of various terms in the action plays a subtle but crucial role here. The scaling properties of the effective fermionic field and hence of the couplings depends on the quadratic action whose form can be fixed reliably only in the large $N$ limit provided all couplings have the proposed large $N$ scalings. Otherwise, the geometry will suffer backreaction and this will in turn affect the quadratic terms themselves.} where $\mathbf{k}_\perp$ denotes the component of the momentum perpendicular to the Fermi surface and $\mathbf{k}_{\vert \vert}$ denotes the tangential component. With this scaling, it is easy to see that all interaction terms are irrelevant. For instance, a coupling $g \chi^4$ has dimension $3 - d -2/\nu$ and is always irrelevant in $d \geq 1$. These arguments were also supported by explicit two-loop self-energy  calculations in the patch approximation, which give a leading behavior $\Sigma \sim \omega^2$, as in the Fermi liquid case and in agreement with the scaling argument.  It would be thus appropriate to say that the semi-holographic non-Fermi liquid allows the notion of \textit{generalised quasi-particle} excitations of the Fermi surface in the sense that the interactions of these generalised quasi-particles are irrelevant in the low-energy limit.

One fundamental issue is that although the infrared semi-holographic theory is a generalisation of Landau's Fermi liquid theory, it is not well-defined in the ultraviolet. Here, we will merely assume that the semi-holographic models provide an \textit{effective non-perturbative infrared description} of an appropriate material. The crucial question of what kind of material physics can lead to such an infrared limit will be left for future investigations. Nevertheless, it is worth asking whether the semi-holographic models described above are of the most general kind to be realised in a class of materials that can be prepared in laboratories. This will be the subject of discussion of the following subsection.

\subsection{More general constructions}\label{gen-mod}

In this subsection we discuss more general constructions of semi-holographic models for both conceptual clarity and completeness. This discussion is not essential for what follows, so the reader can skip this subsection on the first reading.

The arguments presented in the previous subsection depend on the strong assumption that the critical fermions to which the perturbative electrons are linearly coupled live in a dual $AdS_2$ geometry. Naively, one would expect that the backreaction on the $AdS_2$ geometry can originate only from perturbative electrons at the boundary and the bulk critical fermions, and this should be suppressed because these fermions have $\mathcal{O}(1)$ density as opposed to the $\mathcal{O}(N^2)$ density needed for a significant backreaction effect (see also footnote 5). In more general constructions however, this may not be true and then the concept of a stable generalised quasi-particle at the Fermi surface discussed earlier would need to be revisited.

The need to generalise our construction originates in the observation that the background $AdS_2$ geometry represents non-perturbative dynamical effects of the lattice particularly in generating long-range correlations in the fermionic sector. Nevertheless, the lattice itself can have degrees of freedom which must be taken into account perturbatively. In turn they can affect the non-perturbative long range correlations which should be generated in the fermionic sector -- implying a modification of the background $AdS_2$ geometry of the bulk fermions. As a concrete example, one can consider the effect of impurities in the lattice. For this one can introduce a bulk scalar field which provides a non-perturbative counterpart to the density of impurities at the boundary by explicitly coupling to it. In this case, one would need to solve the lattice and bulk dynamics self-consistently. This bulk scalar field can not only change the background  $AdS_2$ geometry, but also dynamically modify the effective mass of the critical bulk fermion through a bulk Yukawa coupling. Both of these will lead to a modification of the leading scaling exponent of the self-energy which has been assumed to be fixed by the choice of parameters in the previous subsection. Furthermore, certain interactions at the Fermi surface can now become relevant. 

The generic construction of the semi-holographic framework as a generalisation of the effective field theory framework including non-perturbative effects has been developed recently in \cite{Banerjee:2017ozx} where a concrete proposal has been made to construct it for the case of Quantum Chromodynamics (QCD). The basic principles of the construction of the framework are as follows. First, we set the coupling rules between the perturbative and the non-perturbative (holographic) sectors such that there exist a local conserved energy-momentum tensor of the full system \cite{Mukhopadhyay:2015smb,Banerjee:2017ozx}. Then we determine the parameters of the holographic gravity theory dual to the non-perturbative sector and the additional couplings between the two sectors in terms of the usual perturbative couplings. This is done by demanding that the ambiguities generated by the lack of Borel resummability of the perturbation series vanish. The feasibility of such a construction has been demonstrated in a toy example \cite{Banerjee:2017ozx}. 

It is not clear to us at this stage how such a semi-holographic framework  can be derived from first principles in case of a specific class of strongly correlated materials. Nevertheless, some natural generalisations discussed above should be worth pursuing in the future particularly for investigating general consequences for the low energy dynamics at the Fermi surface.

\section{How to compute the generalised Lindhard Function} \label{Lindhard Function}\label{sec:GenLindFunc}
The main object of this paper is the generalized Lindhard function $\mathcal{L}(\Omega, \mathbf{q})$, defined as the time-ordered density-density correlation function.\footnote{We call this the generalized Lindhard function from now on, because the term Lindhard function is used in literature in the context of Fermi liquids.} It also gives the medium-induced correction to the photon self-energy at one-loop order. Explicitly, it is given by: 
\begin{eqnarray}\label{LindhardDef}
\mathcal{L}(\Omega, \mathbf{q}) = -2i \int_{\mathbf{k}}\int_{\omega} G_F \left(\omega _+ , \mathbf{k}_+\right) G_F \left(\omega_-,\mathbf{k}_-\right),
\end{eqnarray}
where
\begin{equation}
\omega_\pm = \omega \pm\frac{\Omega}{2}, \quad  \mathbf{k}_\pm = \mathbf{k} \pm\frac{\mathbf{q}}{2}.
\end{equation}
Also, $G_F$ denotes the fermionic Feynman propagator.
We will establish a few properties that will be useful later. 
First, it is easy to see that $\mathcal{L}(\Omega, \mathbf{q}) = \mathcal{L}(-\Omega, \mathbf{q})$, so we can take $\Omega >0$. We will only consider isotropic systems, so $\mathcal{L}$ it is only a function of $|{\bf q}| \equiv q$.  

In order to preserve analytic properties of correlation functions in the Schwinger-Keldysh contour, it is convenient to rewrite the Feynman propagator in terms of the retarded propagator as follows:\footnote{To derive the relation below, we recall that $G_F(\mathbf{x}-\mathbf{x}', t-t') =-i G^> (\mathbf{x}-\mathbf{x}', t-t') \theta(t-t') - iG^< (\mathbf{x}-\mathbf{x}', t-t') \theta(t'-t)$. To go to Fourier space, we can use the convolution theorem, and that $ G^>(\omega, \mathbf{k}) = {\rm Im}G_R(\omega, \mathbf{k})(1- n_F(\omega))$ and $G^<(\omega, \mathbf{k}) = -{\rm Im}G_R(\omega, \mathbf{k})n_F(\omega)$. We also use the Kramers-Kronig relation between ${\rm Re}G_R(\omega, \mathbf{k})$ and ${\rm Im}G_R(\omega, \mathbf{k})$. }
\begin{equation}\label{Feynman}
G_F (\omega, \mathbf{k}) = {\rm Re}G_R(\omega, \mathbf{k}) + i {\rm Im}G_R(\omega, \mathbf{k})(1- 2 n_F(\omega)).
\end{equation}
with $n_F$ denoting the Fermi-Dirac distribution function at finite temperature. Using this it is easy to show that:
\begin{eqnarray}\label{LindhardDef2}
\mathcal{L}(\Omega, q) &=& 2 \int_{\mathbf{k}}\int_{\omega} \Bigg({\rm Re}G_R \left(\omega_-, \mathbf{k}_-\right){\rm Im}G_R \left(\omega_+, \mathbf{k}_+\right)\left(1- 2 n_F\left(\omega_+\right)\right) \\\nonumber &&
\qquad\quad+ {\rm Re}G_R \left(\omega_+, \mathbf{k}_+\right){\rm Im}G_R \left(\omega_-, \mathbf{k}_-\right)\left(1- 2 n_F\left(\omega_-\right)\right)\Bigg)\\\nonumber &&
-2i \int_{\mathbf{k}}\int_{\omega} \Bigg({\rm Re}G_R \left(\omega_-, \mathbf{k}_-\right){\rm Re}G_R \left(\omega_+, \mathbf{k}_+ \right)\\\nonumber &&- {\rm Im}G_R \left(\omega_+,\mathbf{k}_+\right){\rm Im}G_R \left(\omega_-, \mathbf{k}_-\right)\left(1- 2 n_F\left(\omega_+\right)\right)\left(1- 2 n_F\left(\omega_-\right)\right)\Bigg).
\end{eqnarray}
For reasons to be clear soon, it is convenient to note that since $G_R(\omega, \mathbf{k})$ is analytic in $\omega$ in UHP, assuming that $G_R(\omega, \mathbf{k})$ decays sufficiently fast at large values of $\omega$, we should have
\begin{equation}\label{GRid}
\int_{\omega} G_R \left(\omega_-, \mathbf{k}_-\right)G_R \left(\omega_+, \mathbf{k}_+ \right) = 0.
\end{equation}
The real part of the above identity implies
\begin{equation}\label{id1}
\int_{\omega} {\rm Re}G_R \left(\omega_-, \mathbf{k}_-\right){\rm Re}G_R \left(\omega_+, \mathbf{k}_+\right) = \int_{\omega} {\rm Im}G_R \left(\omega_-, \mathbf{k}_-\right){\rm Im}G_R \left(\omega_+, \mathbf{k}_+\right).
\end{equation}

Combining the above with \eqref{LindhardDef2}, we obtain
\begin{subequations}
\begin{align}\label{LindhardDef31}
{\rm Re}\mathcal{L}(\Omega, q) &= 2 \int_{\mathbf{k}}\int_{\omega} \Bigg({\rm Re}G_R \left(\omega_-, \mathbf{k}_-\right){\rm Im}G_R \left(\omega_+, \mathbf{k}_+ \right)\left(1- 2 n_F\left(\omega_+\right)\right) \nonumber\\ &
\qquad\quad+ {\rm Re}G_R \left(\omega_+, \mathbf{k}_+ \right){\rm Im}G_R \left(\omega_-, \mathbf{k}_-\right)\left(1- 2 n_F\left(\omega_-\right)\right)\Bigg),\\
\label{LindhardDef32}
{\rm Im}\mathcal{L}(\Omega, q) &= -2 \int_{\mathbf{k}}\int_{\omega} {\rm Im}G_R \left(\omega_-, \mathbf{k}_-\right){\rm Im}G_R \left(\omega_+, \mathbf{k}_+ \right)
\left(1-\left(1- 2 n_F\left(\omega_+\right)\right)\left(1- 2 n_F\left(\omega_-\right)\right)\right).
\end{align}
\end{subequations}

In order to go to zero temperature, we need to note that $\lim_{T\rightarrow 0}n_F(\omega) = \theta(-\omega)$ and therefore
\begin{equation}
\lim_{T\rightarrow 0} (1 - 2 n_F(\omega)) = {\rm sgn}(\omega). 
\end{equation}
It then follows from \eqref{Feynman} that at zero temperature,
\begin{equation}
{\rm Re}G_F(\omega, \mathbf{k}) = {\rm Re}G_R(\omega, \mathbf{k}), \quad {\rm Im}G_F(\omega, \mathbf{k}) = {\rm Im}G_R(\omega, \mathbf{k}){\rm sgn}(\omega).
\end{equation}
Furthermore, for $\Omega > 0$, the real and imaginary parts of the generalised Lindhard function given by \eqref{LindhardDef31} and \eqref{LindhardDef32} respectively reduce to
\begin{subequations}
\begin{align}\label{LindhardDef41}
{\rm Re}\mathcal{L}(\Omega, q) &= -2 \int_{\mathbf{k}}\int_{-\omega_{\rm c}}^{-\frac{\Omega}{2}} \Bigg({\rm Re}G_R \left(\omega_-, \mathbf{k}_-\right){\rm Im}G_R \left(\omega_+, \mathbf{k}_+\right)+ {\rm Re}G_R \left(\omega_+, \mathbf{k}_+ \right){\rm Im}G_R \left(\omega_-, \mathbf{k}_-\right) \Bigg) \nonumber\\&
+2 \int_{\mathbf{k}}\int_{-\omega_{\rm c}}^{-\frac{\Omega}{2}} \Bigg({\rm Re}G_R \left(\omega_-, \mathbf{k}_-\right){\rm Im}G_R \left(\omega_+, \mathbf{k}_+\right)- {\rm Re}G_R \left(\omega_+, \mathbf{k}_+ \right){\rm Im}G_R \left(\omega_-, \mathbf{k}_-\right) \Bigg) \nonumber\\&
+2 \int_{\mathbf{k}}\int_{-\omega_{\rm c}}^{-\frac{\Omega}{2}} \Bigg({\rm Re}G_R \left(\omega_-, \mathbf{k}_-\right){\rm Im}G_R \left(\omega_+, \mathbf{k}_+\right)+ {\rm Re}G_R \left(\omega_+, \mathbf{k}_+ \right){\rm Im}G_R \left(\omega_-, \mathbf{k}_-\right) \Bigg),\\
\label{LindhardDef42}
{\rm Im}\mathcal{L}(\Omega, q) &= -4 \int_{\mathbf{k}}\int_{-\frac{\Omega}{2}}^{\frac{\Omega}{2}} {\rm Im}G_R \left(\omega_+, \mathbf{k}_+ \right){\rm Im}G_R \left(\omega_-, \mathbf{k}_-\right) .\end{align}
\end{subequations}
Above we have also implemented the cut-off $\omega_{\rm c}$ in the loop integral as should be done in an effective field theory. Furthermore, we have assumed that the external frequency $\Omega$ is less than the cut-off $\omega_{\rm c}$. We find that ${\rm Im}\mathcal{L}(\Omega, q)$ gets contribution only from $-\Omega/2 < \omega < \Omega/2$ and is therefore independent of the cut-off both in the exact theory (where $\omega_{\rm c} = \infty$) and in the effective theory. This justifies the manipulation of the imaginary part of ${\rm Im}\mathcal{L}(\Omega, q)$ in the form \eqref{LindhardDef32} from \eqref{LindhardDef2} using the identity \eqref{id1}. 

Note that if we impose a cut-off $\omega_c$ on both sides of \eqref{id1}, the identity will not be strictly valid. Nevertheless, if the effective theory can be interpolated to the right material physics in the UV, the violation should be suppressed by powers of $\Omega/\omega_{\rm c}$ as argued before. Here we will simply assume this to be the case and therefore the Kramers-Kronig relations between the imaginary and retarded parts of the retarded propagator should be satisfied in the same spirit. 
It is not hard to show that the retarded self-energy $\mathcal{L}_R(\Omega, q)$ is related to $\mathcal{L}(\Omega, q)$ just as the retarded propagator is related to the Feynman propagator, i.e.
\begin{equation}\label{LRL}
{\rm Re}\mathcal{L}(\Omega, q) = {\rm Re}\mathcal{L}_R(\Omega, q), \quad {\rm Im}\mathcal{L}(\Omega, q) = {\rm Im}\mathcal{L}_R(\Omega, q){\rm sgn}(\Omega)
\end{equation}
at zero temperature.

Since the retarded correlator does not decay sufficiently fast for large $\omega$  when $0 \leq \nu \leq 1/2$, the identity \eqref{GRid} and hence the identity \eqref{id1} is not even approximately valid in the effective theory up to positive powers of $\Omega/\omega_{\rm c}$ (for $\nu = 1/2$ the identities are violated by ${\rm log}(\Omega)$ terms as for instance). This leads to a problem in arriving at a consistent prescription where ${\rm Im}\mathcal{L}$ is independent of $\omega_{\rm c}$ for $\Omega <\omega_{\rm c}$. The case of $\nu = 1/2$ is discussed in Appendix \ref{nu1by2} in detail.  In particular, we suspect that for $\nu < 1/2$ the infrared theory does not make sense as it does not decouple from the UV physics. This seems to contradict the intuition from the scaling argument 
where the value $\nu =1/2$ does not appear be special. This issue deserves further investigation. In this paper we will restrict ourselves to $\nu > 1/2$.

From \eqref{LindhardDef42} we can read many things:  the imaginary part of the generalised Lindhard function gets contribution only from a bounded interval of length proportional to $\Omega$, so in particular it is a convergent integral.  Furthermore, in the limit $\Omega \rightarrow 0$, the region of integration shrinks away, so $\mathcal{L}(\Omega = 0 ,\mathbf{q})$ is purely real. It is also clear that ${\rm Im}\mathcal{L}(\Omega, q) < 0$ since ${\rm Im}G_R(\omega, \mathbf{k}) < 0$. Note that $\mathcal{L}$ refers to the bosonic (photonic) self-energy correction. Therefore ${\rm Im}\mathcal{L}_R(\Omega, q)\Omega < 0$ should be satisfied. This is indeed the case as is clear from Eq. (\ref{LRL}) and that ${\rm Im}\mathcal{L}(\Omega, q) < 0$.

For our specific case:
\begin{eqnarray}\label{explicit}
& &{\rm Im}\mathcal{L}(\Omega,\mathbf{q}) =-4  \int{\rm d}^2 k \int_{-\frac{\lvert\Omega\rvert}{2}}^{\frac{\lvert\Omega\rvert}{2}}{\rm d}\omega \\&&\nonumber  \frac{ \zeta_I \tilde \zeta_I(\omega+\frac{\lvert\Omega\rvert}{2})^\nu (\frac{\lvert\Omega\rvert}{2}-\omega)^\nu}{\left( 
\lvert \zeta \rvert^2 (\omega+\frac{\lvert\Omega\rvert}{2})^{2\nu} - 2 \zeta_R \epsilon_{\mathbf{k}+\frac{\mathbf{q}}{2}} (\omega+\frac{\lvert\Omega\rvert}{2})^{\nu} + \epsilon_{\mathbf{k}+\frac{\mathbf{q}}{2}}^2 
\right) \left( 
\lvert \zeta \rvert^2 (\frac{\lvert\Omega\rvert}{2}-\omega)^{2\nu} - 2 \tilde \zeta_R \epsilon_{\mathbf{k}-\frac{\mathbf{q}}{2}} (\frac{\lvert\Omega\rvert}{2}-\omega)^{\nu} + \epsilon_{\mathbf{k}-\frac{\mathbf{q}}{2}}^2,
\right)}
\end{eqnarray}
where $\zeta_R = {\rm Re}\zeta$,  $\zeta_I = {\rm Im}\zeta$, etc. We can change to following dimensionless variables:
\begin{eqnarray}
x = \frac{\omega}{\lvert\Omega\rvert}, \quad  \mathbf{y} = \frac{\mathbf{k}}{\sqrt{2m\lvert\zeta\rvert\lvert\Omega\rvert^\nu}}
\end{eqnarray}
and define the following dimensionless parameters
\begin{eqnarray}
y_F = \frac{k_F}{\sqrt{2m\lvert\zeta\rvert\lvert\Omega\rvert^\nu}}, \quad  \hat{\mathbf{q}} = \frac{\mathbf{q}}{\sqrt{2m\lvert\zeta\rvert\lvert\Omega\rvert^\nu}}, \quad \hat\zeta= \frac{\zeta}{\vert\zeta\rvert} = e^{i\phi}, \quad\hat{\tilde\zeta} = \hat\zeta\,e^{i\pi\nu} = e^{i(\phi +\pi\nu)}.
\end{eqnarray}
Then,
\begin{equation}\label{scaling}
{\rm Im}\mathcal{L}(\Omega, q) = 2m\frac{\lvert\Omega \rvert^{1-\nu}}{\lvert\zeta\rvert}K\left(\hat{q}, y_F, \nu, \phi\right),
\end{equation}
with $K$ being the dimensionless integral
\begin{eqnarray}\label{smallOmega}
K\left(\hat{q}, y_F, \hat{\zeta}, \nu\right) =-4  \int_{0}^\infty{\rm d} y\, y\int_0^{2\pi}{\rm d}\theta \int_{-\frac{1}{2}}^{\frac{1}{2}}{\rm d}x \\\nonumber  \frac{ \sin(\phi) \sin(\phi + \pi\nu)(x+\frac{1}{2})^\nu (\frac{1}{2}-x)^\nu}{\left( 
 (x+\frac{1}{2})^{2\nu} - 2 \cos(\phi) \, \epsilon_1 (x+\frac{1}{2})^{\nu} + \epsilon_1^2 
\right) \left( 
 (\frac{1}{2}-x)^{2\nu} - 2 \cos(\phi + \pi\nu)\, \epsilon_2\,(\frac{1}{2}-x)^{\nu} + \epsilon_2^2
\right)},\\\nonumber
\text{with}\quad \epsilon_1 = y^2 + \frac{\hat{q}^2}{4} + y\hat{q}\,\cos\theta -y_F^2, \quad  \epsilon_2 = y^2 + \frac{\hat{q}^2}{4} - y\hat{q}\,\cos\theta - y_F^2\,.
\end{eqnarray}
The above shows that to study ${\rm Im}\mathcal{L}(\Omega, q)$ qualitatively we can set $\vert \zeta \vert = 1$ and $m = 1/2$ without loss of generality. We will see that as long we keep $\phi$ within holographic bounds, the qualitative features of ${\rm Im}\mathcal{L}(\Omega, q)$ do not change much unless $\phi$ is extremal. Therefore, qualitative features of ${\rm Im}\mathcal{L}(\Omega, q)$ will depend only on $y_F$, $\hat{q}$ and $\nu$ if $\omega_{\rm c} > \Omega$. Similar conclusions will also hold for ${\rm Re}\mathcal{L}(\Omega, q)$ except that it should also depend on $\Omega/\omega_{\rm c}$ as mentioned before. For practical purposes we will choose other dimensionless variables in the next section.

Let us understand ${\rm Im}\mathcal{L}(\Omega, q)$ in the limit of small $\Omega$. In this limit, $\hat{q}, y_F \rightarrow \infty$ with $\hat{q}/y_F$ held fixed. The $y-$integral in \eqref{smallOmega} then will get its contribution maximally from $0< y < y_F$ and the integrand behaves as $y_F^{-4}$.

Therefore,
\begin{equation}
\lim_{\hat{q}, y_F \rightarrow \infty, \hat{q}/y_F = \text{constant}}K\left(\hat{q}, y_F, \hat{\zeta}, \nu\right) \approx \frac{y_F^2}{y_F^4}\int_{-1/2}^{1/2}dx (x+\frac{1}{2})^\nu (\frac{1}{2}-x)^\nu.
\end{equation}
So, in this limit
\begin{equation}
\lim_{\hat{q}, y_F \rightarrow \infty, \hat{q}/y_F = \text{constant}}K\left(\hat{q}, y_F, \hat{\zeta}, \nu\right) \approx \frac{1}{y_F^2}
\end{equation}
Combining above with Eq. (\ref{scaling}) implies that for small $\Omega$, 
\begin{equation}
\lim_{\Omega\rightarrow 0}\frac{1}{\Omega}{\rm Im}\mathcal{L}(\Omega, q) = M(q, k_F, \zeta, \nu).
\end{equation}
Therefore for small $\Omega$, ${\rm Im}\mathcal{L}(\Omega, q)$ has to be proportional to $\Omega$.

\section{Generalised Lindhard function and comparison with the Fermi liquid}\label{GenL}

\subsection{Imaginary part of the generalised Lindhard function}\label{ImL}

\paragraph{The case of the Fermi liquid:} Let us begin with the Fermi liquid. The density response function (aka Lindhard function) in case of the $D-$dimensional Fermi liquid takes the form:
\begin{eqnarray}\label{LFL}
\mathcal{L}^{\rm FL}(\Omega, q) &=& 2\int \frac{{\rm d}^D k}{(2\pi)^D}\,n_{\mathbf{k}-\mathbf{q}/2}(1-n_{\mathbf{k}+\mathbf{q}/2})\nonumber\\&&\times\left(\frac{1}{\Omega -\epsilon_{\mathbf{k}+\mathbf{q}/2}+\epsilon_{\mathbf{k}-\mathbf{q}/2}+i\eta} -\frac{1}{\Omega +\epsilon_{\mathbf{k}+\mathbf{q}/2}-\epsilon_{\mathbf{k}-\mathbf{q}/2}-i\eta}\right),
\end{eqnarray}
where
\begin{eqnarray}\label{nep}
n_{\mathbf{k}} = \theta(k_F -k), \quad \epsilon_{\mathbf{k}} = \frac{k^2}{2m} - \epsilon_F, \quad \epsilon_F =  \frac{k_F^2}{2m}.
\end{eqnarray}
The explicit integrations can be done for $D=1,2$ and $3$ -- the exact results can be found in \cite{Mihaila}. Our case of interest is $D=2$ specifically. Instead of reproducing the exact forms here, we present the key features and necessary plots for the sake of comparison with the semi-holographic non-Fermi liquid. 

\begin{figure}[htbp]
\centering
\includegraphics[width=0.85\textwidth]{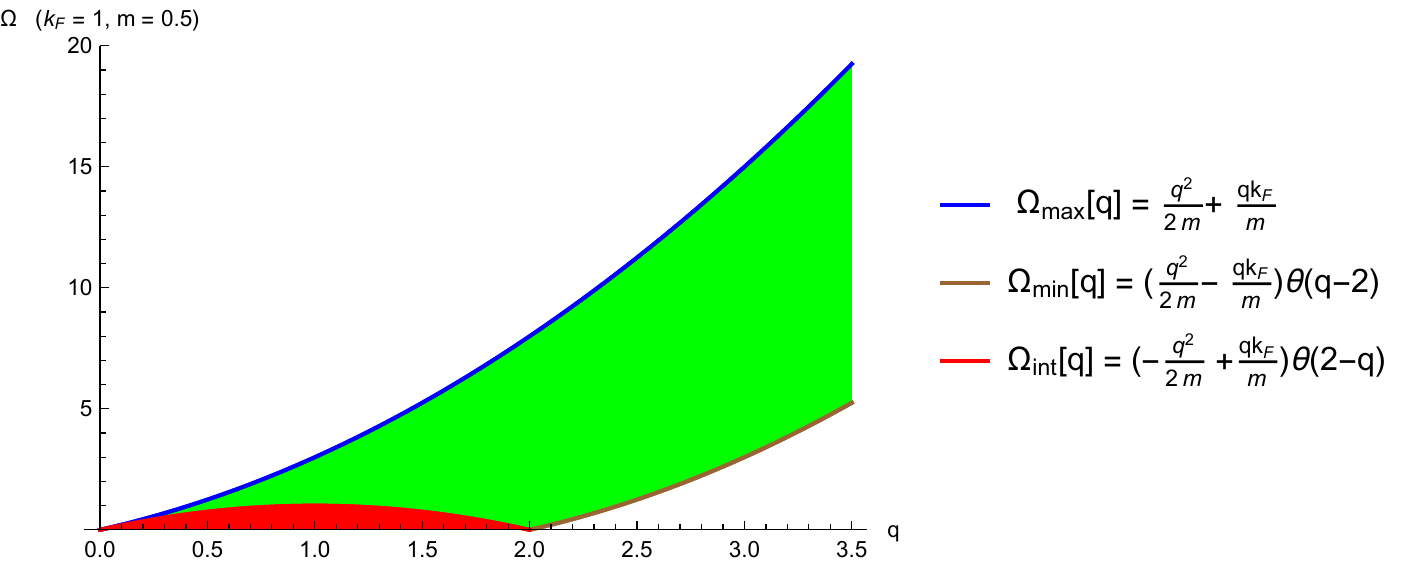}
\caption{The kinematic region in $\Omega-q$ plane where ${\rm Im}\mathcal{L}(\Omega, q)$ is non-zero for the Fermi liquid is the combined green and red regions in the plot shown above. For convenience, we have chosen $k_F = 1$ and $m = 0.5$. This region is essentially the particle-hole continuum. The boundaries of this region are given by $\Omega_{\rm max}(q)$ and $\Omega_{\rm min}(q)$. The red region is what we will call as the \textit{inner core} of the particle-hole continuum and is bounded by the curve $\Omega_{\rm int}(q)$.}
\label{fig:kinematic}
\end{figure}

Let us first examine ${\rm Im}\mathcal{L}(\Omega,q)$, the imaginary part of the Lindhard function. In the $\Omega-q$ plane, ${\rm Im}\mathcal{L}(\Omega,q)$ is supported in the green and red regions of the plots presented in Fig. \ref{fig:kinematic}. These regions combined form the particle-hole continuum, i.e. the range of allowed values of $\Omega$ and $q$ for which an \textit{on-shell} particle-hole pair can have total energy $\Omega$ and carry total momentum of magnitude $q$. The red region in Fig. \ref{fig:kinematic}, which we will refer to as the \textit{inner} core,  will be of special significance for us because we will see that \textit{only} in this region the semi-holographic non-Fermi liquid preserves Fermi-liquid like features. 

In case of the Fermi liquid, the boundaries of the kinematic region can be understood geometrically. For a fixed value of $q$, the allowed values of $\Omega$ (for $\Omega > 0$) is simply the total energy of an on-shell particle-hole pair carrying total momentum $q$:
\begin{equation}
\Omega =  \epsilon_{\mathbf{k}+\mathbf{q}/2}-\epsilon_{\mathbf{k}-\mathbf{q}/2} = \frac{\mathbf{q}\cdot\mathbf{k}}{m} \,\, {\rm with}\,\, 
k \in \{\epsilon_{\mathbf{k}-\mathbf{q}/2} < 0\, \& \,\epsilon_{\mathbf{k}+\mathbf{q}/2} >0\}.
\end{equation}
Without loss of generality, we can choose $\mathbf{q}$ along the (positive) $x-$axis. Then $\mathbf{q}\cdot\mathbf{k}= qk_x$. Therefore, for a fixed value of $q$, the total energy of the particle-hole pair is given by $\Omega = q k_x/m$ when the hole carries momentum $\{k_x - q/2, k_y\}$ and the particle carries momentum $\{k_x+q/2, k_y\}$. 

\begin{figure}[ht]
\begin{minipage}[t]{0.45\linewidth}
\centering
\includegraphics[width=\textwidth]{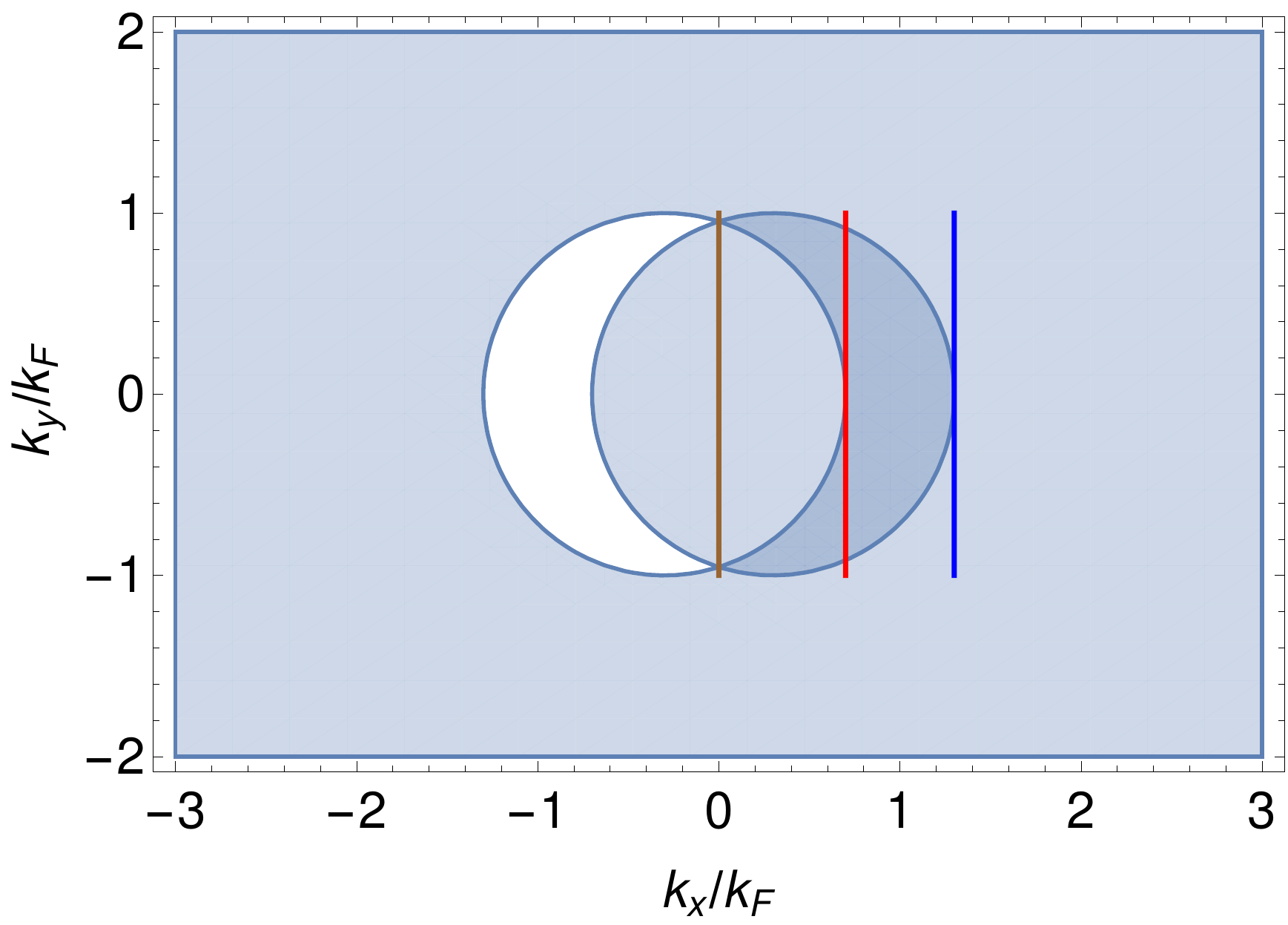}
\caption{\label{fig:Kinematics2}The allowed values of $k_x/k_F$ and $k_y/k_F$ when $q = 0.6 k_F < 2k_F$ is shown above in dark blue. This dark blue region is simply the intersection of the complement of the region outside the circle of unit radius centred at $\{-0.3k_F, 0\}$ (representing the on-shell particle in the pair) and the region bounded by the circle of unit radius centred at  $\{0.3k_F, 0\}$ (representing the on-shell hole in the pair). The allowed values of $\Omega$ are simply those for which $\Omega = qk_x/m$ with $k_x$ lying within in the dark blue region. Clearly, the smallest possible value of $\Omega$ is $0$ which is realised when $k_x =0$ (the brown line) and the largest possible value of $\Omega$ is $\Omega_{\rm max}$ given by Eq. (\ref{Ommax}) which is realised when $k_x = k_F + q/2$. When $k_F + q/2 \geq k_x \geq k_F - q/2$, i.e. $\Omega_{\rm max} \geq\Omega \geq \Omega_{\rm int}$ with $\Omega_{\rm int}$ latter given by Eq. (\ref{Omint}), the allowed values of $k_y$ forms a continuous line instead of disconnected segments. The red vertical line represents $k_x = k_F - q/2$ for which $\Omega = \Omega_{\rm int}$.}
\end{minipage}
\hspace{0.5cm}
\begin{minipage}[t]{0.45\linewidth}
\centering
\includegraphics[width=\textwidth]{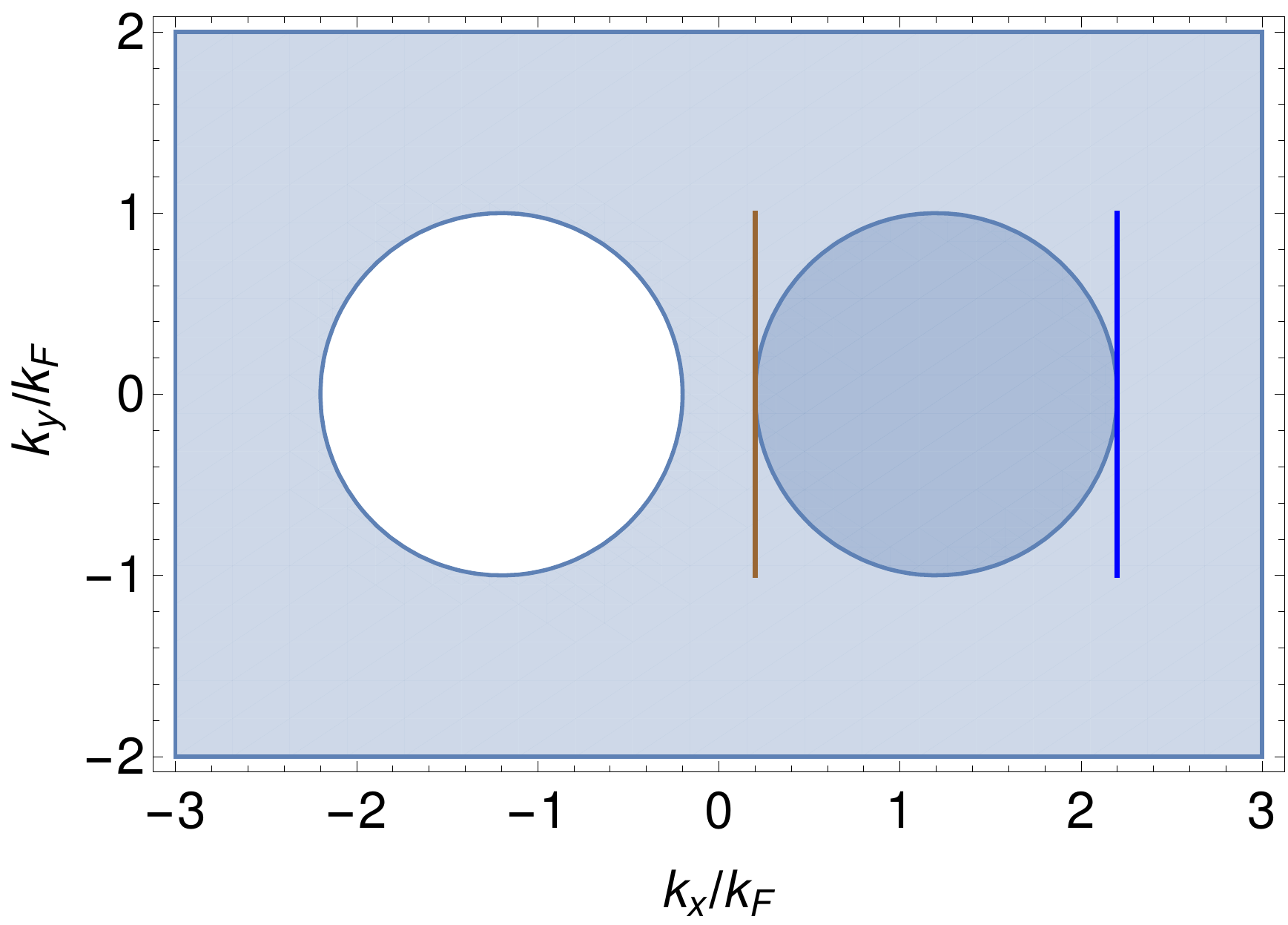}
\caption{\label{fig:Kinematics3} The allowed values of $k_x/k_F$ and $k_y/k_F$ when $q = 2.4 k_F > 2 k_F$ is shown above in dark blue. The region bounded by the unit circle centred at $\{1.2k_F, 0\}$ (representing the hole-like excitation) now lies entirely in the complement of the unit circle centred at $\{-1.2k_F, 0\}$ (representing the particle-like excitation) -- the dark blue region simply then coincides with the first region representing the hole excitations. The allowed values of $\Omega$ are simply those for which $\Omega = qk_x/m$ with $k_x$ lying within the dark blue region. The smallest allowed value of $\Omega = \Omega_{\rm min}$ is given by Eq. (\ref{Ommin}) for which $k_x = q/2 - k_F$ (the brown vertical line) and the largest allowed value of $\Omega = \Omega_{\rm max}$ is given by Eq. (\ref{Ommax}) for which $k_x = q/2 + k_F$ (the blue vertical line). There is no analogue of $\Omega_{\rm int}$ because the allowed values of $k_y$ for fixed $k_x$ (i.e. $\Omega$) and $q$ always form a continuous line segment.}
\end{minipage}
\end{figure}

For $q < 2k_F$, the allowed values of $\Omega$ can be readily inferred from the allowed values of $k_x$ and $k_y$ as shown in Fig. \ref{fig:Kinematics2}. The smallest possible value of $\Omega$  is $0$, corresponding to $k_x = 0$,  and the largest is
\begin{equation}\label{Ommax}
\Omega_{\rm max} (q) = \frac{q^2}{2m} + \frac{qk_F}{m}
\end{equation}
corresponding to $k_x = k_F + q/2$. Furthermore, when 
\begin{equation}
k_F - q/2 \leq k_x  \leq k_F + q/2,
\end{equation}
i.e.
\begin{equation}\label{Omint}
\Omega_{\rm int}(q) \leq \Omega  \leq \Omega_{\rm max}(q), \,\, {\rm with}\,\, \Omega_{\rm int}(q) = -\frac{q^2}{2m} + \frac{qk_F}{m}
\end{equation}
then the allowed values of $k_y$ form a single connected line segment instead of two disconnected short line segments as evident from Fig. \ref{fig:Kinematics2}. Thus $\Omega = \Omega_{\rm int}(q)$ represents a change in topology of the allowed $k_y$ region with fixed $k_x$ (i.e. $\Omega$) and $q$. \textit{The region $0 < \Omega < \Omega_{\rm int}(q)$  forms the inner core of the particle-hole continuum} (marked in red in Fig. \ref{fig:kinematic}). \textit{As evident from Figs. \ref{fig:kinematic} and \ref{fig:Kinematics2}, the inner core corresponds to the region of kinematically allowed values of $\Omega$ and $q$ for which $k_x$ and $k_y$ lie close to the particle Fermi surface.}

For $q > 2k_F$, the allowed values of $k_x$ and $k_y$ are shown in Fig. \ref{fig:Kinematics3}. The smallest allowed value of $\Omega$ is
\begin{equation}\label{Ommin}
\Omega_{\rm min} (q) = \frac{q^2}{2m} - \frac{qk_F}{m}
\end{equation}
for which $k_x = q/2 -k_F$ and the largest allowed value of $\Omega$ is $\Omega_{\rm max}(q)$ given by Eq. (\ref{Ommax}) for which $k_x = k_F + q/2$. There is no analogue of $\Omega_{\rm int}(q)$ because the allowed values of $k_y$ for a fixed value of $k_x$ (i.e. $\Omega$) and $q$ always form a continuous line segment.

\begin{figure}[htbp]
\centering
\includegraphics[width=\textwidth]{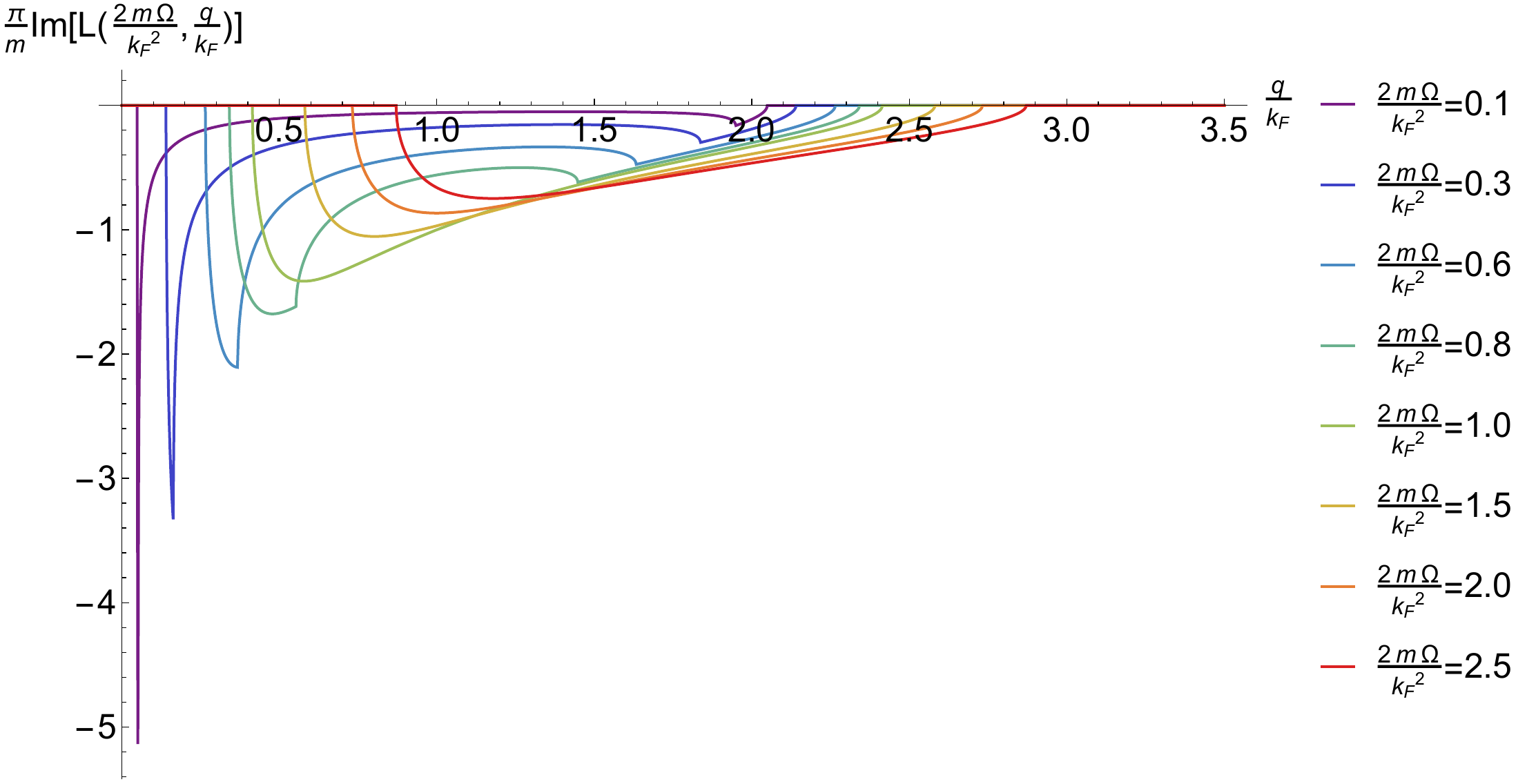}
\caption{Plots of ${\rm Im}\mathcal{L}^{\rm FL}(q)$ for various fixed values of $\Omega$ are shown above. Note $\mathcal{L}^{\rm FL}(q, \Omega) = (m/\pi) f(q/k_F, \Omega/\epsilon_F)$ with $\epsilon_F = k_F^2/2m$. Therefore, we have used the dimensionless variables $q/k_F$, $ \Omega/\epsilon_F$ and $(\pi/m){\rm Im}\mathcal{L}$ in the plots above. Note that the intermediate plateau and the two intermediate kinks where $\partial {\rm Im}\mathcal{L}^{\rm FL}(q)/\partial q$ is discontinuous appear for $\Omega/\epsilon_F < 1$ and disappears when $\Omega/\epsilon_F  \geq 1$. }
\label{fig:ImLqFL}
\end{figure}

The geometric structure of the particle-hole continuum governs the behaviour of ${\rm Im}\mathcal{L}^{\rm FL}$ as a function of $q$ and $\Omega$. For reasons to become clear later, it is instructive to look first at ${\rm Im}\mathcal{L}^{\rm FL}(q)$ of the Fermi liquid at fixed values of $\Omega$. The plots are shown in Fig. \ref{fig:ImLqFL}. Referring to Fig. \ref{fig:kinematic}, we can readily see that a horizontal line at fixed $\Omega$, for $\Omega/\epsilon_F <  1$, will have four special points: when it intersects $\Omega_{\rm max}(q)$ at $q = q_1$, $\Omega_{\rm min}$ at $q =q_4$, and $\Omega_{\rm int} (q)$ at $q=q_2, q=q_3$.  These four special values of $q$ can be readily recognised in each of the the plots in Fig. \ref{fig:ImLqFL} for $\Omega/\epsilon_F = 0.1$, $0.3$ and $0.6$. Firstly, ${\rm Im}\mathcal{L}$ vanishes for $q< q_1$ and $q> q_4$. At the intermediate points $q = q_2$ and $q = q_3$, ${\rm Im}\mathcal{L}$ has peaks where also the derivative $\partial{\rm Im}\mathcal{L}/\partial q$ becomes discontinuous. In the region $q_2 \leq q \leq q_3$, ${\rm Im}\mathcal{L}(q)$ is quite flat after a sharp ascent near $q = q_3$ -- this is a very specific characteristic which will be remarkably preserved in case of the semi-holographic non-Fermi liquid. This intermediate region corresponding to the inner core shrinks continuously with increasing $\Omega$ and disappears when $\Omega/\epsilon_F = 1$. For $\Omega/\epsilon_F  >1$, ${\rm Im}\mathcal{L}(q)$ does not have any kink or flat regions within its domain of support which is bounded by values of $q$ for which $\Omega$ coincides with $\Omega_{\rm max}$ and $\Omega_{\rm min}$ as evident from Fig. \ref{fig:ImLqFL}.

\begin{figure}[htbp]
\centering
\includegraphics[width=\textwidth]{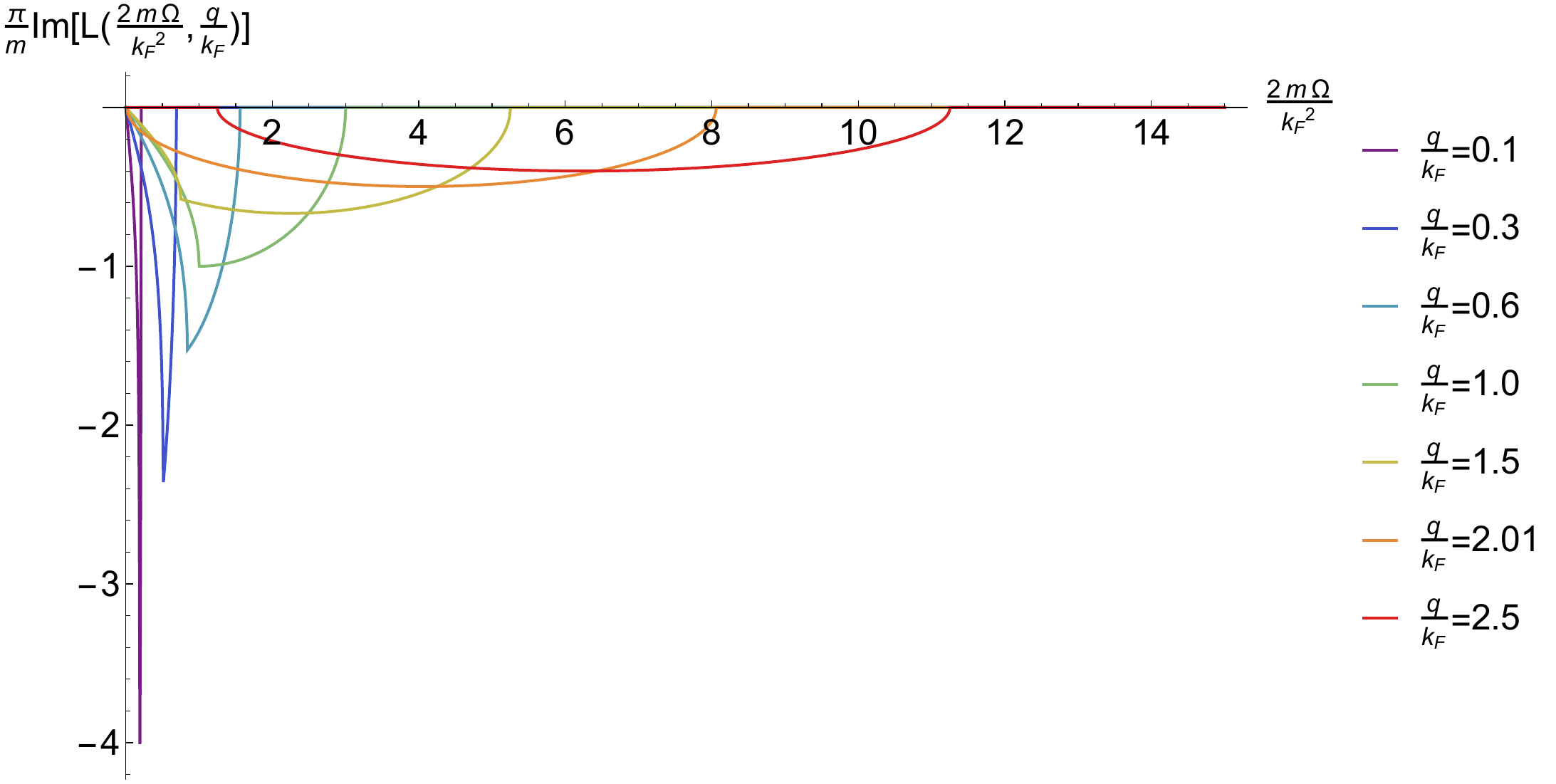}
\caption{Plots of ${\rm Im}\mathcal{L}^{\rm FL}(\Omega)$ for various fixed values of $q$ are shown above. We have used dimensionless variables for plotting as in Fig. \ref{fig:ImLqFL}.  Note the behaviour for $q < 2k_F$ is different from that for $q  >2k_F$. In particular, the minimum value of $\Omega$ for the latter case for which ${\rm Im}\mathcal{L}^{\rm FL}(\Omega)$ is non-vanishing is shifted from the origin. Also the intermediate kink where $\partial {\rm Im}\mathcal{L}^{\rm FL}(q)/\partial \Omega$ is discontinuous appears only for $q < 2k_F$.}
\label{fig:ImLOmFL}
\end{figure}

Similarly, ${\rm Im}\mathcal{L}$ as a function of $\Omega$ for fixed values of $q$ shows a transition in its behaviour as $q$ crosses $2k_F$. This can be readily understood by drawing vertical lines corresponding to fixed values of $q$ through the particle-hole continuum depicted in Fig. \ref{fig:kinematic}. When $q < 2k_F$, the vertical line will have two special points  $\Omega_1$ and $\Omega_2$  corresponding to intersections with $\Omega_{\rm int}(q)$, the boundary of the inner core, and $\Omega_{\rm max}(q)$ respectively.   On the other hand, for $q > 2k_F$, a vertical line at  fixed  $q$ still has two special points corresponding to intersections with $\Omega_{\rm min}$ and $\Omega_{max}$, however it does not intersect the inner core. The plots in Fig. \ref{fig:ImLOmFL} clearly show that there is a maximum with discontinuous derivative at $\Omega = \Omega_1$, but when $q/k_F >2$ the kink disappears and the curves have support only for $\Omega > \Omega_1$. 

 The features of ${\rm Im}\mathcal{L}^{\rm FL}$ Fermi liquid mentioned above will be of particular importance for us to understand the semi-holographic non-Fermi liquid case to be discussed below in relation to both -- the features which are kept intact and also the features which are \textit{blurred out} via incoherent quasi-normal mode fermionic excitations of the $AdS_2$ black hole.

\paragraph{The case of the semi-holographic non-Fermi liquid:} In the case of the semiholographic non-Fermi liquid, apparently $\mathcal{L}(q, \Omega)$ depends on many parameters, namely $\nu$, $\vert \zeta\vert$, $\phi = {\rm arg}(\zeta)$, $m$ and $k_F$. Nevertheless, we can easily show from Eq. (\ref{LindhardDef}) that  $\mathcal{L}(q, \Omega)$ takes the form:
\begin{equation}
\mathcal{L}(q, \Omega) = m f\left(\frac{\Omega}{\vert \zeta \vert^{1-\nu}}, \frac{q}{k_F}, \frac{\epsilon_F}{\vert \zeta \vert^{1-\nu}}, \nu, \phi\right),
\end{equation}
with $\epsilon_F = k_F^2/(2m)$ being the Fermi energy. However, as discussed earlier we must choose $\phi = \pi(1-\nu) - \epsilon$, where $\epsilon$ is a small non-negative number in order to represent a generic case. Therefore, the relevant parameters of the generalised Lindhard function $\mathcal{L}(q, \Omega)$ of our low energy effective theory are 
\begin{equation}\label{pars}
\tilde{\epsilon}_F = \frac{\epsilon_F}{\vert \zeta \vert^{1-\nu}}\quad {\rm and} \quad \nu.
\end{equation}
It turns out that if we fix $\nu$ and vary $\tilde{\epsilon}_F$ (by varying $k_F$ as for instance) the features of $\mathcal{L}(q, \Omega)$ and the phenomenology to be discussed later do not change qualitatively. Therefore, the parameter of interest is actually $\nu$ which for reasons described before should lie between $1/2$ and $1$. We will see that $\nu$ acts the parameter which dials a systematic deformation from the Fermi liquid theory.

From the above discussion, it is clear that we can choose $m = 0.5$ and $\vert \zeta \vert = 1$ without loss of generality. We recall that our effective theory breaks down at energies $\Omega > \omega_{\rm c} = \vert \zeta \vert^{1- \nu}$ (i.e. for $\Omega > 1$ with our choice $\vert \zeta \vert = 1$), therefore we choose $k_F = 0.4$ conveniently so that $\tilde{\epsilon}_F \ll 1$ and we can explore the regime $\Omega /\vert \zeta \vert^{1- \nu} \gg \tilde{\epsilon}_F$. Note that with $k_F  = 0.4$ and $m = 0.5$, $\epsilon_F = \tilde{\epsilon}_F = 0.16 \approx 0.2$.
\begin{figure}[htbp]
\centering
\includegraphics[width=\textwidth]{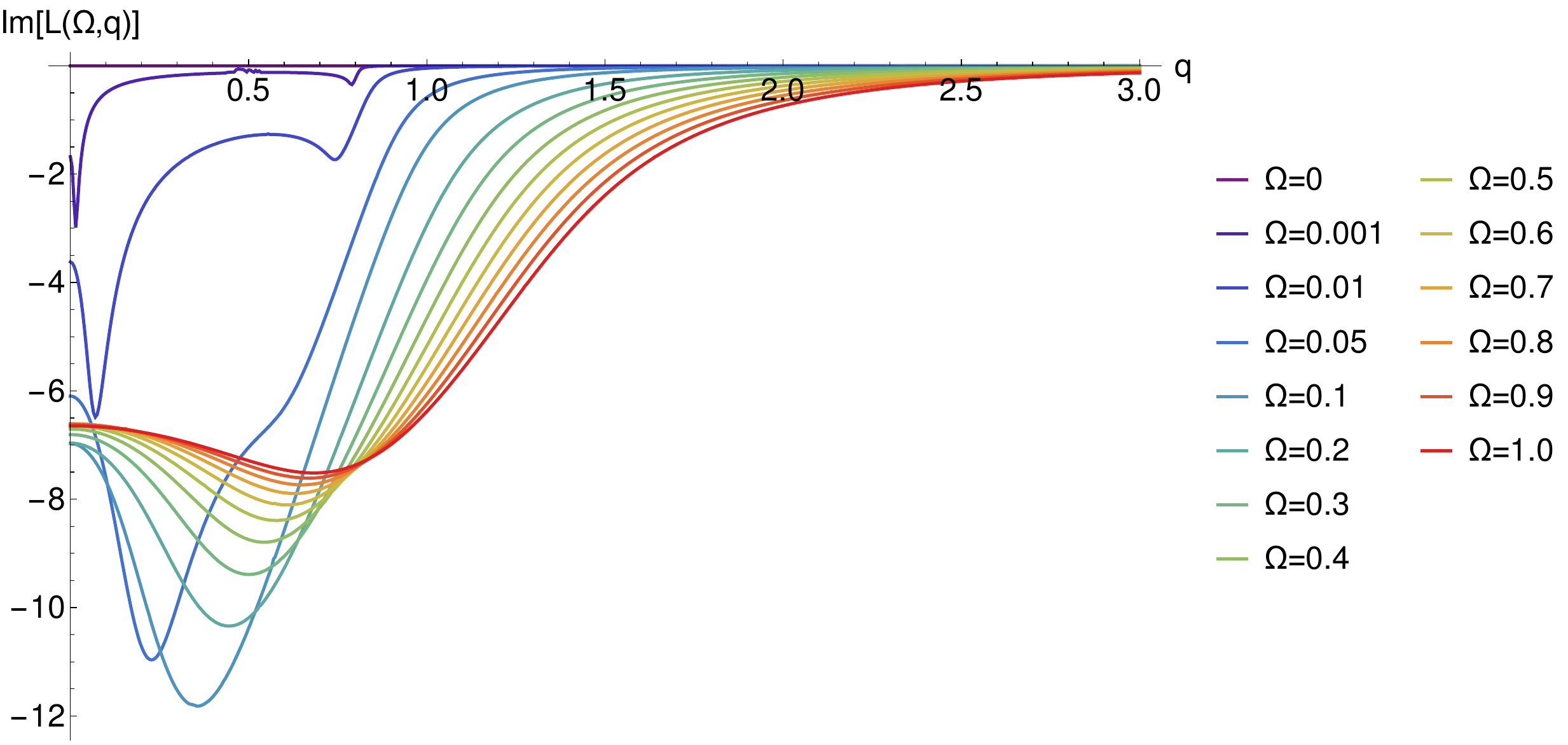}
\caption{Plots of ${\rm Im}\mathcal{L}(q)$ of the semi-holographic non-Fermi liquid for $\nu = 2/3$ and $\phi = \pi/4$ for various fixed values of $\Omega$. In order to compare with plots of ${\rm Im}\mathcal{L}^{\rm FL}(q)$ shown in Fig. \ref{fig:ImLqFL}, we need to take into account that for the above plots we have chosen $k_F = 0.4$, $m = 0.5$ (i.e. $\epsilon_F = 0.16$) and $\vert\zeta \vert = 1$.}
\label{fig:ImLq}
\end{figure}
Let us first study ${\rm Im}\mathcal{L}(q)$ for various fixed values of $\Omega$ for the case $\nu = 2/3$ and $\phi = \pi/4$ with the above choice of parameters. The plots are shown in Fig. \ref{fig:ImLq}. The trivial result is that we find that ${\rm Im}\mathcal{L}(q)$ vanishes at $\Omega = 0$ as expected on general grounds (see Section \ref{Lindhard Function}). We find the following non-trivial features:
\begin{enumerate}
\item For $\Omega <  \epsilon_F$, i.e. $\Omega < 0.2$, we find a sharp descent and a plateau type region squeezed between two peaks/kinks remarkably similar to the features in the Fermi liquid case governed by the \textit{inner core} of the particle hole continuum (particle-hole excitations lying close to the particle Fermi surface) which also appears when $\Omega/\epsilon_F < 1$. To see this distinctly, one can readily compare the curves in Fig. \ref{fig:ImLq} for $\Omega < 0.1$ to those in Fig. \ref{fig:ImLqFL} for $\Omega/\epsilon_F <  0.8$. Just like in the Fermi liquid case, this region between the twin peaks(kinks) shrinks with increasing $\Omega$ and disappears at $\Omega = \epsilon_F$.

\item Just like in the Fermi liquid case, the extent of the inner core features is governed by kinematics (see Figs. \ref{fig:kinematic} and  \ref{fig:Kinematics2}). For small $\Omega$, the extent of this region starts close to $q = 0$ and ends just short of $q = 2k_F$ (i.e. $q = 0.8$). The two end points move away from $q = 0$ and $q= 2k_F$ with increasing $\Omega$ merging at $q = k_F$ (i.e. $q = 0.4$) when $\Omega \approx \epsilon_F$ (i.e. $\Omega \approx 0.16$)\footnote{Here and elsewhere one must resist exact comparisons with Fermi liquid because there is some inherent ambiguity in determining the locations of $q$ corresponding to $\Omega_{\rm int}$, etc. Nevertheless the comparisons do hold approximately even quantitatively.}.

\item Unlike the Fermi liquid case, the plots in Fig. \ref{fig:ImLq} demonstrate no special values of $q$ corresponding to $\Omega_{\rm max}$ and $\Omega_{\rm min}$ (see Figs. \ref{fig:kinematic}, \ref{fig:Kinematics2} and \ref{fig:Kinematics3}) where ${\rm Im}\mathcal{L}(q)$ vanishes at fixed values of $\Omega$ as visible clearly in Fig. \ref{fig:ImLqFL}. Nevertheless, these values of $q$ corresponding to $\Omega_{\rm max}$ and $\Omega_{\rm min}$ get replaced by regions where $\partial^2{\rm Im}\mathcal{L}(q)/\partial q^2$ varies rapidly as happens near inflexion points. There are indeed two such regions for any value of $\Omega$ in the plots in Fig. \ref{fig:ImLq}, one centred at the value of $q$ where $\Omega_{\rm max}(q)$ is supposed to be in the Fermi liquid case and similarly another centred at the value of $q$ where $\Omega_{\rm min}(q)$ is supposed to be.
\end{enumerate}
To summarise, we can conclude that the Fermi liquid features arising from the inner core of the particle-hole continuum remain sharply defined in the semi-holographic non-Fermi liquid case in the same ranges of $\Omega$ and $q$. However, the rest of the continuum including the boundaries $\Omega_{\rm max}(q)$ and $\Omega_{\rm min}(q)$ are \textit{blurred out} -- in particular these boundaries get replaced by regions where $\partial^2{\rm Im}\mathcal{L}(q)/\partial q^2$ varies rapidly at fixed values of $\Omega$. The size of these regions (and hence the degree of blurring of the boundaries) increases with increasing $\Omega$. Physically the blurring out of the particle-hole continuum away from the inner core (i.e. the part of the continuum away from the particle Fermi surface) arises from the incoherent fermionic quasinormal mode excitations of the black hole geometry with which the electron hybridises and which also leads to particle-hole asymmetry.

\begin{figure}[htbp]
\centering
\includegraphics[width=\textwidth]{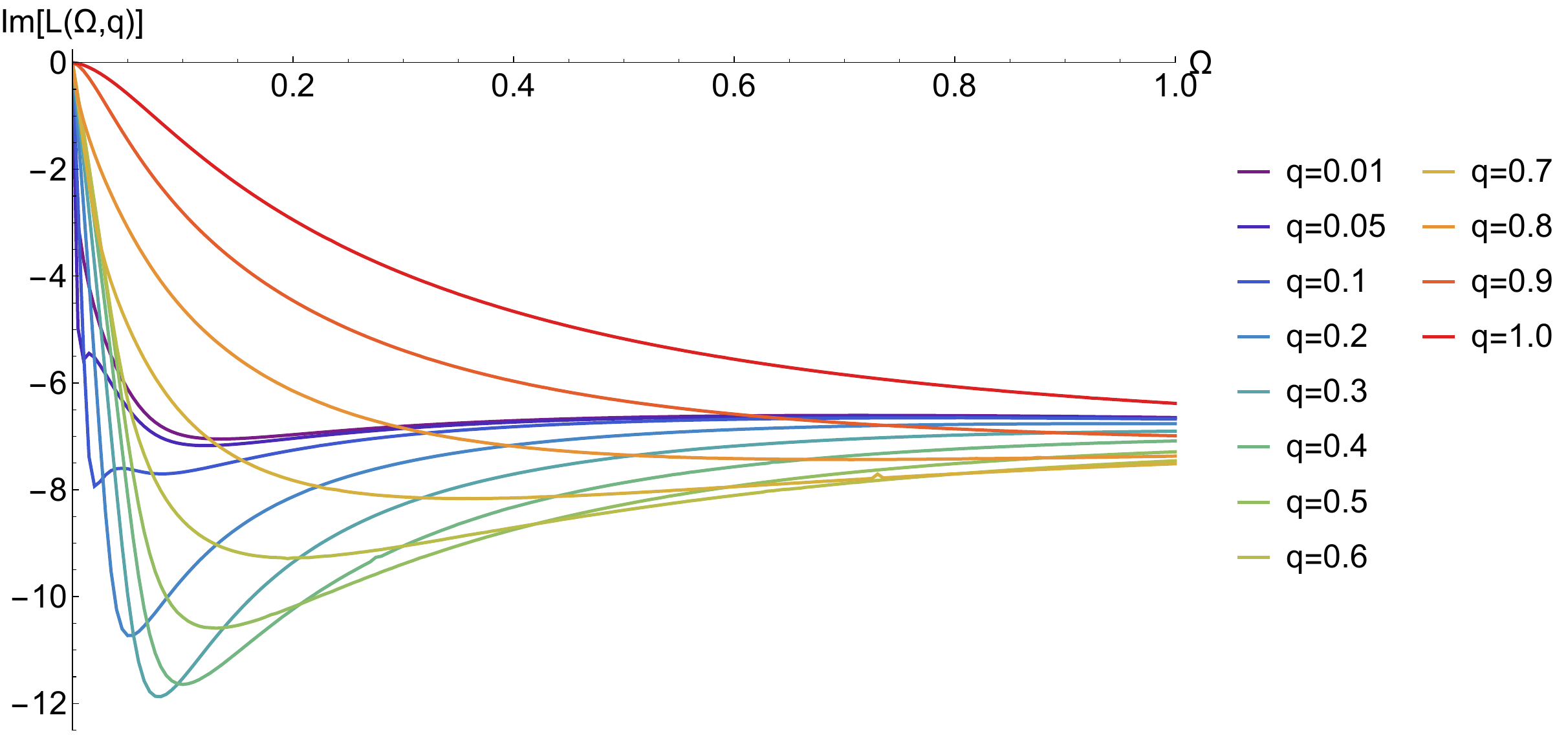}
\caption{Plots of ${\rm Im}\mathcal{L}(\Omega)$ of the semi-holographic non-Fermi liquid for $\nu = 2/3$ and $\phi = \pi/4$ for various fixed values of $q$. In order to compare with plots of ${\rm Im}\mathcal{L}^{\rm FL}(\Omega)$ shown in Fig. \ref{fig:ImLOmFL}, we need to take into account that for the above plots we have chosen $k_F = 0.4$, $m = 0.5$ (i.e. $\epsilon_F = 0.16$) and $\vert\zeta \vert = 1$.}
\label{fig:ImLOm}
\end{figure}

Our conclusions are also validated by the plots of ${\rm Im}\mathcal{L}(\Omega)$ at various fixed values of $q$ with the same choice of parameters as presented in Fig. \ref{fig:ImLOm}. Nevertheless some of these features are complicated by the presence of the energy cut-off $\omega_{\rm c}$. Firstly, as expected on general grounds (see Section \ref{Lindhard Function}), we find that ${\rm Im}\mathcal{L}(\Omega) \approx$ $\Omega$, for small $\Omega$ and small fixed values of $q$. The slope at $\Omega = 0$, i.e. ${\rm Im}\mathcal{L}(\Omega)/\Omega$ in the limit $\Omega \rightarrow 0$ grows with decreasing $q$ as in the Fermi liquid case which is evident from comparison with Fig. \ref{fig:ImLOmFL}. The other features are:
\begin{enumerate}
\item For fixed value of $q < k_F$ (i.e. $q < 0.4$), we find a peak for $\Omega = \Omega_{\rm int}(q)$, corresponding to the inner core boundary -- this peak also involves a sharp change in $\partial{\rm Im}\mathcal{L}(\Omega)/\partial \Omega$ for small $q$ as in the Fermi liquid case (see Fig. \ref{fig:ImLOmFL}). When $q= k_F = 0.4$, the peak appears close to $\epsilon_F = 0.16$ as expected. This inner core feature however is not so prominent for $q > k_F$ perhaps due to the presence of the energy cut-off $\omega_{\rm c} = 1$ although it is expected to persist for $k_F < q <2 k_F$.
\item For fixed values of $q < k_F (= 0.4)$, we find another region centred at the place where $\Omega_{\rm max}(q)$ is supposed to be where $\partial^2{\rm Im}\mathcal{L}(\Omega)/\partial\Omega^2$ changes rapidly.  Unlike the Fermi liquid case ${\rm Im}\mathcal{L}(\Omega)$ does not vanish for larger values of $\Omega$ indicating presence of spectral weight arising from fermionic quasinormal mode excitations of the black hole. The presence of the energy cut-off $\omega_{\rm c} =1$ perhaps makes this region less prominent for $q > k_F$.
\item For $q$ slightly greater than $2k_F$ (see the plots for $q = 0.9$ and $1.0$ in Fig. \ref{fig:ImLOm}), we find another region appearing close to $\Omega= 0$ where $\Omega_{\rm min}(q)$ is supposed to be and where $\partial^2{\rm Im}\mathcal{L}(\Omega)/\partial\Omega^2$ changes rapidly.
\end{enumerate}
Especially for fixed $q > k_F$, the presence of the energy cut-off complicates some of the features of ${\rm Im}\mathcal{L}$ as a function of $\Omega$. The comparison with the Fermi liquid is therefore best revealed when we study ${\rm Im}\mathcal{L}$ as function of $q$ for fixed $\Omega$ rather than the other way round.

\begin{figure}[htbp]
\centering
\includegraphics[width=\textwidth]{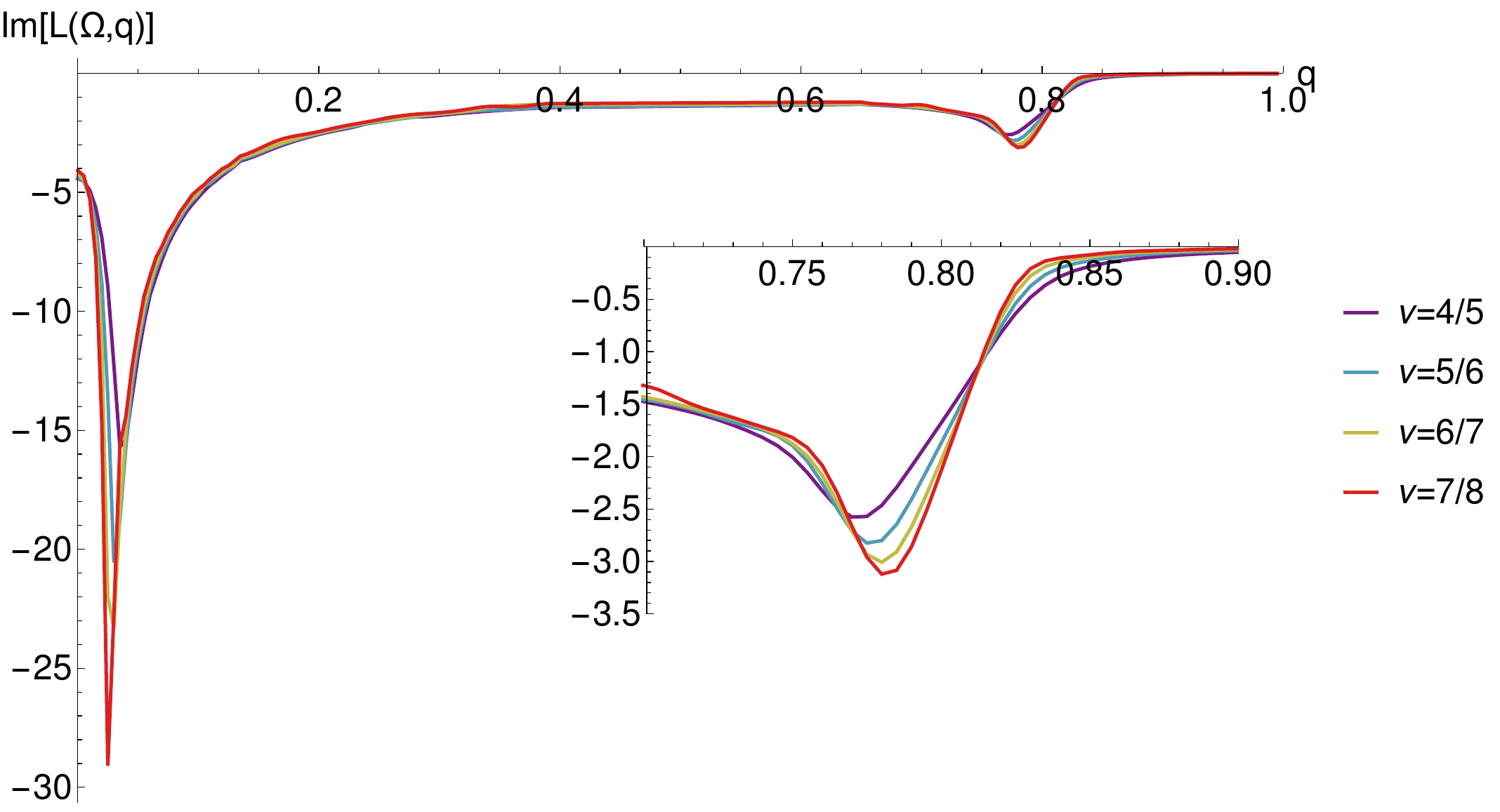}
\caption{Plots of ${\rm Im}\mathcal{L}(q)$ for $\Omega = 0.01$ for $\nu = n/(n+1)$ and $\phi = \pi/(n+2)$ with $n = 4, 5, 6$ and $7$. The other parameters are kept the same as in previous plots, i.e. $k_F = 0.4$, $m = 0.5$ and $\vert \zeta \vert = 1$. We see that as we increase $\nu$ keeping $\phi$ \textit{near-extremal}, ${\rm Im}\mathcal{L}(q)$ becomes closer to the Fermi liquid. The inlay plots demonstrate the behaviour near $q*$  for which $\Omega = \Omega_{\rm min}(q*) = 0.01$ (corresponding to the second boundary of the particle-hole continuum) where $\partial^2{\rm Im}\mathcal{L}(q)\partial q^2$ varies very rapidly. It is clear that with increasing $\nu$ the decay of ${\rm Im}\mathcal{L}(q)$ for $q>q*$ occurs faster. }
\label{fig:ImLqnu}
\end{figure}
As mentioned before, changing $\tilde{\epsilon}_F$ of the two effective parameters in Eq. (\ref{pars}) of the low energy theory does not change the qualitative features discussed above. Therefore, we report specifically on how the features of ${\rm Im}\mathcal{L}(q, \Omega)$ change when we increase $\nu$ while keeping $\phi$ \textit{near-extremal} (i.e. $\phi = \pi(1-\nu)- \epsilon$ with $\epsilon$ being a small non-negative number) for reasons described before and other parameters the same.\footnote{As mentioned before, choosing $\phi$ to be the extremal value exactly (i.e. imposing $\phi = \pi (1-\nu))$ gives rise to numerical noise because the singularities in the propagators approach the real axis, so we avoid doing this.} To do this, we can choose $\nu = n/(n+1)$ and $\phi = \pi/(n+2)$, and study the cases $\nu = 4, 5, 6$ and $7$. The plots of ${\rm Im}\mathcal{L}(q)$ are presented in Fig. \ref{fig:ImLqnu} for a representative value of $\Omega = 0.01$ (we recall that we have set $k_F = 0.4$, $m = 0.5$ and $\vert \zeta \vert = 1$). 

It is clear from  Fig. \ref{fig:ImLqnu} that as we increase $\nu$, the spectral weight in the inner core region squeezed between the two peaks increases -- therefore the spectral weight of the coherent particle-hole continuum is enhanced compared to that of the incoherent quasinormal mode excitations. The extent of the inner core region is given by the kinematics of $k-$space and is therefore independent of $\nu$. However the decay of the spectral weight away from this core inner regions occurs faster as $\nu$ becomes closer to $1$ -- in particular the regions centred around the values of $q$ where $\Omega_{\rm min}$ and $\Omega_{\rm max}$ become more prominent so that the spectral weights away from these boundaries become more and more insignificant thus approaching Fermi liquid type behaviour. Similar features are seen particularly for $\Omega < \epsilon_F$. Therefore, $\nu$ can be thought of as a deformation parameter -- by decreasing $\nu$ from $1$ towards $1/2$ keeping $\phi$ near-extremal, we can interpolate between Fermi-liquid-like behaviour and non-Fermi liquid behaviour particularly for $\Omega < \epsilon_F$ (which is needed for the inner core region to exist). \footnote{We believe that this interpolation to Fermi liquid as $\nu \rightarrow 1$ can be made perfectly if we keep the phase strictly extremal; however as discussed before, it is not easy to achieve this numerically.} It is remarkable that the features governed by the inner core of the particle-hole continuum corresponding to the pair excitations closer to the particle Fermi surface remain sharply defined as we change $\nu$.

\subsection{Real part of the generalised Lindhard function}

The real part of the Lindhard function is related to the imaginary part of the Lindhard function via the Kramers-Kronig relations\footnote{The Kramers-Kronig relations for the retarded propagator imposes similar relations for the Feynman propagator.} as discussed in Section \ref{Lindhard Function}. Since we need to impose a frequency cut-off $\omega_{\rm c}$ in our effective semi-holographic non-Fermi liquid framework for reasons discussed before, the Kramers-Kronig relations will be valid only up to corrections involving $\Omega/\omega_{\rm c}$, where $\Omega$ is the frequency at which we are observing the response.\footnote{Recall that to do this we need to find the right type of UV completion, i.e. the right type of material properties that ensures appropriate large frequency behaviour of the fermionic Green's functions. Since the exact Kramers-Kronig relations dictate that UV (i.e. large $\Omega$) behaviour of the imaginary part of the the Lindhard function can affect the real part in the IR (i.e. small $\Omega$), we need to ensure appropriate UV behaviour of the fermionic propagators in order that the effective semi-holographic theory results hold in the IR.} 

\begin{figure}[ht]
\begin{minipage}[b]{0.8\linewidth}
\centering
\includegraphics[width=\textwidth]{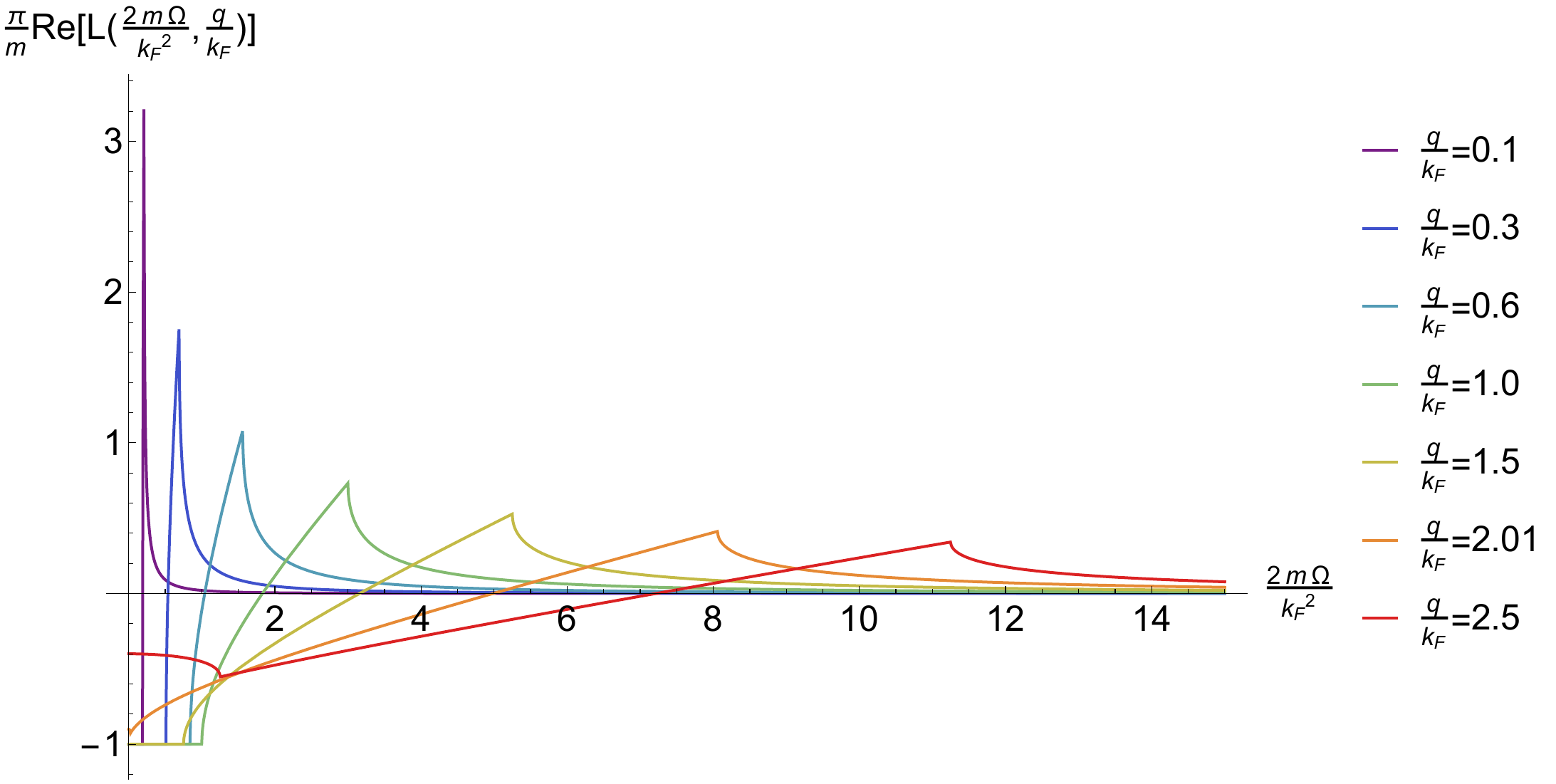}
\caption{\label{fig:ReLOmFL} Plots of ${\rm Re}\mathcal{L}^{\rm FL}(\Omega)$ for various fixed values of $q$ are shown above.}
\end{minipage}
\hspace{0.5cm}
\begin{minipage}[b]{0.8\linewidth}
\centering
\includegraphics[width=\textwidth]{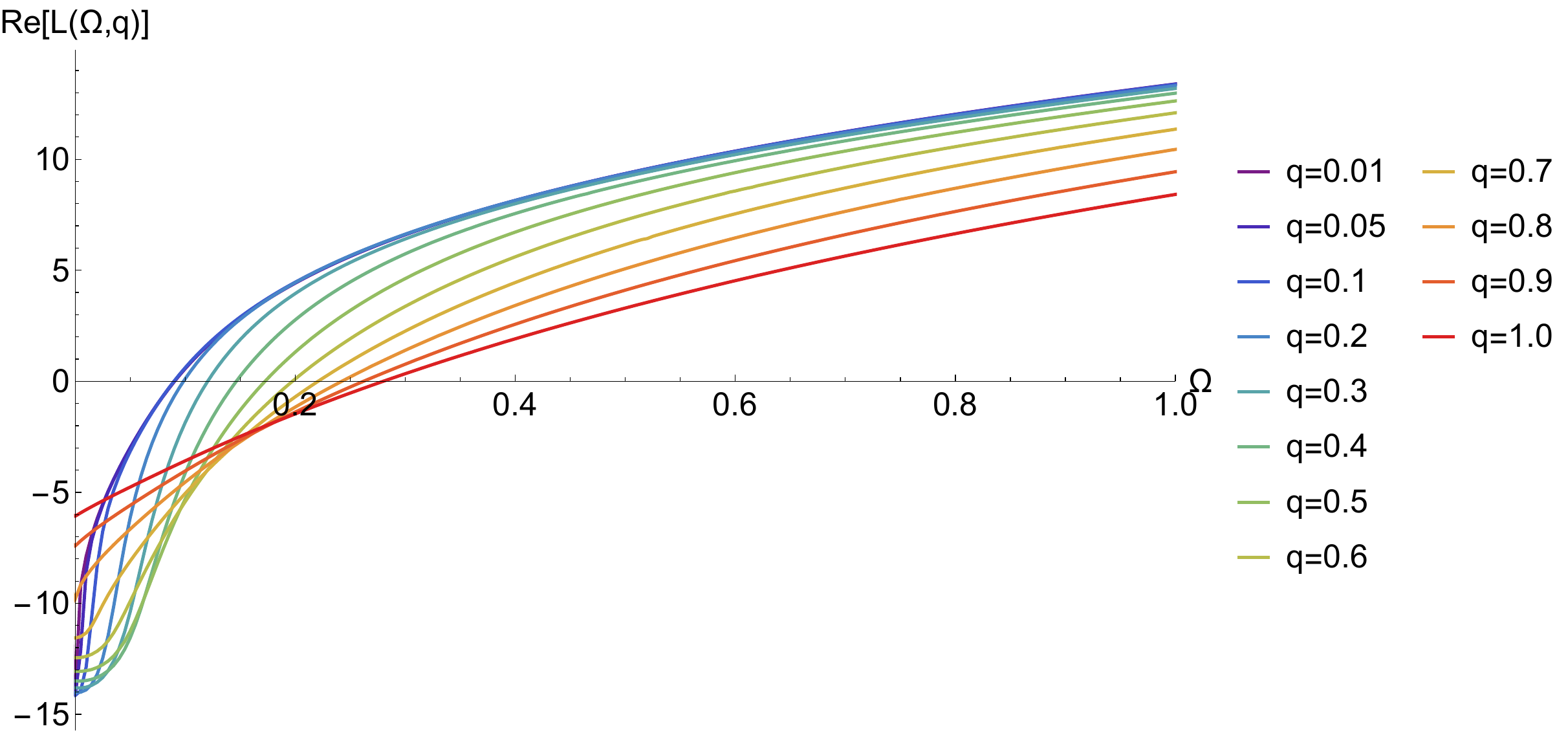}
\caption{\label{fig:ReLOmega} Plots of ${\rm Re}\mathcal{L}(\Omega)$ for various fixed values of $q$ are shown above.  We have made the same choices of parameters as for ${\rm Im}\mathcal{L}(q,\Omega)$ in Fig. \ref{fig:ImLOm}, i.e. for $\nu = 2/3$, $\phi = \pi/4$, $k_F = 0.4$, $m=0.5$ and $\vert\zeta\vert = 1$.}
\end{minipage}
\end{figure}

Examining ${\rm Re}\mathcal{L}^{\rm FL}(\Omega, q)$ as a function of $\Omega$ for various fixed values of $q$ (see Fig. \ref{fig:ReLOmFL}), we can readily see how it correlates with ${\rm Im}\mathcal{L}^{\rm FL}(\Omega, q)$ (see Fig. \ref{fig:ImLOmFL}).
\begin{enumerate}
\item For fixed $q < 2k_F$, ${\rm Re}\mathcal{L}^{\rm FL}(\Omega/\epsilon_F)$ is almost constant and negative for $\Omega < \Omega_{\rm int}(q)$ (see plots for $q/k_F= 0.1$, $0.3$, $0.6$, $1.0$ and $1.5$ in Fig. \ref{fig:ReLOmFL}).  Note that $\Omega_{\rm int}(q)$ attains its maximal value $\epsilon_F$ when $q/k_F = 1$, so the (negative) plateau has the largest extension for $q = k_F$. In the domain $\Omega_{\rm int}(q) < \Omega < \Omega_{\rm max}(q)$, ${\rm Re}\mathcal{L}^{\rm FL}(\Omega/\epsilon_F)$ is monotonically increasing for increasing values of $\Omega$ and changes sign at an intermediate point. ${\rm Re}\mathcal{L}^{\rm FL}(\Omega/\epsilon_F)$ peaks at $\Omega = \Omega_{\rm max}(q)$. For $\Omega > \Omega_{\rm max}(q)$, ${\rm Re}\mathcal{L}^{\rm FL}(\Omega/\epsilon_F)$ monotonically decreases to zero for increasing values of $\Omega$ while staying positive definite in sign. Recall that ${\rm Im}\mathcal{L}(\Omega)$ vanishes in this domain.
\item For fixed $q > 2k_F$, ${\rm Re}\mathcal{L}^{\rm FL}(\Omega/\epsilon_F)$ does not have any plateau-like flat region for small values of $\Omega$ (see plots for $q/k_F= 2.01$ and $2.5$ in Fig. \ref{fig:ReLOmFL}). We recall that for $q > 2k_F$, there is no $\Omega_{\rm int}(q)$ because the inner core of the particle-hole continuum does not extend here. Furthermore, ${\rm Im}\mathcal{L}(\Omega, q)$ is supported only between $\Omega_{\rm min}(q) < \Omega < \Omega_{\rm max}(q) $. For $\Omega < \Omega_{\rm min}(q)$, ${\rm Re}\mathcal{L}^{\rm FL}(\Omega/\epsilon_F)$ stays negative and decreases monotonically reaching a minima at $\Omega = \Omega_{\rm min}(q)$. In the domain $\Omega_{\rm min}(q) < \Omega < \Omega_{\rm max}(q)$, ${\rm Re}\mathcal{L}^{\rm FL}(\Omega/\epsilon_F)$ monotonically increases changing sign at an intermediate point and reaching a maxima at $\Omega = \Omega_{\rm max}(q)$. ${\rm Re}\mathcal{L}^{\rm FL}(\Omega/\epsilon_F)$ again decreases monotonically to zero for $ \Omega > \Omega_{\rm max}(q)$ staying positive definite in sign.
\end{enumerate}

We study features of ${\rm Re}\mathcal{L}(q,\Omega)$ of the semi-holographic non-Fermi liquid as a function of $\Omega$ for fixed values of $q$. The plots are presented in Fig. \ref{fig:ReLOmega} for the same choices of parameters as for ${\rm Im}\mathcal{L}(q,\Omega)$ in Fig. \ref{fig:ImLOm}, i.e. for $\nu = 2/3$, $\phi = \pi/4$, $k_F = 0.4$, $m=0.5$ and $\vert\zeta\vert = 1$.
\begin{enumerate}
\item We note that for fixed $q < 2k_F$ (i.e. for $q < 0.8$) there is a small flat region of ${\rm Re}\mathcal{L}(\Omega)$ for small values of $\Omega$ which is reminiscent of the Fermi liquid features appearing when $\Omega < \Omega_{\rm int} (q)$ corresponding to the inner core of the particle-hole continuum. As in the case of the Fermi liquid, when $\Omega < \Omega_{\rm max}(q)$, ${\rm Re}\mathcal{L}(\Omega)$ monotonically increases for increasing values of $\Omega$ and changes sign at an intermediate value of $\Omega$. Furthermore, $\partial^2{\rm Re}\mathcal{L}(\Omega)/\partial\Omega^2$ varies rapidly near the value $\Omega$  corresponding to where $\Omega_{\rm max}(q)$ is supposed to be. However, instead of peaking at $\Omega = \Omega_{\rm max}(q)$ and then decreasing for increasing $\Omega$ as in the case of the Fermi liquid, ${\rm Re}\mathcal{L}(\Omega)$ increases and saturates to its maximal value at the cut-off $\Omega = \omega_{\rm c} \approx 1$.
\item For fixed $q> 2k_F$ (i.e. for $q > 0.8$), ${\rm Re}\mathcal{L}(\Omega)$ has no flat region for small values of $\Omega$ as in the case of the Fermi liquid. The features corresponding to the boundaries of the particle-hole continuum, i.e. $\Omega = \Omega_{\rm min}(q)$ and $\Omega = \Omega_{\rm max}(q)$ which are sharply visible in the Fermi liquid case are also blurred out and replaced by an approximate linear growth in the entire regions $0 < \Omega < \omega_{\rm c}$. Nevertheless, the change in sign of ${\rm Re}\mathcal{L}(\Omega)$ occurs at an intermediate point between $\Omega_{\rm min}(q)$ and  $\Omega_{\rm max}(q)$ as in the case of the Fermi liquid.
\end{enumerate}

As we have observed in the case of ${\rm Im}\mathcal{L}(\Omega)$ for fixed values of $q$, the comparisons between the Fermi liquid and the semi-holographic non-Fermi liquid are complicated by the presence of the frequency cut-off $\omega_{\rm c}$. Nevertheless, as in the case of ${\rm Im}\mathcal{L}(\Omega,q)$, we will find the comparisons with the Fermi liquid are easier to do when we study ${\rm Re}\mathcal{L}(\Omega,q)$ as a function of $q$ for fixed values of $\Omega$. 

\begin{figure}[ht]
\begin{minipage}[b]{0.8\linewidth}
\centering
\includegraphics[width=\textwidth]{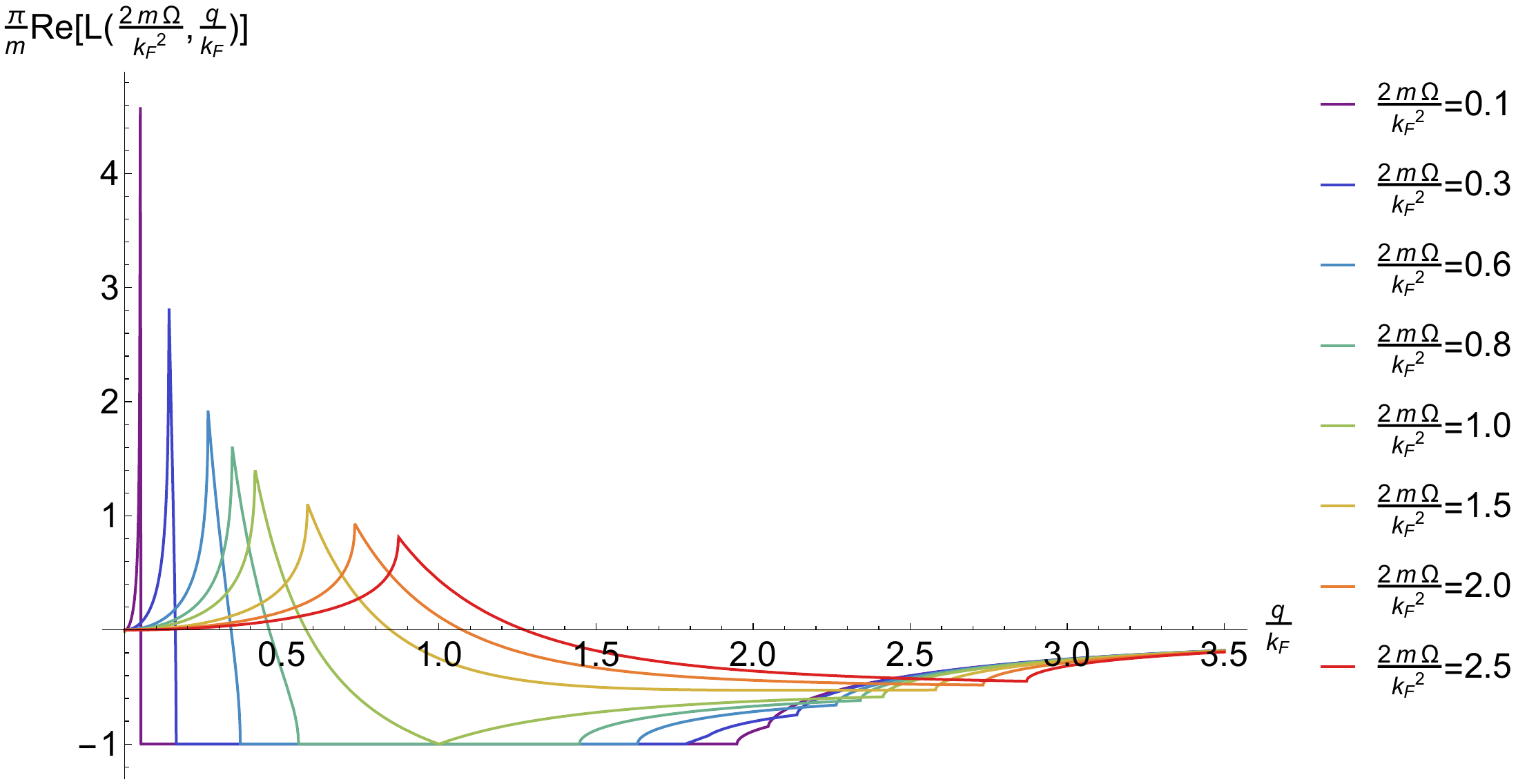}
\caption{\label{fig:ReLqFL} Plots of ${\rm Re}\mathcal{L}^{\rm FL}(q)$ for various fixed values of $\Omega$ are shown above.}
\end{minipage}
\hspace{0.5cm}
\begin{minipage}[b]{0.8\linewidth}
\centering
\includegraphics[width=\textwidth]{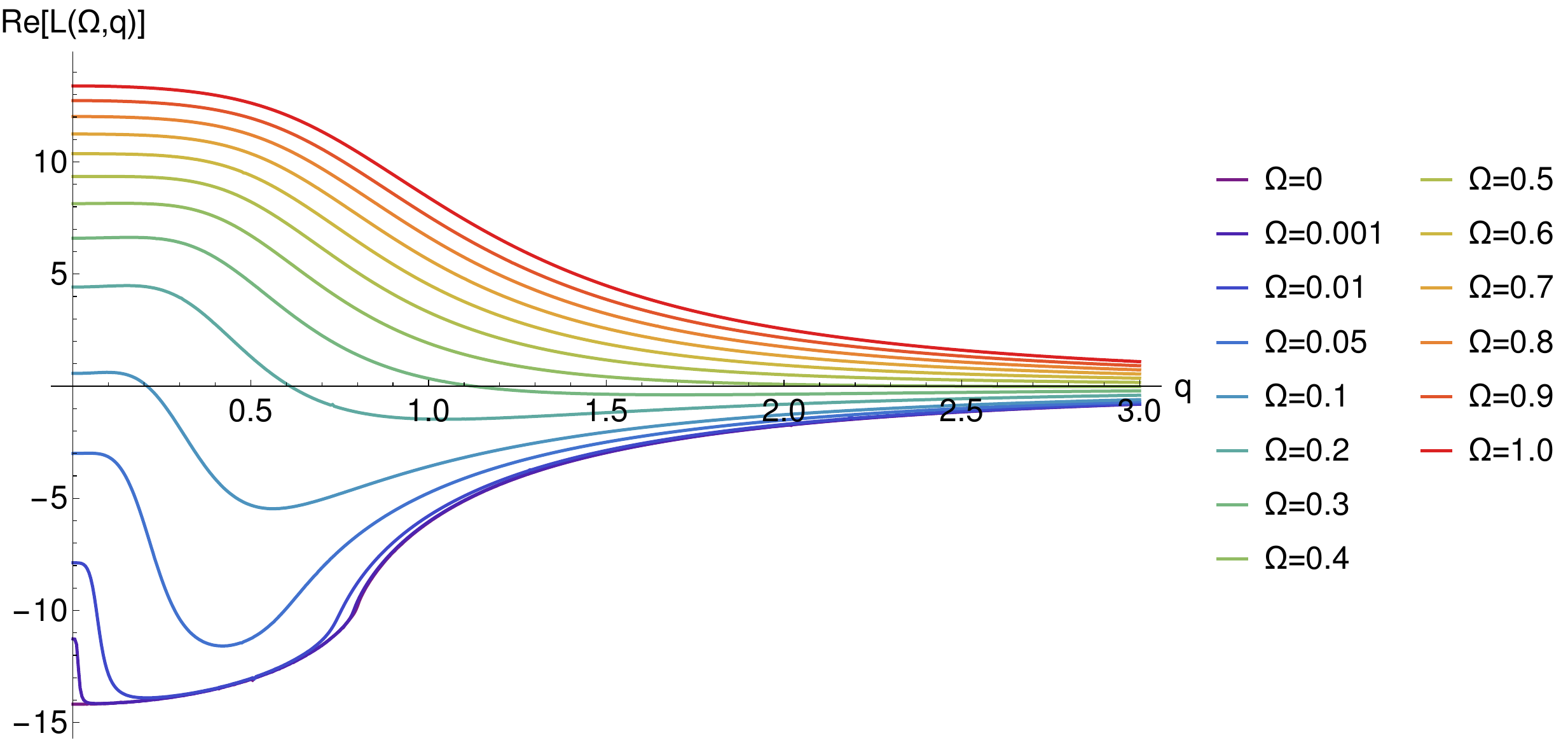}
\caption{\label{fig:ReLq} Plots of ${\rm Re}\mathcal{L}(q)$ for various fixed values of $\Omega$ are shown above.  We have made the same choices of parameters as for ${\rm Im}\mathcal{L}(q,\Omega)$ in Fig. \ref{fig:ImLq}, i.e. for $\nu = 2/3$, $\phi = \pi/4$, $k_F = 0.4$, $m=0.5$ and $\vert\zeta\vert = 1$.}
\end{minipage}
\end{figure}

We can readily compare the Fermi liquid ${\rm Re}\mathcal{L}^{\rm FL}(\Omega,q)$ (see Fig. \ref{fig:ReLqFL}) with ${\rm Re}\mathcal{L}(\Omega,q)$ of the semi-holographic non-Fermi liquid (see Fig. \ref{fig:ReLq}) as a function of $q$ for various fixed values of $\Omega$. We set the values of all parameters of the semi-holographic non-Fermi liquid as just above. 

\begin{figure}[htbp]
\centering
\includegraphics[width=\textwidth]{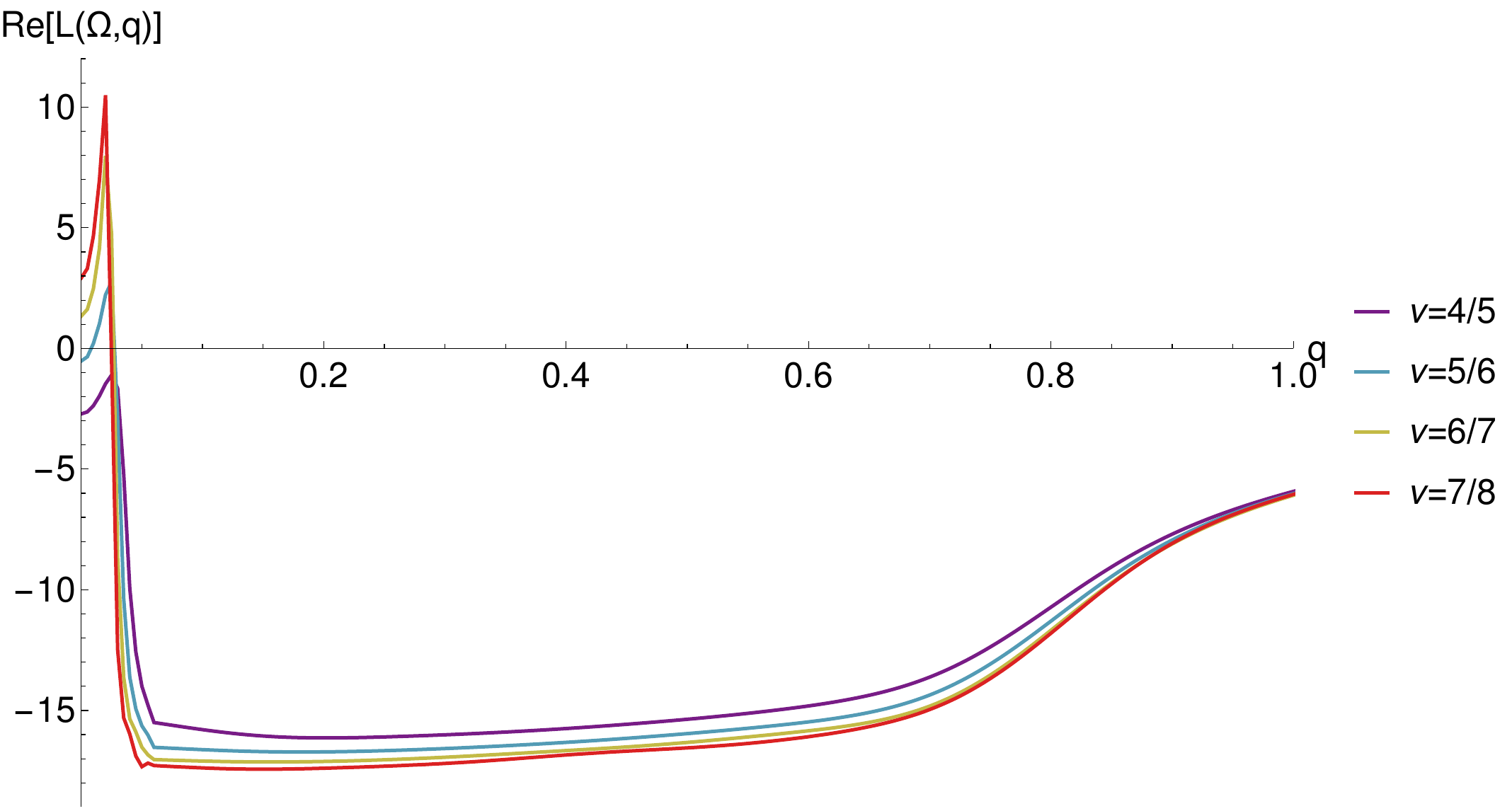}
\caption{Plots of ${\rm Re}\mathcal{L}(q)$ for $\Omega = 0.01$ for $\nu = n/(n+1)$ and $\phi = \pi/(n+2)$ with $n = 4, 5, 6$ and $7$. The other parameters are kept the same as in previous plots, i.e. $k_F = 0.4$, $m = 0.5$ and $\vert \zeta \vert = 1$.}
\label{fig:ReLqnu}
\end{figure}

Let us first consider the case of $\Omega < \epsilon_F$ (i.e. $\Omega < 0.16$ in the case of the semi-holographic non-Fermi liquid). We recall that when $\Omega < \epsilon_F$, a horizontal line corresponding to a fixed value of $\Omega$ passing through the particle-hole continuum has four special points, namely those which hit the boundaries $\Omega_{\rm min}(q)$ and $\Omega_{\rm max}(q)$ at $q_1$ and $q_4$ respectively, and the two points $q_2$ and $q_3$ where it intersects the boundary of the inner core $\Omega_{\rm int}(q)$ such that $q_1 < q_2 < q_3 < q_4$. In the domain $\Omega < \epsilon_F$, the comparisons between the Fermi liquid and the semi-holographic non-Fermi liquid can be based on the following points.\footnote{We again remind the reader that the comparisons can only be approximate because there is an inherent ambiguity in defining $q_1$, $q_2$, $q_3$ and $q_4$ exactly in the case of the semi-holographic non-Fermi liquid.}
\begin{enumerate}
\item In the range $q < q_1$ at fixed $\Omega$, ${\rm Re}\mathcal{L}^{\rm FL}(q)$ increases dramatically in the Fermi liquid case from zero to a large positive value peaking sharply at $q=q_1$ (see Fig. \ref{fig:ReLqFL}). In the semi-holographic case, the behaviour is somewhat flat, however $\partial^2{\rm Re}\mathcal{L}(q)/\partial q^2$ varies rapidly around $q=q_1$ as reflected by change in the curvatures of the plots in Fig. \ref{fig:ReLq}. Furthermore, if we increase $\nu$ keeping $\phi$ near-extremal and all other parameters the same as in Fig. \ref{fig:ImLqnu}, even the semi-holographic behaviour becomes more Fermi-liquid-like. This is evident from Fig. \ref{fig:ReLqnu} where we have plotted ${\rm Re}\mathcal{L}(q)$ for various values of $\nu$ at a fixed representative value of $\Omega$ set to $0.01$. For $\nu = 5/6$, $6/7$ and $7/8$ particularly, 
${\rm Re}\mathcal{L}(q)$ at $\Omega = 0.01$increases sharply with $q$ for $q< q_1$ staying positive definite in sign and at $q=q_1$ there is a peak as in the Fermi liquid case, although it is slightly blurred by the incoherent nature of the quasinormal mode excitations.

\item In the region $q_1 < q <q_2$, both in the cases of the Fermi liquid and semiholographic non-Fermi liquid ${\rm Re}\mathcal{L}(q)$ is monotonically decreasing as we can find from Figs. \ref{fig:ReLqFL} and \ref{fig:ReLq}. Note that the point $q= q_2$ is slightly blurred by the incoherent nature of the quasinormal mode excitations in the semi-holographic case. If we increase $\nu$ towards $1$ keeping $\phi$ near-extremal, ${\rm Re}\mathcal{L}(q)$ also changes sign as in the Fermi liquid case at an intermediate value of $q$ as evident from Fig. \ref{fig:ReLqnu}. 

\item In the inner core region $q_2 < q < q_3$, the behaviour of ${\rm Re}\mathcal{L}(q)$ is relatively flat as in the case of the Fermi liquid ${\rm Re}\mathcal{L}^{\rm FL}(q)$. This flat behaviour is also conserved as we increase $\nu$ keeping $\phi$ near-extremal as evident from Fig. \ref{fig:ReLqnu}. Note that at $\Omega = 0$, $q_2$ coincides with $q_1$ and $q_3$ coincides with $q_4$ making the flat region prominent. As we increase $\Omega$ towards $\epsilon_F$ (which equals $0.16$ in the semi-holographic case), this flat region corresponding to the inner core shrinks to zero.
\item For $q > q_3$, both ${\rm Re}\mathcal{L}(q)$ and ${\rm Re}\mathcal{L}^{\rm FL}(q)$ increase monotonically with increasing $q$ from a negative value towards $0$ (see Figs. \ref{fig:ReLqFL} and \ref{fig:ReLq}). However, in case of the Fermi liquid, there is a sharp kink at $q= q_4$ corresponding to the other boundary of the particle-hole continuum. This kink is replaced by a region where $\partial^2{\rm Re}\mathcal{L}(q)/\partial q^2$ varies rapidly in the semi-holographic case. This feature is also preserved as we increase $\nu$ towards $1$ keeping $\phi$ near-extremal and all other parameters the same as evident from Fig.  \ref{fig:ReLqnu}.
\end{enumerate}
To summarise, one can conclude that the features of the Fermi liquid corresponding to the inner core region are preserved in the semi-holographic case but the boundaries of the particle-hole continuum are blurred out with peaks/kinks replaced by small regions where $\partial^2{\rm Re}\mathcal{L}(q)/\partial q^2$ varies rapidly particularly for values of $\nu$ closer to $1/2$ (eg. $\nu = 2/3$). Increasing $\nu$ keeping $\phi$ near-extremal and all other parameters the same, makes the first boundary $\Omega = \Omega_{\max}(q_1)$ more visible as a peak with a positive value as in the Fermi liquid case, but the other boundary $\Omega = \Omega_{\min}(q_4)$ remains blur in the above sense. These conclusions are thus very similar to those we made for ${\rm Im}\mathcal{L}(q)$ at fixed values of $\Omega$.

Finally, we can compare ${\rm Re}\mathcal{L}^{\rm FL}(q)$ (see Fig. \ref{fig:ReLqFL}) with ${\rm Re}\mathcal{L}(q)$ of the semi-holographic non-Fermi liquid (see Fig. \ref{fig:ReLq})  for various fixed values of $\Omega > \epsilon_F$ where the inner core region is absent. In the case of the Fermi liquid, there are two sharp kinks corresponding to values of $q$ where $\Omega = \Omega_{\rm max}(q_1)$ and $\Omega = \Omega_{\rm min}(q_2)$, i.e. where the horizontal line at fixed $\Omega$ intersects the boundaries of the particle-hole continuum (see Fig. \ref{fig:ReLqFL}). These sharp kinks are replaced by regions where $\partial^2{\rm Re}\mathcal{L}(q)/\partial q^2$ varies rapidly in the semi-holographic case (see Fig. \ref{fig:ReLq}) as near inflexion points. However, ${\rm Re}\mathcal{L}(q)$ changes sign at an intermediate value of $q$ like the Fermi liquid  ${\rm Re}\mathcal{L}^{\rm FL}(q)$ only if $\Omega$ is not much larger than $\epsilon_F$. ${\rm Re}\mathcal{L}(q)$ stays positive definite for fixed $\Omega \gg \epsilon_F$ retaining the features of the particle-hole continuum boundaries in the blurred form of small regions where $\partial^2{\rm Re}\mathcal{L}(q)/\partial q^2$ varies rapidly. Once again, our conclusions here agree with those we made for the case of ${\rm Im}\mathcal{L}(q)$ at fixed values of $\Omega > \epsilon_F$.

\section{Random phase approximation for effective Coulomb interactions}\label{collective}
The medium modifications of the Coulomb interactions between the electrons can be understood via the Random Phase Approximation which actually stands for summing over \textit{ring diagrams} involving the bare Coulomb interaction. 

The RPA has been extensively used in many-body physics since the 1950's. In metals, it is
justified by the long range character of the bare Coulomb repulsion, which makes it possible
to treat the response of the electronic fluid to a slowly varying perturbation by a self-consistent
field approach. A detailed discussion of RPA in this context can be found in chapter 5 of the book by
Pines and Nozi\`eres~\cite{Pines66}.

\begin{figure}[htbp]
\centering
\includegraphics[width=0.85\textwidth]{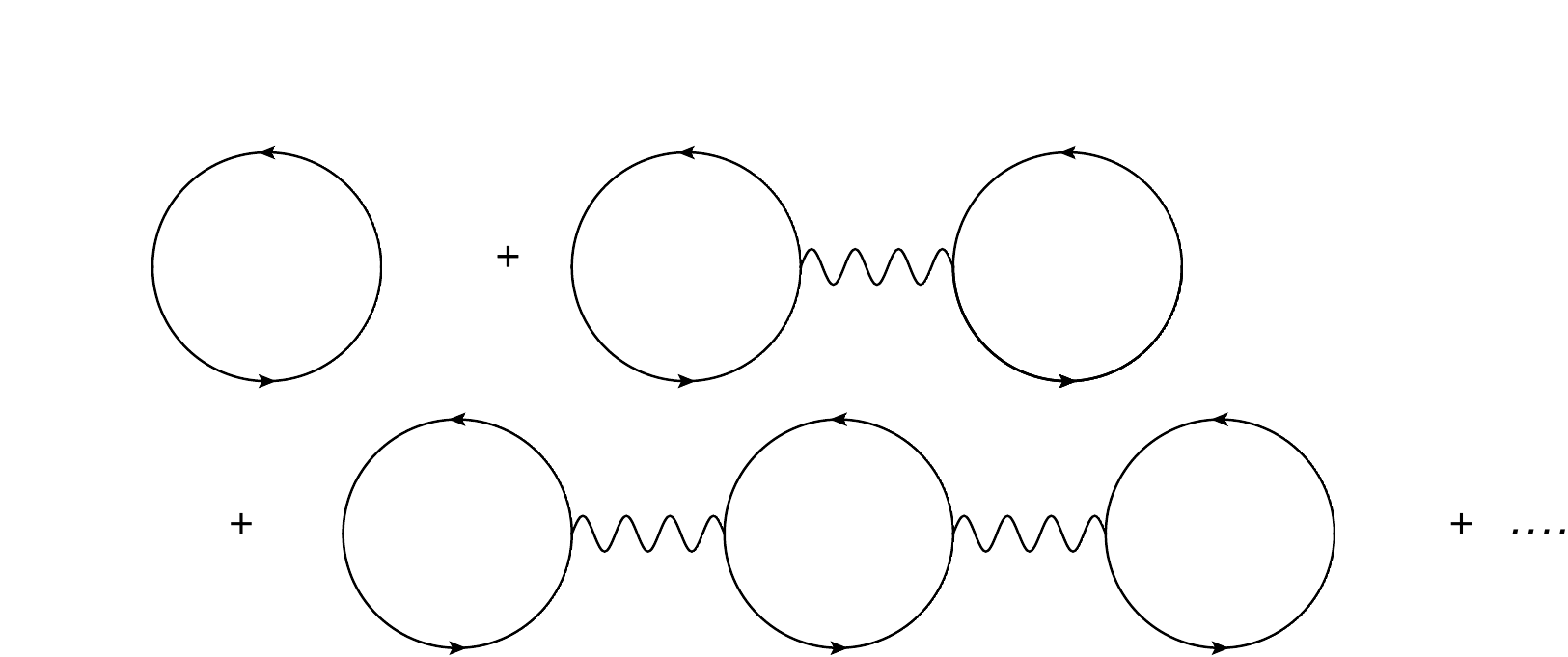}
\caption{The ring diagrams summing which we obtain $\mathcal{L}^{\rm imp}(q, \Omega)$.}
\label{fig:RPA1}
\end{figure}

\begin{figure}[htbp]
\centering
\includegraphics[width=0.85\textwidth]{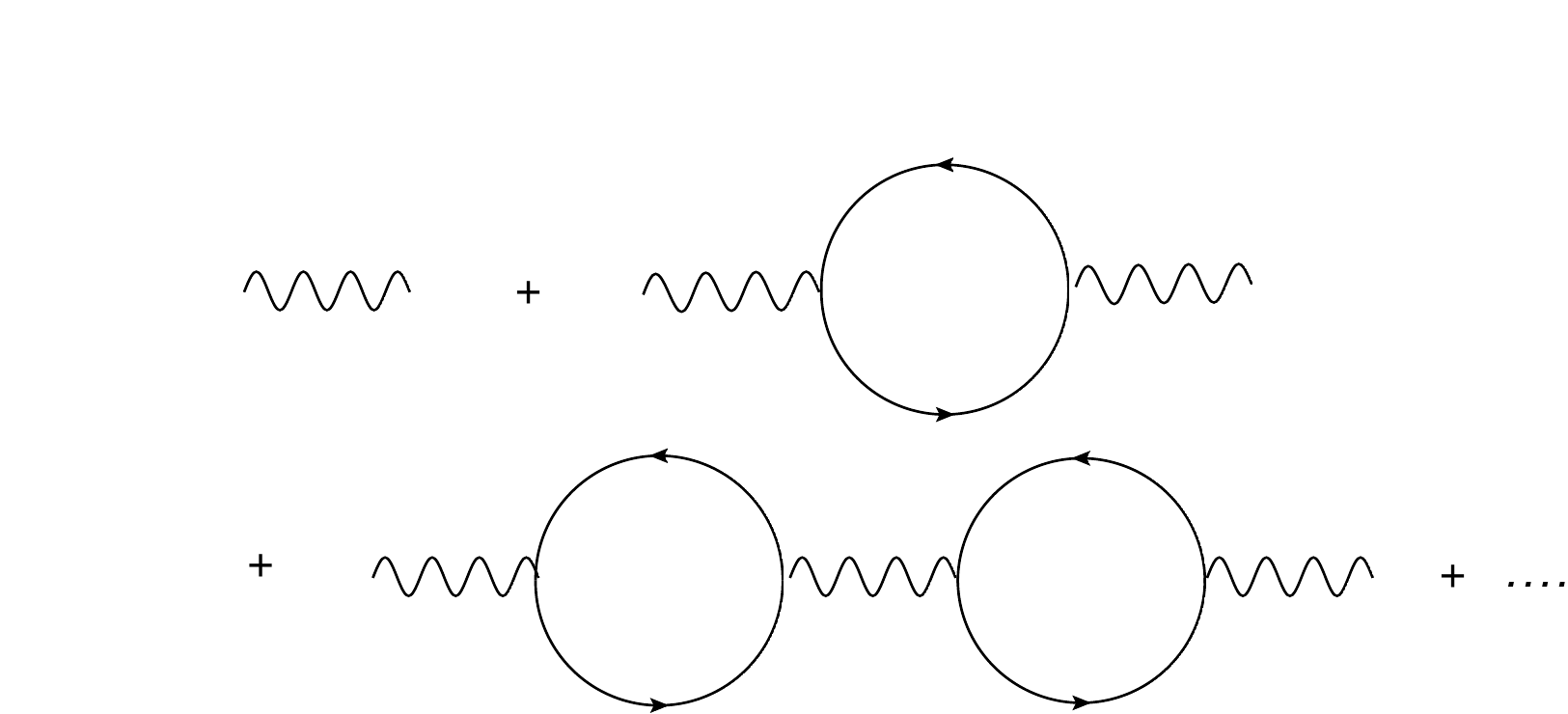}
\caption{The ring diagrams summing which we obtain $V_s(q, \Omega)$.}
\label{fig:RPA2}
\end{figure}

Two objects of interest here are (i) the improved  generalised Lindhard function which we will denote as $\mathcal{L}^{\rm imp}$ and (ii) the dynamically screened Coulomb potential which we will denote as $V_s$. Both of these objects can be obtained by summing over ring diagrams shown in Figs. \ref{fig:RPA1} and \ref{fig:RPA2} respectively. The bubbles in these diagrams stand for the one-loop generalised Lindhard function which we have studied in the previous section and the wavy lines denote the  Coulomb potential. 

It is clear from Figs. \ref{fig:RPA1} and \ref{fig:RPA2} that summing over the ring diagrams we obtain
\begin{eqnarray}
\mathcal{L}^{\rm imp}(q, \Omega) &=&  \frac{\mathcal{L}(q,\Omega)}{1- V(q)\mathcal{L}(q,\Omega)},\label{Limp} \\
V_s(q, \Omega)&=&\frac{V(q)}{1- V(q)\mathcal{L}(q,\Omega)}\label{Vs}.
\end{eqnarray}
Since we are interested in particular in 2D non-Fermi liquids, we consider a potential $V(q) = e^2/(2\epsilon_b q)$.\footnote{In the cuprates, for example, the electrons are mechanically confined in a 2D plane. Although the electric field lines are not confined to this plane, we need to do a 2D Fourier transform of the Coulomb potential $e^2/(4\pi \epsilon_b \vert \mathbf{r}\vert)$. This gives the $1/q$ potential with a prefactor determined by the dielectric constant $\epsilon_b$ of the material.} 

In order to investigate collective response it will be useful to define the retarded $\mathcal{L}^{\rm imp}_R$ as below:
\begin{equation}
\mathcal{L}^{\rm imp}_R(q, \Omega) =  \frac{\mathcal{L}_R(q,\Omega)}{1- V(q)\mathcal{L}_R(q,\Omega)}.\label{LimpR}
\end{equation}
It follows that analogous to $\mathcal{L}_R$, $\mathcal{L}^{\rm imp}_R$ also satisfies:
\begin{equation}
{\rm Re}\mathcal{L}^{\rm imp}_R (q, \Omega)= {\rm Re}\mathcal{L}^{\rm imp}(q, \Omega), \quad {\rm Im}\mathcal{L}^{\rm imp}_R (q, \Omega)= {\rm Im}\mathcal{L}^{\rm imp}(q, \Omega) {\rm sgn}(\Omega)
\end{equation}
at zero temperature. Once again given that $\mathcal{L}(q, \Omega) = \mathcal{L}(q, -\Omega) $, we obtain $\mathcal{L}^{\rm imp}(q, \Omega) = \mathcal{L}^{\rm imp}(q, -\Omega) $, ${\rm Re}\mathcal{L}^{\rm imp}_R(q, \Omega) = {\rm Re}\mathcal{L}^{\rm imp}_R(q, -\Omega) $ and ${\rm Im}\mathcal{L}^{\rm imp}_R(q, \Omega) = -{\rm Im}\mathcal{L}^{\rm imp}_R(q, -\Omega) $.

Below we will study the effective electronic interactions and collective behaviour of our semi-holographic model and compare them to Fermi-liquid behaviour. We will find striking and unexpected contrasts, and we will argue that these contrasts result from the character of the continuum outside of the inner core region.

\subsection{Plasma oscillations}
In order to trigger plasma oscillations in an electronic system, we consider a kick to the system with a time-dependent inhomogeneous external electric potential of the form:
\begin{equation}
\phi^{\rm ext}(\mathbf{x}, t) = (2\pi)^2\phi_0 \cos (\mathbf{q}\cdot \mathbf{x}) \delta(t).
\end{equation}
The above perturbation induces a time-dependent change in the charge density $\delta\rho(\mathbf{x},t)$ which can be obtained from:
\begin{eqnarray}
\delta\rho(\mathbf{x}, t) &=& -e\int\frac{ {\rm d}^2k}{(2\pi)^2}\int_{-\infty}^\infty \frac{{\rm d}\Omega}{(2\pi)} e^{i(\mathbf{k}\cdot \mathbf{x}-\Omega t)}\mathcal{L}^{\rm imp}_R(k,\Omega) \phi^{\rm ext}(\mathbf{k}, \Omega) \nonumber\\
&=& -e\phi_0(2\pi)^{-1}\cos(\mathbf{q}\cdot \mathbf{x})\int_{-\infty}^\infty {\rm d}\Omega\, e^{-i\Omega t}\mathcal{L}^{\rm imp}_R(q,\Omega) \nonumber\\
&=& -e\phi_0(2\pi)^{-1} \cos(\mathbf{q}\cdot \mathbf{x})\int_{-\infty}^\infty {\rm d}\Omega\,\left({\rm Re}\mathcal{L}^{\rm imp}_R(q,\Omega) \cos(\Omega t) - {\rm Im}\mathcal{L}^{\rm imp}_R(q,\Omega) \sin(\Omega t)\right) \nonumber\\
&=& -2e\phi_0(2\pi)^{-1} \cos(\mathbf{q}\cdot \mathbf{x})\int_0^\infty {\rm d}\Omega\,\left({\rm Re}\mathcal{L}^{\rm imp}(q,\Omega) \cos(\Omega t) - {\rm Im}\mathcal{L}^{\rm imp}(q,\Omega) \sin(\Omega t)\right) .
\end{eqnarray}
Due to the translational symmetry of the system, the spatial variation of the response $\delta\rho(\mathbf{x}, t)$ at a fixed moment of time $t$ is the same as that of the external driving $\phi^{\rm ext}$, i.e. of the form $\cos(\mathbf{q}\cdot \mathbf{x})$. Therefore, in order to study the time-dependent response, we can set $\mathbf{x} = 0$ for any fixed value of $q$ and study
\begin{eqnarray}
\delta\rho(\mathbf{x} = 0, t) &=&-2e\phi_0(2\pi)^{-1} \int_0^\infty {\rm d}\Omega\,\left({\rm Re}\mathcal{L}^{\rm imp}(q,\Omega) \cos(\Omega t) - {\rm Im}\mathcal{L}^{\rm imp}(q,\Omega) \sin(\Omega t)\right) .
\end{eqnarray}
It is clear from the above expression that poles in $\mathcal{L}^{\rm imp}_R(\Omega)$ at a fixed value of $q$ will lead to a characteristic oscillation in the induced $\delta\rho(\mathbf{x} = 0, t)$ -- this will be damped if the pole lies far from the imaginary axis. Also it is clear from the definition of $\mathcal{L}^{\rm imp}$ provided in Eq. \eqref{Limp} that a pole of $\mathcal{L}^{\rm imp}$ can arise only from the vanishing of the denominator, i.e. at $\Omega(q) =  \Omega_q - i \gamma_q$ where
\begin{equation}\label{denom}
1 = V(q)\,\mathcal{L}_R(q, \Omega_q - i \gamma_q ).
\end{equation}
When $\gamma_q \ll \Omega_q$, we will call the pole a \textit{proper pole}. In this case, the real and imaginary parts of the above equation can be separated so that, expanding in $\gamma_q$, we can demand
\begin{eqnarray}\label{pp}
1 &=&  V(q){\rm Re}\mathcal{L}_R(q, \Omega_q) \, \,=\,\,  V(q){\rm Re}\mathcal{L}(q, \Omega_q), \nonumber\\
\gamma_q &=& \frac{{\rm Im}\mathcal{L}_R(q, \Omega_q)}{\frac{\partial{\rm Re}\mathcal{L}_R(q, \Omega)}{\partial\Omega}\vert_{\Omega_q}}\, \,=\,\, \frac{{\rm sgn}(\Omega_q){\rm Im}\mathcal{L}(q, \Omega_q)}{\frac{\partial{\rm Re}\mathcal{L}(q, \Omega)}{\partial\Omega}\vert_{\Omega_q}}.
\end{eqnarray}
A proper pole leads to an easily discernible oscillation pattern of $\delta\rho(\mathbf{x} = 0, t)$ with small damping. When $\gamma_q \neq 0$, $\Omega_q$ can be identified numerically as the value of $\Omega$ at which $-{\rm Im}\mathcal{L}^{\rm imp}(\Omega)$ has a narrow (Lorentzian) peak that also coincides with a zero of ${\rm Re}\mathcal{L}^{\rm imp}(\Omega)$ at a fixed value of $q$.  When $\gamma_q = 0$, ${\rm Re}\mathcal{L}^{\rm imp}(\Omega)$ diverges at $\Omega = \Omega_q$ while ${\rm Im}\mathcal{L}^{\rm imp}(\Omega)$ becomes a Delta function. For numerical purposes however, ${\rm Im}\mathcal{L}^{\rm imp}(\Omega)$ becomes a vanishing function.

\begin{figure}[htbp]
\centering
\includegraphics[width=0.9\textwidth]{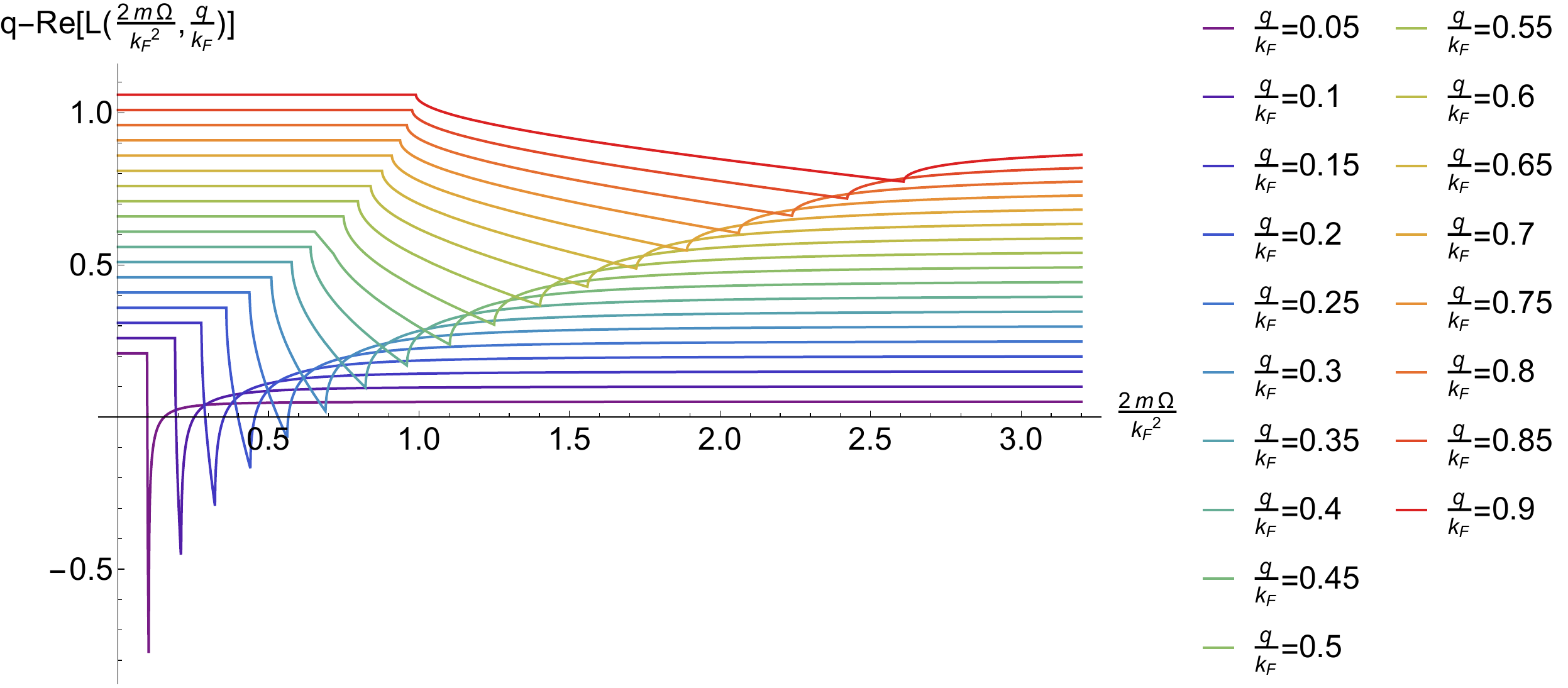}
\caption{We see that in the case of the Fermi liquid, $q -{\rm Re}\mathcal{L}^{\rm imp}(\Omega, q)$ has two zeroes for each value of $q< q_{\rm crit} \approx 0.25 k_F$. }
\label{fig:ReLzeroes}
\end{figure}

Let us first study the case of the Fermi liquid. We set $e^2/(2\epsilon_b) = 1$ for convenience so that for the case of Coulombic interactions $V(q) = 1/q$. To locate proper poles, we need to study first the zeroes of $X(q) = q - {\rm Re}\mathcal{L}^{\rm imp}(\Omega, q)$ as clear from Eq. (\ref{pp}). As shown in Fig. \ref{fig:ReLzeroes}, we find that for $q < q_{\rm crit} \approx 0.25 k_F$, for each value of $q$ there exist two zeroes of $X(q)$. 

\begin{figure}[ht]
\begin{minipage}[b]{0.8\linewidth}
\centering
\includegraphics[width=\textwidth]{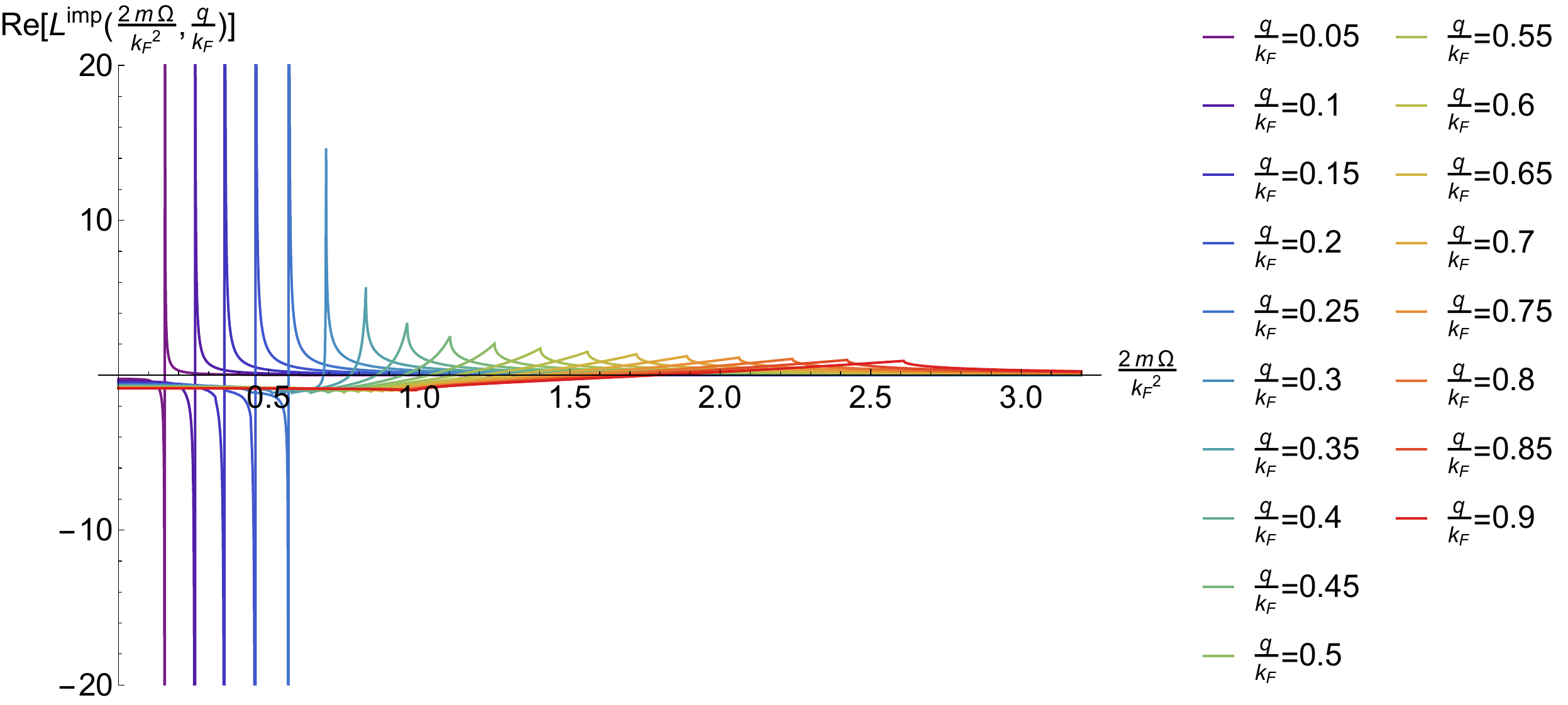}
\caption{\label{fig:ReLimpFL} ${\rm Re}\mathcal{L}^{\rm imp}(\Omega, q)$ for the Fermi liquid as a function of $\Omega$ for various values of $q$.}
\end{minipage}
\hspace{0.5cm}
\begin{minipage}[b]{0.8\linewidth}
\centering
\includegraphics[width=\textwidth]{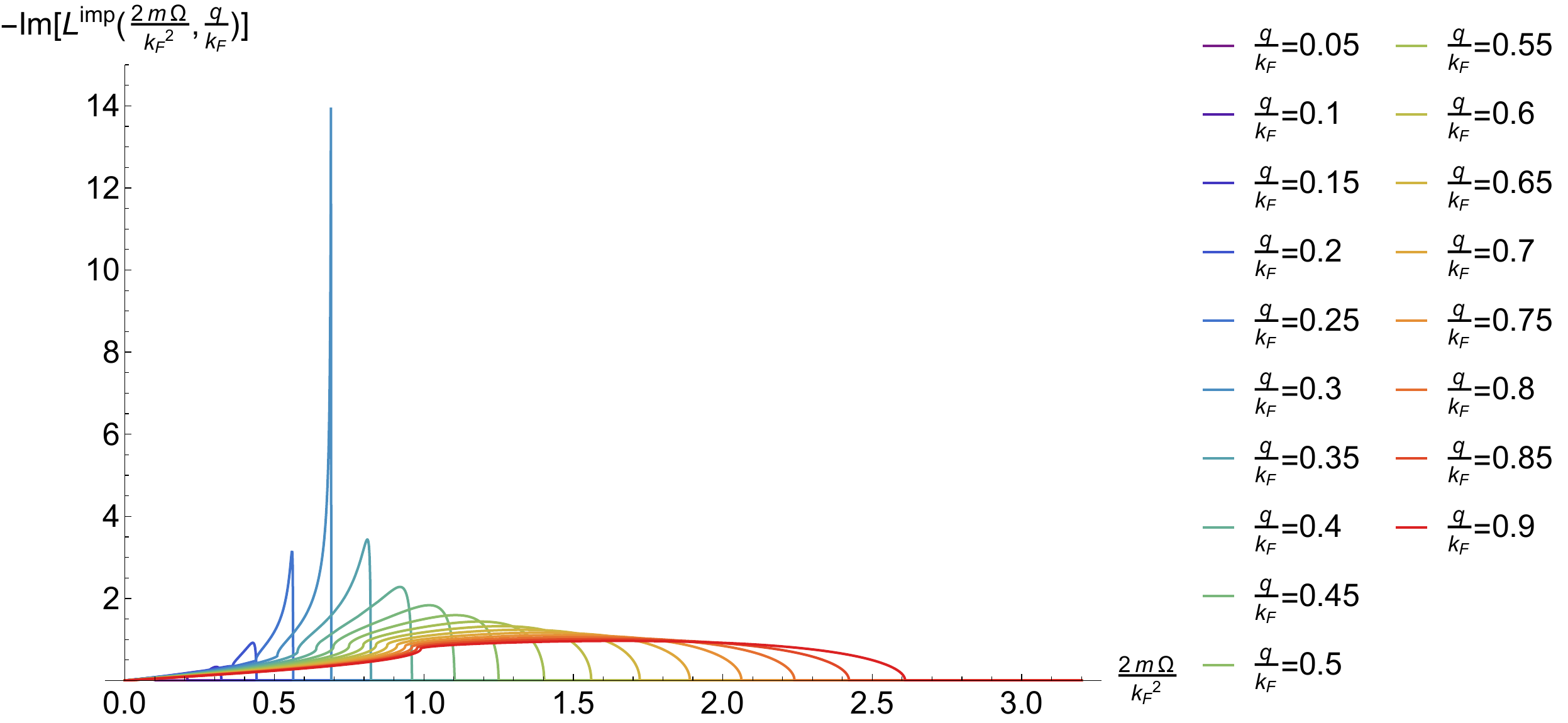}
\caption{\label{fig:ImLimpFL} $-{\rm Im}\mathcal{L}^{\rm imp}(\Omega, q)$ for the Fermi liquid as a function of $\Omega$ for various values of $q$. Note that $-{\rm Im}\mathcal{L}^{\rm imp}(\Omega, q)$ never diverges in the continuum. The delta function peaks outside the continuum corresponding to the plasmonic poles where ${\rm Re}\mathcal{L}^{\rm imp}(\Omega, q)$ also diverges are not shown in the above plots due to numerical limitations.}
\end{minipage}
\end{figure}

However, at the smaller zero $-{\rm Im}\mathcal{L}^{\rm imp}(\Omega)$ turns out to be large giving a large $\gamma_q > \Omega_q$ (note  Eq. (\ref{pp}) then is strictly not valid here), and at the larger zero ${\rm Im}\mathcal{L}^{\rm imp}(\Omega)$ vanishes so that $\gamma_q = 0$ unless $q$ is very near $q_{\rm crit}$. Therefore, the smaller zero of $X(q)$ is not a proper pole, but the larger zero of $X(q)$ is one with $\gamma_q = 0$ and lies mostly \textit{outside} of the particle-hole continuum except when $q \approx q_{\rm crit}$. This is also vindicated in plots shown in Figs. \ref{fig:ReLimpFL} and \ref{fig:ImLimpFL} where one sees that ${\rm Re}\mathcal{L}^{\rm imp}(\Omega)$ diverges at $\Omega_q$ corresponding to the second zero of $X(q)$ for each $q \ll q_{\rm crit}$. Furthermore, $\Omega_q$ lies outside the domain (the particle-hole continuum) where $-{\rm Im}\mathcal{L}^{\rm imp}(\Omega)$ has a support implying $\gamma_q = 0$. Although $-{\rm Im}\mathcal{L}^{\rm imp}(\Omega)$ is a delta function at $\Omega_q$ when $\gamma_q = 0$, numerically however it appears as a vanishing function outside the continuum. This pole $\Omega_q$ which is non-dissipative (since $\gamma_q = 0$) in the RPA approximation is also called the \textit{plasmon} pole and can be shown to have a dispersion relation $\propto\sqrt{q}$ for small $q$. Near $q=q_{\rm crit}$, $\gamma_q$ is non-zero and leads to Landau damping. Here, however $\gamma_q$ is not small and therefore the corresponding peak in $-{\rm Im}\mathcal{L}^{\rm imp}(\Omega)$ is also non-Lorentzian.

These features are expected at weak coupling where typically the collective excitation pole does not co-exist with the continuum. The gapless on-shell particle-hole excitations of the Fermi surface constituting the continuum lead to significant broadening of the collective excitation whose presence becomes hard to detect.

Remarkably, the case of the semi-holographic non-Fermi liquid presents a somewhat different scenario. Naively, since the spectral weight never vanishes for $\Omega \neq 0$ due to presence of incoherent excitations, we do not expect any sharply defined collective excitation of the system to exist. Nevertheless, we have seen in the previous sections that although the continuum in the semi-holographic non-Fermi liquid cannot be thought of as particle-hole type excitations outside of the inner core region, the kinematically determined boundaries denoted as $\Omega_{\rm max}(q)$ and $\Omega_{\rm min}(q)$ (see Fig. \ref{fig:kinematic}) are preserved despite blurring of the sharp Fermi liquid features. We will see below that the plasmonic excitations arise in the semi-holographic non-Fermi liquid because both the real and imaginary parts of the generalised Lindhard functions behave in the right way in the low frequency tail of the continuum for $q > 2k_F$.

\begin{figure}[htbp]
\centering
\includegraphics[width=0.9\textwidth]{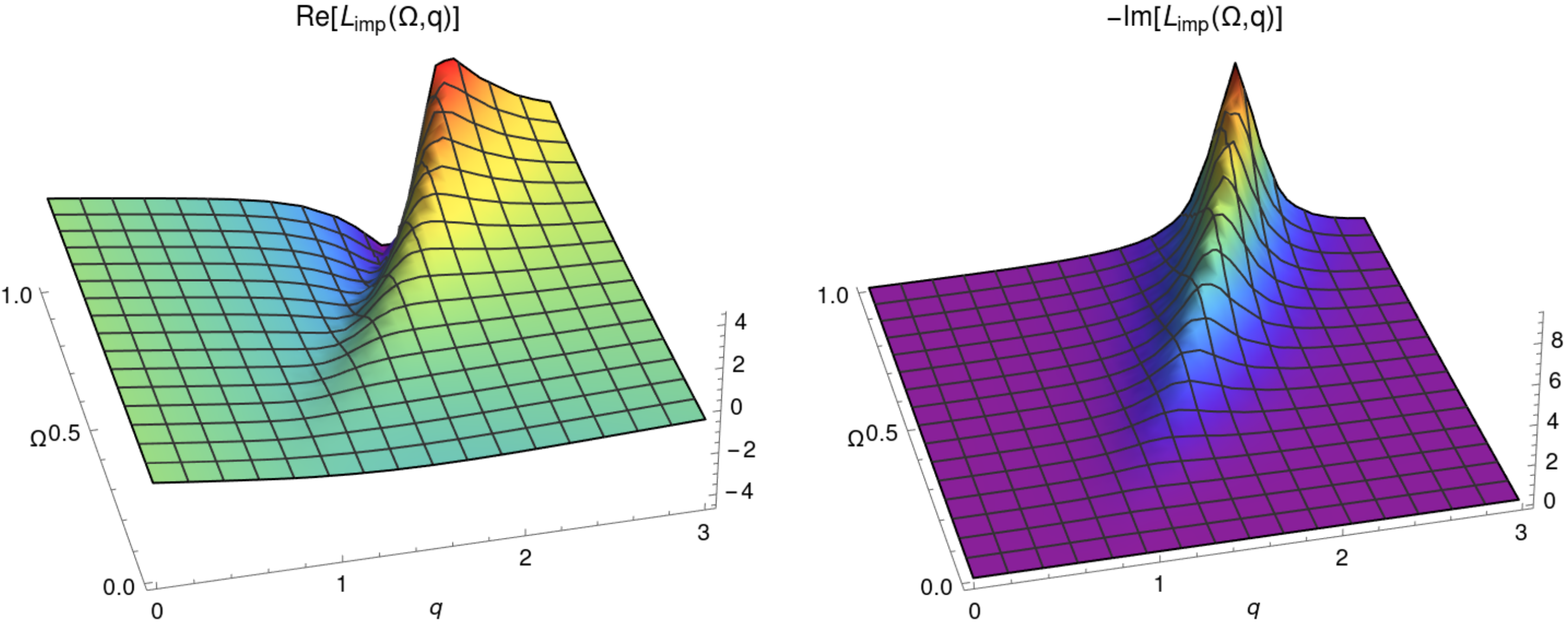}
\caption{Three-dimensional plots of real and imaginary parts of $\mathcal{L}^{\rm imp}(q, \Omega)$ for $e^2/(2 \epsilon_b)= 1$, $\nu = 2/3$, $k_F = 0.4$, ${\rm arg}(\zeta) = \pi/4$ and all other parameters as in the previous section.}
\label{fig:Limp3D}
\end{figure}

Following our earlier discussion, we can readily identify the proper poles of $\mathcal{L}^{\rm imp}$ in the three-dimensional plots of its real and imaginary parts presented in Fig. \ref{fig:Limp3D} from the Lorentzian peaks of $-{\rm Im}\mathcal{L}^{\rm imp}$ which coincide with zeroes of ${\rm Re}\mathcal{L}^{\rm imp}$. We have chosen $e^2/(2\epsilon_b)= 1$, $\nu = 2/3$, $k_F = 0.4$, ${\rm arg}(\zeta) = \pi/4$ and all other parameters as in the previous section. We find that we can clearly identify a proper pole (with narrow width) for each value of $q > q_{\rm crit}\approx 1.4$ which lies \textit{away} from the inner core region. 

\begin{figure}[htbp]
\centering
\includegraphics[width=0.85\textwidth]{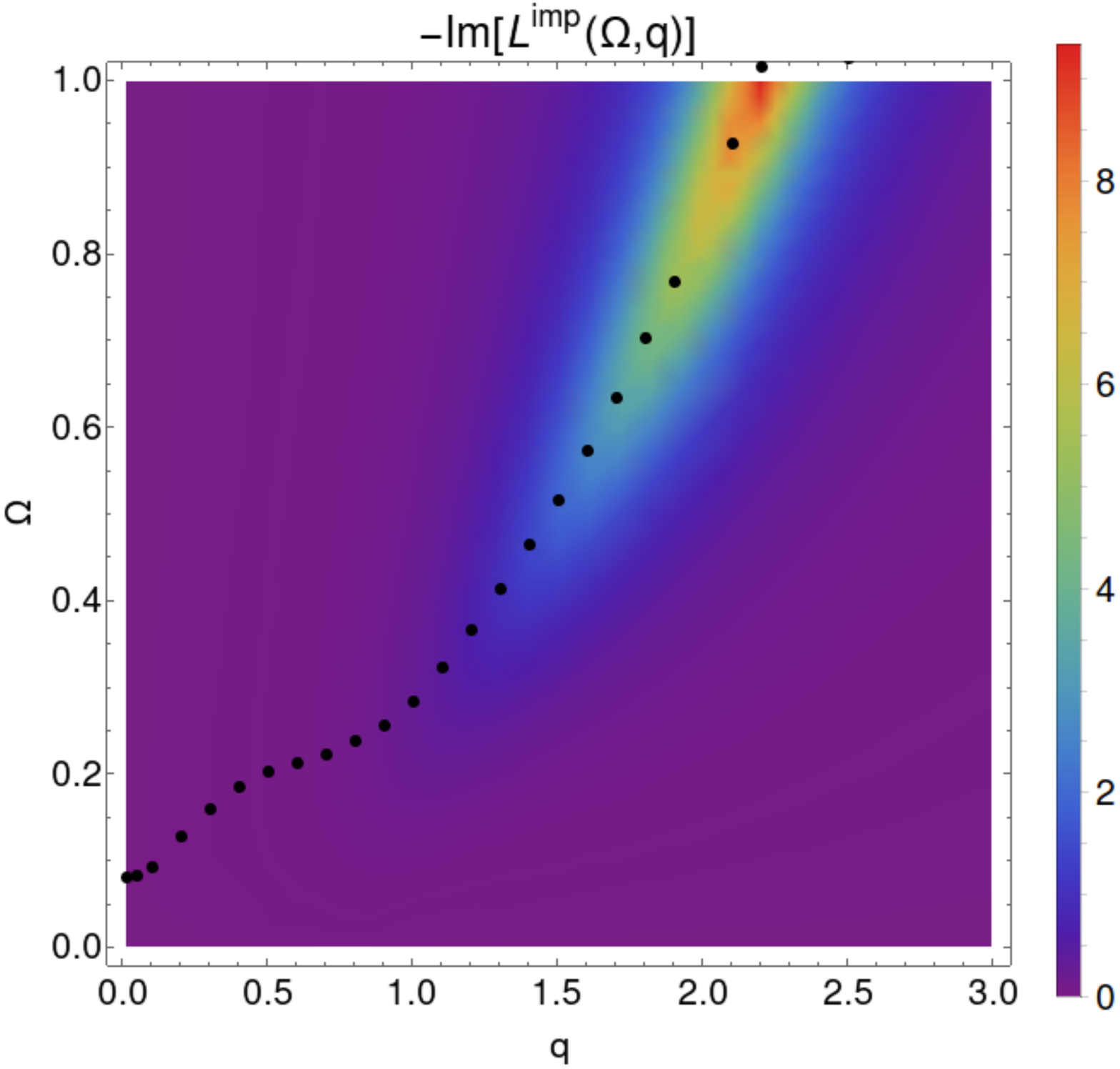}
\caption{A contour plot of ${\rm Im}\mathcal{L}^{\rm imp}(q, \Omega)$ is shown with the black dots indicating the maxima.}
\label{fig:ImLimpDensityPlot}
\end{figure}

For a better understanding of the plasmonic pole, we can refer back to Fig. \ref{fig:ReLq} where ${\rm Re}\mathcal{L}(\Omega, q)$ has been plotted as a function of $q$ respectively for various fixed values of $\Omega < \omega_{\rm c}$. Firstly, it is not hard to see from Fig. \ref{fig:ReLq} that for $\Omega < 0.2$, ${\rm Re}\mathcal{L}(\Omega, q)$ is not appreciably positive in a sufficiently large range of $q$ and that a solution for $X(q) = q - {\rm Re}\mathcal{L}(\Omega, q) = 0$ cannot exist.  However, for sufficiently large $\Omega$, it is also clear from Fig. \ref{fig:ReLq} that ${\rm Re}\mathcal{L}(\Omega, q)$ is positive definite and a solution to $X(q) = q - {\rm Re}\mathcal{L}(\Omega, q) = 0$ does exist.\footnote{The reader can readily see that the line $y= q$ will intersect the curves ${\rm Re}\mathcal{L}(\Omega, q)$.} The contour plot of $-{\rm Im}\mathcal{L}^{\rm imp}(\Omega, q)$ where we have marked the peaks with black dots is shown in Fig. \ref{fig:ImLimpDensityPlot}. It is clear from this plot that well-defined plasmonic poles can be found only above a threshold energy of about $\Omega = 0.2$ corresponding to what we have just inferred from Fig. \ref{fig:ReLq}. Obviously this also implies that plasmonic poles exist only for $q > q_{\rm crit}$ with $q_{\rm crit} \approx 1.4$.

We can also understand why the plasmonic poles are well-defined by referring back to Fig. \ref{fig:ImLq} where ${\rm Im}\mathcal{L}(\Omega, q)$ has been plotted as a function of $q$ respectively for various fixed values of $\Omega < \omega_{\rm c}$. Firstly for a fixed $\Omega > 0.2$, it is easy to identify from Fig. \ref{fig:ImLimpDensityPlot} the value $q_p(\Omega)$ corresponding to a well defined plasmonic pole where $\Omega= \Omega_{q_p}$. As discussed above $q_p > q_{\rm crit} \approx 1.4$ for any $\Omega$. One can verify from Fig. \ref{fig:ImLq} that ${\rm Im}\mathcal{L}(\Omega, q_p(\Omega))$ is small, and furthermore $q_p(\Omega) > q_{\rm min}(\Omega)$, where $q_{\min}$ denotes the value of $q$ for which $\Omega = \Omega_{\rm min} (q_{\rm min})$, the (blurred) outer edge of the kinematically determined continuum where ${\rm Im}\mathcal{L}(q)$ changes its curvature rapidly and becomes a rapidly decaying function. We can readily conclude from here that the plasmonic poles $\Omega_q$ should satisfy $\Omega_q < \Omega_{\rm min}(q)$, i.e. they must lie in the low frequency tail of the blurred outer edge of the continuum (please refer to Fig. \ref{fig:kinematic} for quick visualisation). Since $\Omega_{\rm min}(q)$ exists only for $q > 2 k_F$, clearly the threshold $q_{\rm crit}$ should also satisfy $q_{\rm crit} > 2 k_F$ as indeed is the case \footnote{By looking at fig. \ref{fig:ReLqnu} we can see that there is some value of $\nu$, between 2/3 and 7/8, at which ${\rm Re} \mathcal{L}(\Omega=0.01, q)$ is close to zero around $q=0$. In this cases there will be another solution to $X(q)=0$ at small $q$ and $\Omega$, so there can be another plasmonic mode at low frequency. However this is much more damped than the one we have discussed.}. 

The plasmonic poles are thus a \textit{a relatively low frequency and high momentum feature} of the dynamics and our arguments above indicate that they should exist in generic semi-holographic non-Fermi liquid models. This contrasts with the 2D Fermi liquid case where the plasmons are low momentum (but also low energy) features of the dynamics. From Fig. \ref{fig:ImLimpDensityPlot}, we also see that these poles (corresponding to the well-defined peaks in this figure) satisfy an approximately linear dispersion relation.

It is also clear from Fig. \ref{fig:ImLimpDensityPlot} that for $q < 1.4$, the peaks of $-{\rm Im}\mathcal{L}^{\rm imp}(\Omega)$ are very broad and reminiscent of Landau damping features of the Fermi liquid. For $q<1.4$, we cannot really solve Eq. (\ref{denom}) but we identify the peaks by minimising the modulus of the denominator of $\mathcal{L}^{\rm imp}$, i.e.
\begin{equation*}
\vert X(q)\vert = \left(q-  {\rm Re}\mathcal{L}_R(q, \Omega)\right)^2 +  \left({\rm Im}\mathcal{L}_R(q, \Omega)\right)^2
\end{equation*}
We also note that for $q < 1.4$, the peaks of $-{\rm Im}\mathcal{L}^{\rm imp}$ are non-dispersive meaning that $\partial \Omega_q/\partial q$ is small compared to the average width of the peaks.

\begin{figure}[htbp]
\centering
\includegraphics[width=0.85\textwidth]{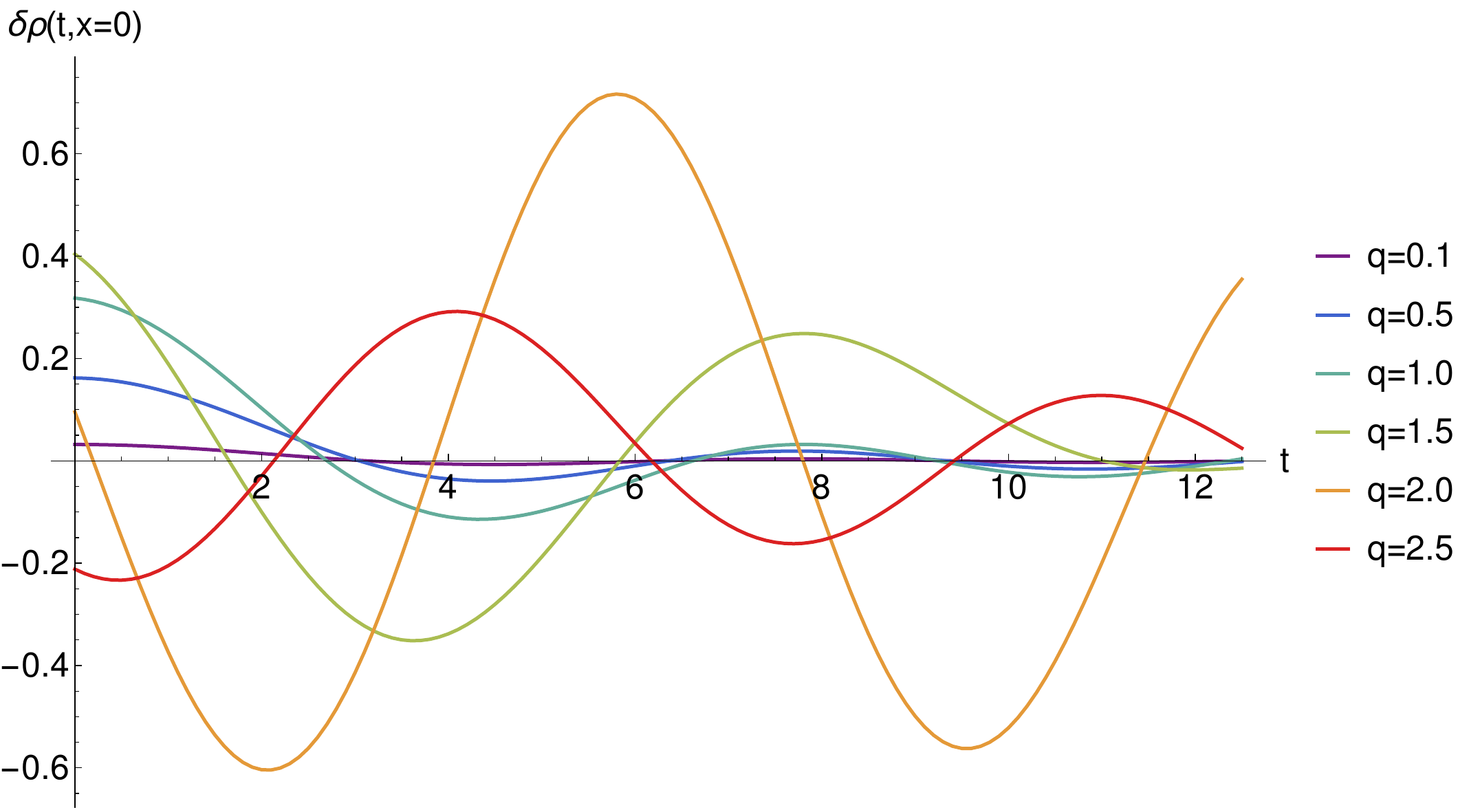}
\caption{The response of the induced $\delta\rho(t)$ at $\mathbf{x}= 0$ for the semi-holographic non-Fermi liquid.}
\label{fig:rhot}
\end{figure}

The corresponding plots for $\delta\rho(t)$ in Fig. \ref{fig:rhot} at $\mathbf{x}= 0$ clearly demonstrate the behavior of the plasma oscillations that result from these features of $\mathcal{L}^{\rm imp}$; one can see that for $q < 1.4$, the oscillation frequencies are nearly independent of $q$ (for $q <1$ they appear to be decaying but this is a cut-off dependent effect). For $q > 1$, the oscillations are damped only slightly over the time-scale of oscillation with the oscillation frequency growing with $q$ approximately in a linear fashion. 
It is to be noted that ${\rm Im}\mathcal{L}^{\rm imp}(\Omega, q)$ can be measured separately via electron energy-loss spectroscopy \cite{FetterWalecka}. Therefore, one can also confirm the existence of the unusual semi-holographic plasmonic poles through this method.

\subsection{Dynamic screening and possible pairing instability}
Dynamic screening can be studied via the effective screened potential $V_s(q, \Omega)$ defined in \eqref{Vs}. For a better physical understanding, it is more useful to study $V_s(\Omega, r)$ defined as the $2-$D Fourier transform of $V_s(q,\Omega)$, i.e.
\begin{equation}
V_s(\Omega, r) = (2\pi)^{-2}\int_0^{2\pi} {\rm d\theta} \int_0^\infty {\rm d}q\,\, qe^{i qr\cos(\theta)} V_s(q, \Omega).
\end{equation}

\begin{figure}[htbp]
\centering
\includegraphics[width=0.85\textwidth]{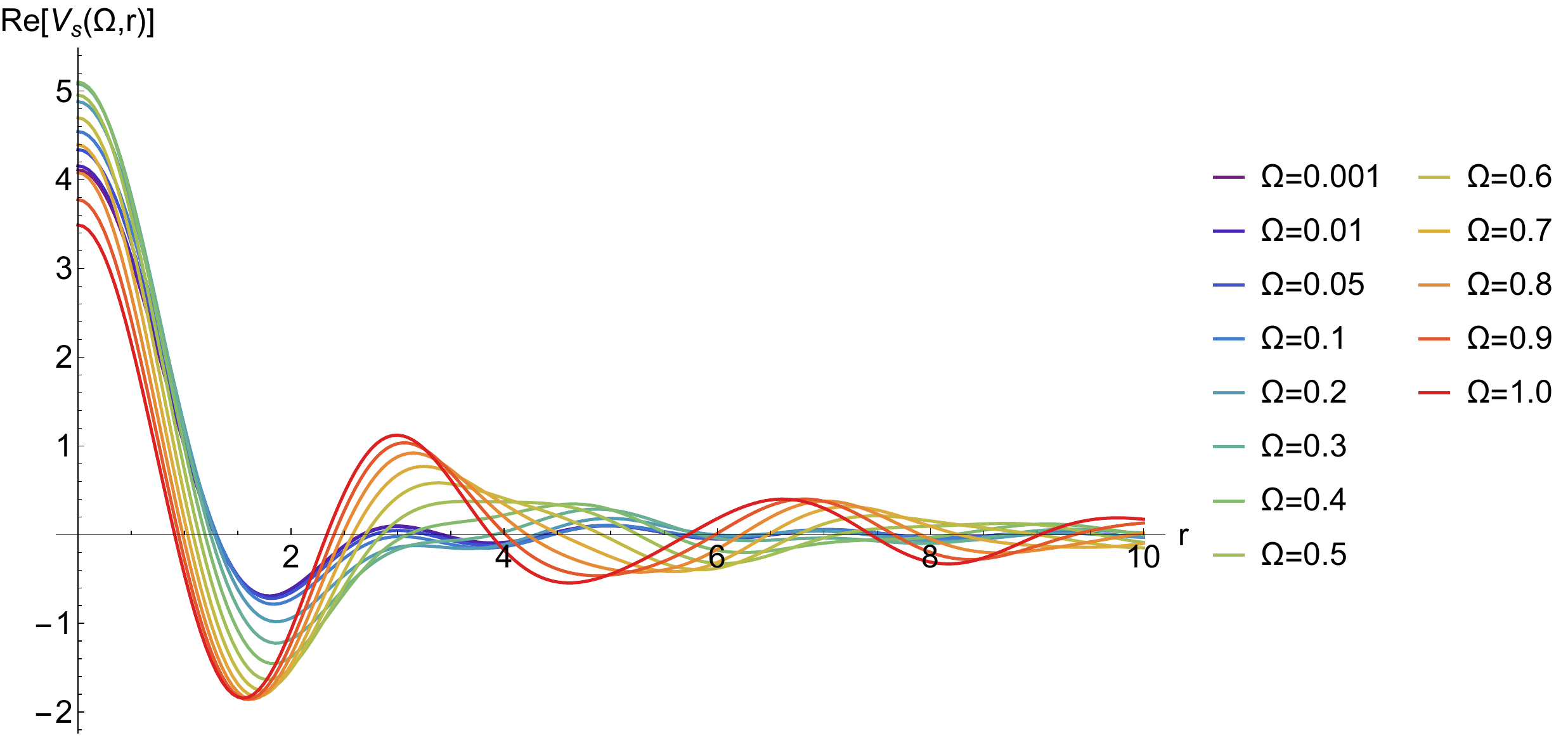}
\caption{${\rm Re}V_s(\Omega,r)$ shown as a function of $r$ for various fixed values of $\Omega$.}
\label{fig:ReVr}
\end{figure}

\begin{figure}[htbp]
\centering
\includegraphics[width=0.85\textwidth]{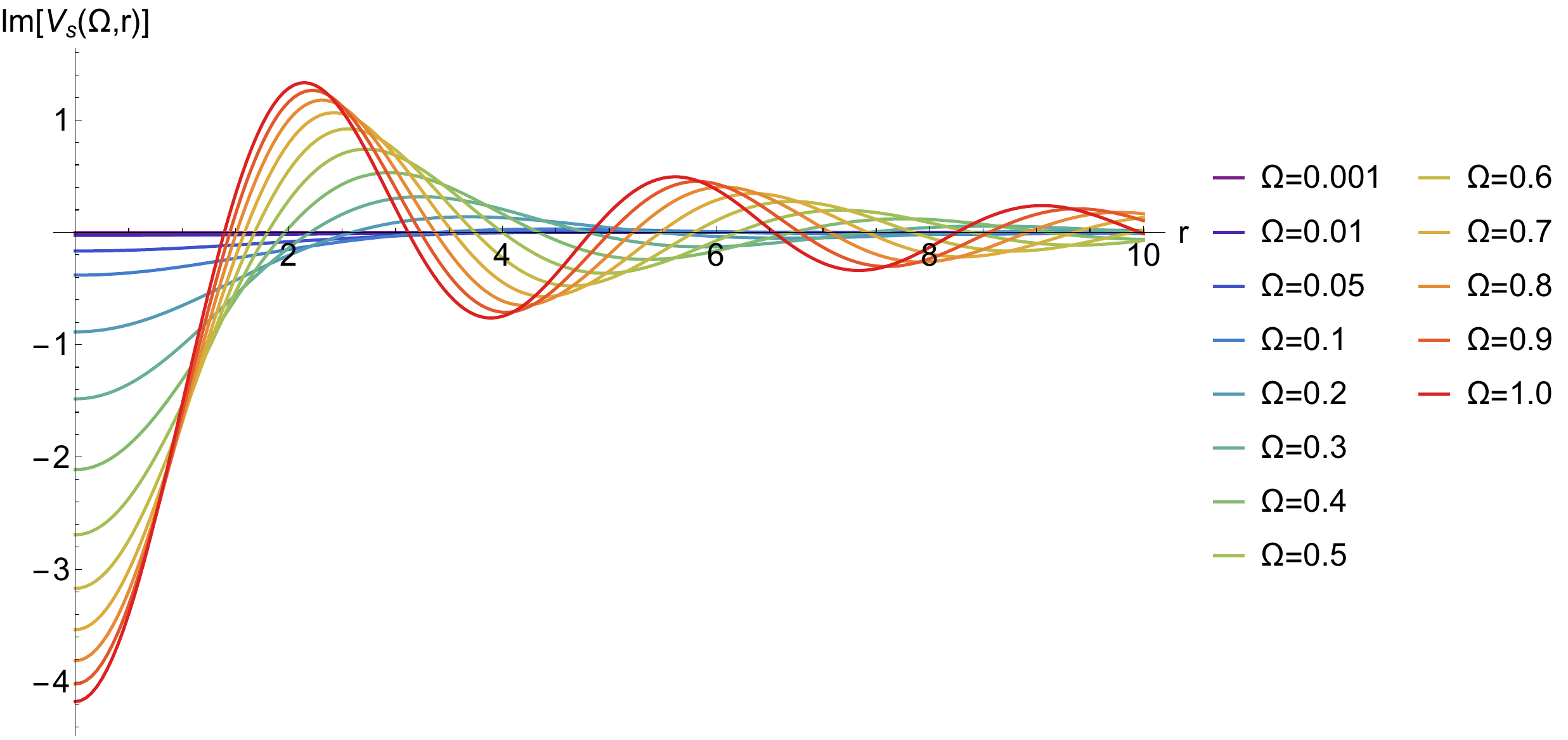}
\caption{${\rm Im}V_s(\Omega,r)$ shown as a function of $r$ for various fixed values of $\Omega$.}
\label{fig:ImVr}
\end{figure}

The plots in Fig. \ref{fig:ReVr} and \ref{fig:ImVr} present the real and imaginary parts of $V_s(\Omega, r)$ for $\nu = 2/3$ and other parameters exactly the same as in the above subsection. Note we cannot trust the computations for small values of $r$ because of our semi-holographic effective model is an effective infrared description. 

The above plots show that for large values of $\Omega$, ${\rm Re}V_s(\Omega,r)$ develops substantially deep wells where it becomes \textit{attractive} -- these wells become \textit{deeper} as the value of $\Omega$ \textit{increases}. Note although ${\rm Im}V_s(\Omega,q) < 0$ for all values of $\Omega$ and $q$ implying we cannot have a runaway linear response of the system to the influence of externally introduced moving charges,  ${\rm Im}V_s(\Omega,r)$ also alternate in sign not exactly in phase with ${\rm Re}V_s(\Omega,r)$. The latter implies that some part of the attractive regions of ${\rm Re}V_s(\Omega,r)$ may not decay and can lead to production of sufficiently long-lived pairs. Therefore, the system can have a non-linear instability particularly in the large $\Omega$ region, i.e. when the externally introduced charges are subjected to oscillations at time-scales comparable to those of the system. We will study the static limit soon where we will find usual but suppressed Friedel oscillations.

The above features of $V_s(\Omega,r)$ are unexpected and they also point towards a novel dynamical mechanism of pair formation from pure effective electronic interactions (for electrons living in a constrained environment imposed by the lattice) without the need for mediation via phonons (i.e. excitations of the lattice itself). Since the quasi-particle type excitations in our semi-holographic effective framework are not sharply defined, they can have substantial spectral weight in large $\Omega$ region for the stable attractive nature of $V_s(\Omega,r)$ to lead to pair formation with long lifetimes. For this mechanism to work, there needs to be sufficient spectral weight of each member in the pair in the large $\Omega$ region whilst the attractive wells of $V_s(\Omega,r)$ need to be sufficiently deep. Therefore, to see whether this mechanism can lead to an unconventional superconducting instability of the system, we need to first examine the dynamical pair susceptibility and then how it is modified by the effective dynamical Coulomb interactions as examined carefully in \cite{Sham} in the context of Fermi liquid. We leave this study for the future. At this point, we merely observe that just as in case of induced plasma oscillations via $\mathcal{L}^{\rm imp}$, the breakdown of quasi-particle picture and the presence of particle-hole asymmetry can together contribute to a novel mechanism of pair formation which cannot exist in a weakly coupled system. A similar argument for a \textit{mid-infrared} scenario where the plasmonic pole can play an important dynamical role has been presented by Leggett earlier in order to explain experimental data \cite{Leggett}.
 
\subsection{Static limit and Friedel-like oscillation}
\begin{figure}[htbp]
\centering
\includegraphics[width=0.85\textwidth]{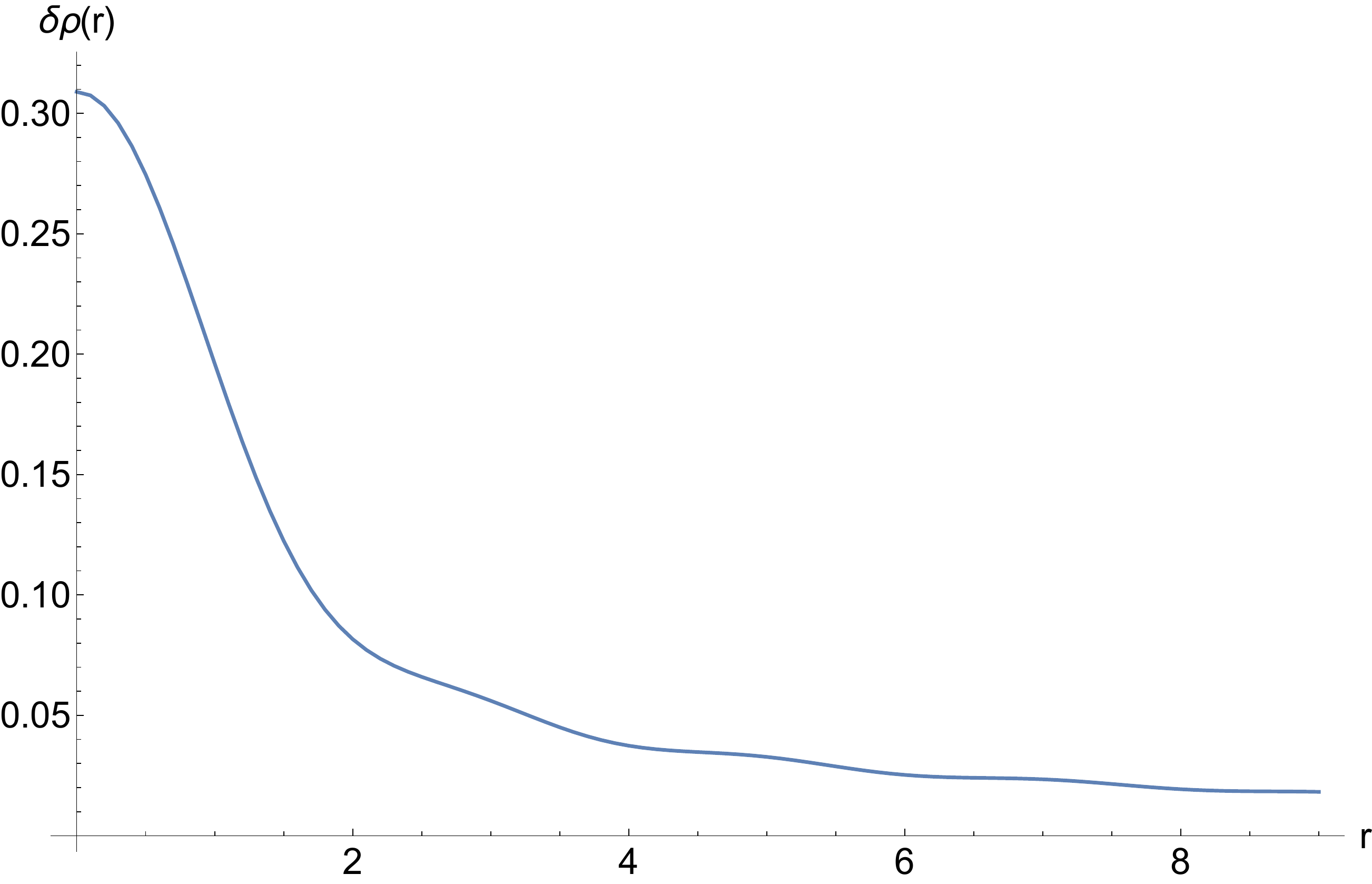}
\caption{The induced charge density as a result of the introduction of static charge shows suppressed Friedel oscillation.}
\label{fig:Friedel}
\end{figure}

Consider the introduction of a static impurity charge $-Ze$ into the system. The change in the electronic density induced by this charge is given by:
\begin{eqnarray}
\delta \rho(r) &=& -(2\pi)^{-2}Z\int_0^\infty {\rm d}q\int_0^{2\pi}{\rm d}\theta \, q e^{i qr\cos(\theta)}\mathcal{L}^{\rm imp}(q, \Omega = 0) \frac{e^2}{2\epsilon_b q}\nonumber\\
&=&-(8\pi^2\epsilon_b)^{-1}Ze^2\int_0^\infty {\rm d}q\int_0^{2\pi}{\rm d}\theta \,  e^{i qr\cos(\theta)}\mathcal{L}^{\rm imp}(q, \Omega = 0),
 \end{eqnarray}
where $r$ denotes the radial distance from the external static charge impurity. We have plotted $\delta\rho(r)$ n Fig. \ref{fig:Friedel} in the case of the semi-holographic non-Fermi liquid with $\nu = 2/3$ $k_F = 0.4$, $Z = 1$ and all other parameters the same as in the previous subsections.

It is clear that $\delta\rho(r)$ shows a Friedel type oscillation although we do not have a Kohn singularity in $\mathcal{L}(q, \Omega = 0)$ or $\mathcal{L}^{\rm imp}(q, \Omega = 0)$ at $q= 2k_F$ -- in our case the $q-$derivative of $\mathcal{L}(q, \Omega = 0)$ is large but not divergent. However, the reasonably sharp transition in $\mathcal{L}(q, \Omega = 0)$ at $q= 2k_F$ as visible clearly in Fig. \ref{fig:ReLq} does lead to oscillations in $\delta\rho(r)$ as in a Fermi liquid although these are very suppressed. It is worthwhile to note that $\delta\rho(r)$ is always greater than zero signifying that the oscillations in sign in $V_s(q, \Omega = 0)$ shown in Fig. \ref{fig:ReVr} do not have sufficiently high amplitude to change the sign of $\delta\rho(r)$ in the static limit.

\section{Conclusions and outlook}\label{conclude}

To conclude, we find that although the semi-holographic non-Fermi liquids retain many features of the Fermi liquid particularly in the inner core of the continuum, the phenomenological manifestations are different. The most unexpected finding is the appearance of a well-defined plasmonic collective excitation above a energy (and momentum) threshold set by the boundary of the inner core region and the universal character of the continuum. Furthermore, it has an approximately linear dispersion relation. Such a type of behaviour can arise from a fundamentally new nature of the continuum outside of the inner core region. 

We have also observed that at higher frequencies the dynamic screened potential can lead to formation of pairs with long lifetimes and therefore trigger a superconducting instability. We plan to investigate this in detail in a forthcoming publication. 

Our calculation of the generalised Lindhard function has some similarities with the computation of the DC and optical conductivities presented in \cite{Faulkner:2013bna}. In this work, the authors have used a purely holographic model and have considered the effect of the \textit{bulk} fermionic loop with bulk fermionic propagators that produce non-Fermi-liquid like spectral functions in the dual theory. Even though the bulk loop contribution is $1/N^2$ suppressed compared to the tree level contribution (given by a single photon propagator in the bulk), it turns out that it is responsible for the leading low temperature and low frequency behaviour which is non-analytic and also of a non-Fermi-liquid type. There are several technical differences with our calculations nevertheless: first of all, our computation is at zero temperature and finite momentum, whereas the computation of \cite{Faulkner:2013bna} has been performed at finite temperature and zero momentum. Furthermore, not only the fermionic propagators but also the bulk vertex plays a crucial role in determining the low temperature and low frequency behaviour in the computation of \cite{Faulkner:2013bna}. In our case, the result has been obtained without including the vertex corrections as discussed before. Finally, a certain ad-hoc prescription for the analytic continuation from the Euclidean signature has been utilised in order to derive the real-time bulk loop contribution since it is not known how to compute the bulk loop in the Lorentzian signature directly. In our case we have been able to perform the loop computation employing standard techniques without any need for a specific prescription. Despite all these differences, it would be interesting to see if the sub-leading $1/N^2$ effects in our semi-holographic models arising from bulk loops and vertices can give significant contributions at low temperatures and low frequencies as in case of the holographic computation of \cite{Faulkner:2013bna}.

Another interesting subject for future investigations is the non-equilibrium spectral function and statistical function of the fermionic excitations particularly because the time-evolution of these following a global quench/energy injection can be measured experimentally via a variety of methods. In a recent work, it has been found that the gross features of the non-equilibrium evolution of the holographic spectral function are universal being determined just by the difference between the final and initial temperatures and the quenching/energy-injection time \cite{Joshi:2017ump}. This leads to the exciting possibility that one can construct a very general quantum kinetic theory of the generalised quasi-particles of the semi-holographic non-Fermi liquid which are stable from interactions at the Fermi surface.\footnote{Note at strong coupling the statistical and spectral functions evolve in a non-trivial fashion. The evolution of both of these have to be evaluated separately. For this we need to find a general prescription for obtaining the non-equilibrium statistical function too -- some of the authors are working in this direction.} This quantum kinetic theory can readily include the non-equilibrium evolution of the background geometry representing the holographic IR-CFT degrees of freedom.

The main message of our work is that semi-holographic systems have remarkable features which cannot be obtained purely from a weak coupling kinetic picture or from a strong coupling holographic picture exclusively. Furthermore, these features are robust, generic and surprising. In the context of non-Fermi liquids, we hope that the semi-holographic picture with suitable modifications discussed here can be subjected to experimental confrontation in the near future.

\begin{acknowledgments}
We would like to thank Karsten Held, Niko Jokela, Ville Ker\"anen, Florian Preis, Anton Rebhan, and Aleksi Vuorinen for usefull discussion. C. E. is supported by the Austrian Science Fund (FWF), project no. P27182-N27 and DKW1252-N27. The research of A.M. is supported by a Lise-Meitner fellowship of the Austrian Science Fund (FWF), project no. M 1893-N27.  C.E. thanks the Helsinki Institute of Physics and the Departament de F\'isica Qu\`antica i Astrof\'isica at the University of Barcelona for the warm hospitality while part of this work was completed.
 
\end{acknowledgments}

\appendix 
\section{Spectral representation}\label{nu1by2}

In this appendix we consider some issues that arise because of the lack of integrability of the spectral function; these issues can make the model sensitive to 
the UV completion, but we could not perform a complete analysis; for this reason we mostly restricted our investigation to the case $\nu > 1/2$. 

In section \ref{sec:GenLindFunc} we derived the representation (\ref{LindhardDef42}) for the imaginary part of the Lindhard function as a convolution of spectral functions; the textbook derivation of this relation starts from the spectral representation of the propagator: 
\begin{subequations}
\begin{align}
G_R(\omega, \epsilon) = & \int {A(E, \epsilon) \over \omega - E + i \eta } dE\,, \\
A  (E, \epsilon) & = - {\textrm{Im}} G_R = |E|^{1/2} \left[ \theta(E) {\zeta_2 \over (\zeta_1 E^{1/2} - \epsilon)^2 + \zeta_2^2 E)} + 
\theta(-E) {\zeta_1 \over (\zeta_2 |E|^{1/2} + \epsilon)^2 + \zeta_1^2 |E|)} \right] \,.
\end{align}
\end{subequations}
Inserting this expression in the loop integral, and performing the integration over $\omega$ first, one finds again (\ref{LindhardDef42}). However, the exchange of the $\omega$ and the $E$ integral is not always legitimate; in the case $\nu = 1/2$, one can perform the $\omega$ integration analytically and the difference can be seen explicitly. 

It is simplest to consider the case $\Omega=0$; the loop integral gives   (notice that $\tilde \zeta = i \zeta^*$) 

\begin{eqnarray} \label{zeroomegaint}
\int d\omega & G_F(\omega, \epsilon_1) G_F(\omega, \epsilon_2)=& -2i \int_0^\infty \, d\omega  \left( {1 \over (\zeta \sqrt{\omega} - \epsilon_1)(\zeta \sqrt{\omega} - \epsilon_2)} + 
 {1 \over ( i \zeta^* \sqrt{\omega} + \epsilon_1)(i \zeta^* \sqrt{\omega} + \epsilon_2)} \right)  = \nonumber \\
 &&= - {4 i \over \epsilon_1-\epsilon_2} \int_0^\infty \, du \left(   {u \over \zeta u - \epsilon_1} - {u \over \zeta u - \epsilon_2} - {u \over i \zeta^* u + \epsilon_1} + 
 {u \over i \zeta^* u + \epsilon_2} 
 \right) \nonumber \\
 &&= - {4 \over \epsilon_1-\epsilon_2 } ( \epsilon_1 f(\epsilon_1) - \epsilon_2 f(\epsilon_2) ) 
\end{eqnarray}  
where in the middle line we changed the integration variable to $u = \sqrt{\omega}$, and we have defined in the last line 
\begin{equation}\label{functionf} 
f(\epsilon) = {1 \over \lvert \zeta \lvert^2}\int_0^\infty \, du \left(  {i \zeta^* \over \zeta u - \epsilon} + {\zeta \over i \zeta^* u + \epsilon}
\right) \,. 
\end{equation}
The divergence of the integral in $f(\epsilon)$ is purely real. The imaginary part is convergent and well-defined, and one can easily see that it is in fact independent 
of $\epsilon$, so that the energy dependence cancels in (\ref{zeroomegaint}).  It can be evaluated: 
\begin{equation*} 
\textrm{Im} \,  f = \int_0^\infty \, du {  (\zeta_1 - \zeta_2) (\zeta_1^2 + 4 \zeta_1 \zeta_2 + \zeta_2^2) u^2 -4 \zeta_1 \zeta_2 u - (\zeta_1  - \zeta_2) \over  \lvert \zeta \lvert^2 (\lvert \zeta \lvert^2 u^2 -2 \zeta_1 u + 1 )(\lvert \zeta \lvert^2 u^2 + 2 \zeta_2 u + 1) }   = {\pi \over 2 |\zeta|^2 } \textrm{sin} (2 \phi) \,.
\end{equation*}

Clearly this constant will create an infrared divergence when inserted in the momentum integral, whereas there is no divergence if the $\omega$ integral is exchanged with the $E$ integral. 

This problem appears only for $\nu \leq 1/2$; given the simplicity of the result, it is possible that for  $\nu=1/2$ case one could devise a simple subtraction procedure, but it is not clear if it possible to deal with the $\nu < 1/2$ case.

The problem disappears if we regularize the high-frequency behaviour by taking a crossover to a normal Fermi liquid. This is achieved by taking the retarded propagator as follows: 
\begin{equation}
G_R(\omega, \mathbf{k}) = \frac{1}{\zeta \omega^{\frac{1}{2}} + \omega - \epsilon_ \mathbf{k} } \, .
\end{equation}

In this case, again at $\Omega=0$ we find an expression like (\ref{zeroomegaint}) but with $f$ replaced by  
\begin{equation}\label{functionF}
F(\epsilon) = \int_0^\infty \, du \left(  {i u \over \zeta u + u^2  - \epsilon} - {i u \over i  \zeta^* u + u^2 + \epsilon} \right) = - \frac{\pi}{2} + 2 \, \textrm{Re} \, \int_0^\infty du \, \frac{i u}{\zeta u+u^2-\epsilon}  \,,
\end{equation} 
This is finite, and the imaginary part is zero, as evident from the last expression. 
The same considerations apply at $\Omega \neq 0$, we refrain from giving the expressions that do not add any new insight. 

\bibliographystyle{apsrev4-1}

\end{document}